\documentclass[11pt,urlcolor=blue, linkcolor=blue]{article} 
\usepackage{cite}

\usepackage{float}

\usepackage{multicol}
\usepackage{tablefootnote}
\usepackage{amsmath, amsthm, amssymb,slashed,mathtools}
\usepackage{ifpdf}
\ifpdf
  \usepackage[pdftex]{graphicx}
  \usepackage{epstopdf}
\else
  \usepackage[dvips]{graphicx}
\fi
\parskip=\baselineskip

\usepackage[usenames, dvipsnames]{color}

\allowdisplaybreaks[1]

\numberwithin{equation}{section}

\usepackage{booktabs}

\theoremstyle{definition}

\definecolor{mygray}{gray}{0.6}

\usepackage{upgreek}
\usepackage{bbm}

\newenvironment{myfont}[2][]{\csname#2\endcsname[#1]}{}

\usepackage{slashed}
\usepackage[makeroom]{cancel}
\usepackage[normalem]{ulem}
\usepackage{soul}
\newcommand{\stkout}[1]{\ifmmode\text{\sout{\ensuremath{#1}}}\else\sout{#1}\fi}

\usepackage{sseq}
\usepackage[all,cmtip]{xy}
\usepackage{tikz-cd}
\usepackage{tikz}
\usetikzlibrary{matrix}

\newcommand{\bea}{\begin{eqnarray}}
\newcommand{\eea}{\end{eqnarray}}
\def\be{\begin{equation}}
\def\ee{\end{equation}}

\newcommand{\ii}{\hspace{1pt}\mathrm{i}\hspace{1pt}}

\newcommand{\nn}{\nonumber}

\usepackage{color}
\usepackage[colorlinks,citecolor=blue]{hyperref}
\definecolor{red}{rgb}{1,0,0}
\definecolor{blue}{rgb}{0,0,1}
\definecolor{dblue}{rgb}{0,0,0.4}
\definecolor{green}{rgb}{0,1,0}
\definecolor{black}{rgb}{0,0,0}
\definecolor{white}{rgb}{1,1,1}

\definecolor{brn}{rgb}{.8,.4,.0}
\definecolor{redo}{rgb}{1,.5,.0}
\definecolor{ddgrn}{rgb}{0,0.4,0}
\definecolor{dgrn}{rgb}{0,0.55,0}
\definecolor{dbl}{rgb}{0,0,0.5}

\usepackage[bbgreekl]{mathbbol}
\usepackage{amscd}

\newcommand{\Z}{\mathbb{Z}}

\newcommand{\dd}{\hspace{1pt}\mathrm{d}}

\newcommand{\Ref}[1]{Ref.~\cite{#1}}
\newcommand{\Eq}[1]{(\ref{#1})}

\newcommand{\Eqn}[1]{Eq.~(\ref{#1})}

\newcommand{\up}{\uparrow} 
\newcommand{\down}{\downarrow}

\newcommand{\bpm}{\begin{pmatrix}}
\newcommand{\epm}{\end{pmatrix}}
\newcommand{\bmm}{\begin{matrix}}
\newcommand{\emm}{\end{matrix}}

\def\Z{{\mathbb{Z}}}

\def \H{\operatorname{H}}

\def \Z{\mathbb{Z}}
\def \Pin{\mathrm{Pin}}
\def \A{\mathcal{A}}

\def\Ext{\operatorname{Ext}}

\newcommand{\Sec}[1]{Sec.~\ref{#1}}

\usetikzlibrary{decorations.markings}
\usetikzlibrary{positioning}
\usetikzlibrary{shadings}

\newcommand{\PSU}{{\rm PSU}}

\newcommand{\SO}{{\rm SO}}
\newcommand{\Spin}{{\rm Spin}}
\newcommand{\U}{{\rm U}}
\newcommand{\SU}{{\rm SU}}

\renewcommand{\O}{{\rm O}}

\def\Sq{\mathrm{Sq}}

\def\B{\mathrm{B}}

\usepackage{datetime}
\usepackage{enumitem} 
\usepackage{moreenum}

\newcommand{\sharpfootnote}[1]{%
\let\oldthefootnote=\thefootnote%
\stepcounter{mpfootnote}%
\addtocounter{footnote}{-1}%
\renewcommand{\thefootnote}{{$\sharp$}}
\footnote{#1}%
\let\thefootnote=\oldthefootnote%
}

\newcommand{\naturalfootnote}[1]{%
\let\oldthefootnote=\thefootnote%
\stepcounter{mpfootnote}%
\addtocounter{footnote}{-1}%
\renewcommand{\thefootnote}{{$\natural$}}
\footnote{#1}%
\let\thefootnote=\oldthefootnote%
}

\newcommand{\Wplusfootnote}[1]{%
\let\oldthefootnote=\thefootnote%
\stepcounter{mpfootnote}%
\addtocounter{footnote}{-1}%
\renewcommand{\thefootnote}{{W$^+$}}
\footnote{#1}%
\let\thefootnote=\oldthefootnote%
}

\newcommand{\Wminusfootnote}[1]{%
\let\oldthefootnote=\thefootnote%
\stepcounter{mpfootnote}%
\addtocounter{footnote}{-1}%
\renewcommand{\thefootnote}{{W$^-$}}
\footnote{#1}%
\let\thefootnote=\oldthefootnote%
}

\usepackage{comment}
\def\bZ{{\mathbf{Z}}}

\usepackage{esint}

\newcommand{\Fig}[1]{Fig.~\ref{#1}} 
\begin{document}
\begin{titlepage}
\begin{flushright}
\end{flushright}
\begin{center}

{\bf
 \LARGE{Higher Anomalies, Higher Symmetries, 
 and\\[3.75mm]  
 Cobordisms III:}\\[4.75mm]
\LARGE{
QCD Matter Phases Anew}\\[5.75mm]
}

\vskip0.5cm 
\Large{Zheyan Wan$^1$\Wminusfootnote{e-mail: {\tt wanzheyan@mail.tsinghua.edu.cn}} 
and Juven Wang$^{2,3}$\Wplusfootnote{e-mail: {\tt jw@cmsa.fas.harvard.edu} (Corresponding Author) \href{http://sns.ias.edu/~juven/}{http://sns.ias.edu/$\sim$juven/} 
} 
\\[2.75mm]  
} 
\vskip.5cm
{\small{\textit{$^1${Yau Mathematical Sciences Center, Tsinghua University, Beijing 100084, China}\\}}
}
 \vskip.2cm
 {\small{\textit{$^2${Center of Mathematical Sciences and Applications, Harvard University,  Cambridge, MA 02138, USA}}\\}}
 \vskip.2cm
 {\small{\textit{$^3${School of Natural Sciences, Institute for Advanced Study, Einstein Drive, Princeton, NJ 08540, USA} \\}}
}

\end{center}

\vskip1.5cm
\baselineskip 16pt
\begin{abstract}


We explore QCD$_4$ quark matter, the $\mu$-T (chemical potential-temperature) phase diagram, possible 't Hooft anomalies, and topological terms,
via non-perturbative tools of cobordism theory and higher anomaly matching.
We focus on quarks in 3-color and 3-flavor on bi-fundamentals of SU(3), then
analyze the continuous and discrete global symmetries and pay careful attention to finite group sectors.
We input constraints from $T=CP$ or $CT$ time-reversal symmetries, implementing QCD on unorientable spacetimes and distinct topology.
Examined phases include the high T QGP (quark-gluon plasma/liquid),
the low T ChSB (chiral symmetry breaking), 2SC (2-color superconductivity) and CFL (3-color-flavor locking superconductivity) at high density.
We introduce a possibly useful but only \emph{approximate} higher anomaly, involving discrete 0-form axial and 1-form mixed chiral-flavor-locked center symmetries,
matched by the above four QCD phases.
We also enlist as much as possible, but without identifying all of, 't Hooft anomalies and topological terms relevant to 
Symmetry Protected/Enriched Topological states (SPTs/SETs)
of  gauged SU(2) or SU(3) QCD$_d$-like matter theories in general in any spacetime dimensions $d=2,3,4,5$
via cobordism.



\end{abstract}

\end{titlepage}

  \pagenumbering{arabic}
    \setcounter{page}{2}
    

\tableofcontents


\section{Introduction and Summary}

\subsection{Physics in QCD quark matter}
\label{sec:QCD}

We are made of atoms, which are made out of quarks, the particles\footnote{In particle physics,
quarks are \emph{elementary particles}. In condensed matter viewpoint, it may be beneficial to alternatively view the quarks as \emph{quasiparticles}, quasi-excitations out of certain vacuum.} 
of Quantum Chromodynamics (QCD) vacuum, plus some electrons.
The majority of our mass is from the mass of nuclei.
While about the 2\% of our mass is from Higgs condensate, the surprising significant 98\% of our mass is from QCD chiral condensate.
Meanwhile, we live in the chiral symmetry breaking (ChSB) phase of QCD vacuum. 
In order to investigate the nature of QCD matter and its vacuum structure,
it is helpful to move out from this particular vacuum (ChSB) to other new foreign phases. 
In condensed matter language, we try to explore other unfamiliar foreign phases outside the familiar domestic phase, away from the ground state (i.e., vacuum) we live in, 
by tuning parameters in the QCD phase diagram (see recent selected reviews \cite{0011333Rajagopal, 0709.4635RMPQCD, 1005.4814Fukushima}).
Namely, we should explore different new vacua or ground state structures and their excitation spectra.

In this work,  we will look at some simplified ideal toy models. One model is that quarks 
are nearly massless and on bi-fundamentals of SU(3).
Here the quarks are the Dirac spinors in 3+1 dimensional spacetime (we denoted as 3+1D or 4d). We will consider various types of curved spacetime
manifolds with different topology and with Spin structure, Pin$^+$, or Pin$^-$ or other twisted structures.\footnote{We consider the smooth differentiable manifolds with a metric g as spacetime -- if the fermions/spinor can live
on them, we require Spin structure; if we require time-reversal $T=CP$, or $CT$ or other reflection symmetries, we 
require Pin$^+$, Pin$^-$ or other semi-direct ($\ltimes$) product or twisted structures between the spacetime tangent bundle $TM$ and the gauge bundle $E_G$ of the gauge group $G$. See more in the main text and see an overview of our setting in \cite{1711.11587GPW}.}
In physics language, quarks are 4d Dirac fermions, in the fundamental representation 3 of SU(3) color gauge group and 
fundamental representation of SU(3) flavor global symmetry group. 
We denote them 
as the representation (Rep) 3$_c$ in SU(3$_c)_{V}$ for color and  3$_f$ in SU(3$_f)_{V}$ for flavor, where subindex $V$ stands for the vector symmetry. 
Follow the notations in  \cite{1711.11587GPW}, for the Euclidean path integral, we have the schematic partition function
\begin{multline}
\label{eq:ZQCD}
{
\int {
[{\cal D} {a}] [{\cal D} {\psi}] [{\cal D}\bar{\psi}] }\exp\Big(
-S_{\text{E,Dirac},(\slashed{D}_a +m_q+ \mu)}- S_{\text{YM}}
-S_{{\theta}}
) 
\Big)
}\\
\hspace{-10mm}
\equiv
\int [{\cal D} {a}][{\cal D} {\psi}] [{\cal D}\bar{\psi}] \exp\Big(
-\int_{M_4} \dd^4x_{\text{E}}\, \sqrt{ \det \text{g}}
\big( \sum_q \bar\psi_q (\slashed{D}_a +m_q+ \mu)\psi_q \big)
- \int_{M_4} (\frac{1}{4g^2}\text{Tr}\,(F_a\wedge \star F_a)
+ \frac{\rm i \theta}{8 \pi^2}  \text{Tr}\,(F_a\wedge F_a)
) 
\Big).
\end{multline}
This path integral describes an $\SU(N_c) = \SU(3_c)$ gauge theory with 1-form gauge field $a$ and 2-form field strength $F_a$ for the (exact) color $N_c = 3$ with three colors in fundamental: red ($r$), green ($g$), and blue ($b$). 
It also describes the (approximate) $N_f=3$-flavor in fundamental for fermions $\psi_q$,
where we choose the lightest bare quarks in nature: the up quark ($u$), 
the down quark ($d$), and the strange quark ($s$).
At the massless limit $m_q=0$, the topological $\theta$-term can be absorbed by axial  U(1)$_A$ rotations of quarks, and we may set $\theta=0$.
The fermion $\psi_q$ are quarks that carry quantum number (denoted the quark quantum number $q$) of color ($c$), flavor ($f$), the spacetime self-rotational spin ($s$), 
whose quark-pairing can also carry the spatial momentum $\vec{k}$, angular momentum $\vec{L}$, and parity quantum number $P$, 
etc.\footnote{
For convenience, sometime we specify the color ($c$)  quantum number as the first subindex $\alpha$ and the 
 flavor ($f$) quantum number as the second subindex $i$:
\bea
\psi_q \equiv \psi_{\alpha i}.
\eea
See Table \ref{table:quarkPairing1} and \ref{table:quarkPairing2} for some examples of quantum numbers for different quark-pairing condensates.
}
See Table \ref{table:quarkPairing1} and \ref{table:quarkPairing2} for some examples of quantum numbers for different quark-pairing condensates.
In \Eqn{eq:ZQCD}, we can tune the temperature T $=\beta_{\rm E}=\frac{1}{L_{\rm E}}$ proportional to the inverse size of the Euclidean time circle ${L_{\rm E}}$,
and we can also tune the chemical potential $\mu$, which changes the density of quark matter.
 
The standard lore from the pioneer studies of quark matter \cite{0011333Rajagopal, 0709.4635RMPQCD} teaches us that
four dominant phases occur at different regions of QCD phase diagram drawn in the $\mu$-T (chemical potential v.s. temperature) axes.
Below we aim to revisit some of these four phases, and applying modern perspectives of symmetries and anomalies to constrain these quantum systems.\\[-10mm]

\begin{enumerate}[leftmargin=2mm, label=\textcolor{blue}{(\Roman*)}., ref={(\Roman*)}]

\item \emph{Global Symmetries}: For symmetries, we explore and exhaust both continuous and discrete global symmetries and pay special attention to finite group sectors.
\item \emph{Higher Symmetries}: For gauge theories, there are extended operators of lines and surfaces, etc. They can also carry quantum numbers thus also \emph{charged} under the
higher generalized global symmetries  \cite{Gaiotto:2014kfa}. There are also corresponding symmetry generators as \emph{charge} operators. We also need to pay attention to higher symmetries.

\item \emph{Anomalies and Higher Anomalies}: Given the global symmetry,
there can be potential obstructions to couple the symmetry to background gauge field or to subsequently gauging the symmetry --- the phenomena are known as the 't Hooft anomalies \cite{H8035}. For higher symmetries, there are also associated higher  't Hooft anomalies.
By anomalies, we mean to include both


\begin{itemize}
\item
{\bf Perturbative local anomalies} calculable from perturbative Feynman diagram loop calculations, 
classified by the integer group $\mathbb{Z}$ classes (or the so-called  free classes). 
Selective examples include:
\begin{enumerate} [label=\textcolor{blue}{(\arabic*)}:, ref={(\arabic*)}]
\item
Perturbative fermionic anomalies from chiral fermions with U(1) symmetry, originated from Adler-Bell-Jackiw (ABJ) anomalies \cite{Adler1969gkABJ,Bell1969tsABJ}
with $\mathbb{Z}$ classes.
 \item Perturbative gravitational anomalies \cite{AlvarezGaume1983igWitten1984}.
\end{enumerate}
\item {\bf  Non-perturbative global anomalies}, classified by finite groups such as $\mathbb{Z}_N$  (or the so-called torsion classes).
Some selective examples from QFT or gravity include:
\begin{enumerate} [label=\textcolor{blue}{(\arabic*)}:, ref={(\arabic*)}]
\item An SU(2) anomaly of Witten in 4d or in 5d \cite{Witten1982fp} with a $\mathbb{Z}_2$ class, which is a gauge anomaly. 
\item A new SU(2) anomaly  in 4d or in 5d \cite{WangWenWitten2018qoy1810.00844} with another $\mathbb{Z}_2$ class, which is a mixed gauge-gravity anomaly.
\item Non-perturbative bosonic anomalies from finite-group-symmetry bosonic systems \cite{Wang2014tia1403.5256, Kapustin2014zva1404.3230}.
 
\item Higher 't Hooft anomalies for a pure 4d SU(2) YM theory  
with a second-Chern-class topological term \cite{Gaiotto2017yupZoharTTT1703.00501, Wan2018zqlWWZ1812.11968, Wan2019oyr1904.00994} (the SU(2)$_{\theta =\pi}$ YM):
The higher anomaly involves a discrete 0-form $\mathbb{Z}_2^T$ time-reversal symmetry and a 1-form center $\Z_{2,[1]}$-symmetry.  
The first anomaly is discovered in \cite{Gaiotto2017yupZoharTTT1703.00501}; later the anomaly is refined via a mathematical well-defined 5d co/bordism invariant as its 
invertible 
Topological Quantum Field Theories (iTQFTs)
topological term, 
with additional new anomaly for four siblings of YM \cite{Wan2018zqlWWZ1812.11968, Wan2019oyr1904.00994}.
\item Global gravitational anomalies \cite{Witten1985xe}.
\end{enumerate}
\end{itemize}

\item \emph{Symmetry Protected Topological states (SPTs)/ Symmetry Enriched Topologically ordered states (SETs) or Higher SPTs/SETs}: Quantum systems (usually the quantum vacuum or the ground state) can be protected by
global symmetry in a topological way. These are known as the interacting generalizations of topological insulators (TI) and 
topological superconductors (TSC) \cite{2010RMP_HasanKane, 2011_RMP_Qi_Zhang,RyuSPT,Kitaevperiod,Wen1111.6341}
known as the SPTs  
for interacting bosons and interacting fermions (see the overview \cite{XieSPT4, Senthil2014ooa1405.4015, 1610.03911}).
\Ref{Kapustin1403.1467, Kapustin:2014dxa, Freed2016} propose mathematical theories of cobordism  
classifying these SPTs and their low energy 
iTQFTs.
In the context that we require to apply is the SU(N) and time-reversal symmetry generalization of SPTs studied in \Ref{1711.11587GPW} 
suitable for Yang-Mills and QCD systems. In this work, we will apply a generalized cobordism theory including the higher-SPTs classifications 
(given by higher classifying spaces and higher symmetries) based on the computations and tools in \cite{Wan2018bns1812.11967}.

\end{enumerate}

\begin{figure}[!h]
\centering
\includegraphics[scale=.64]{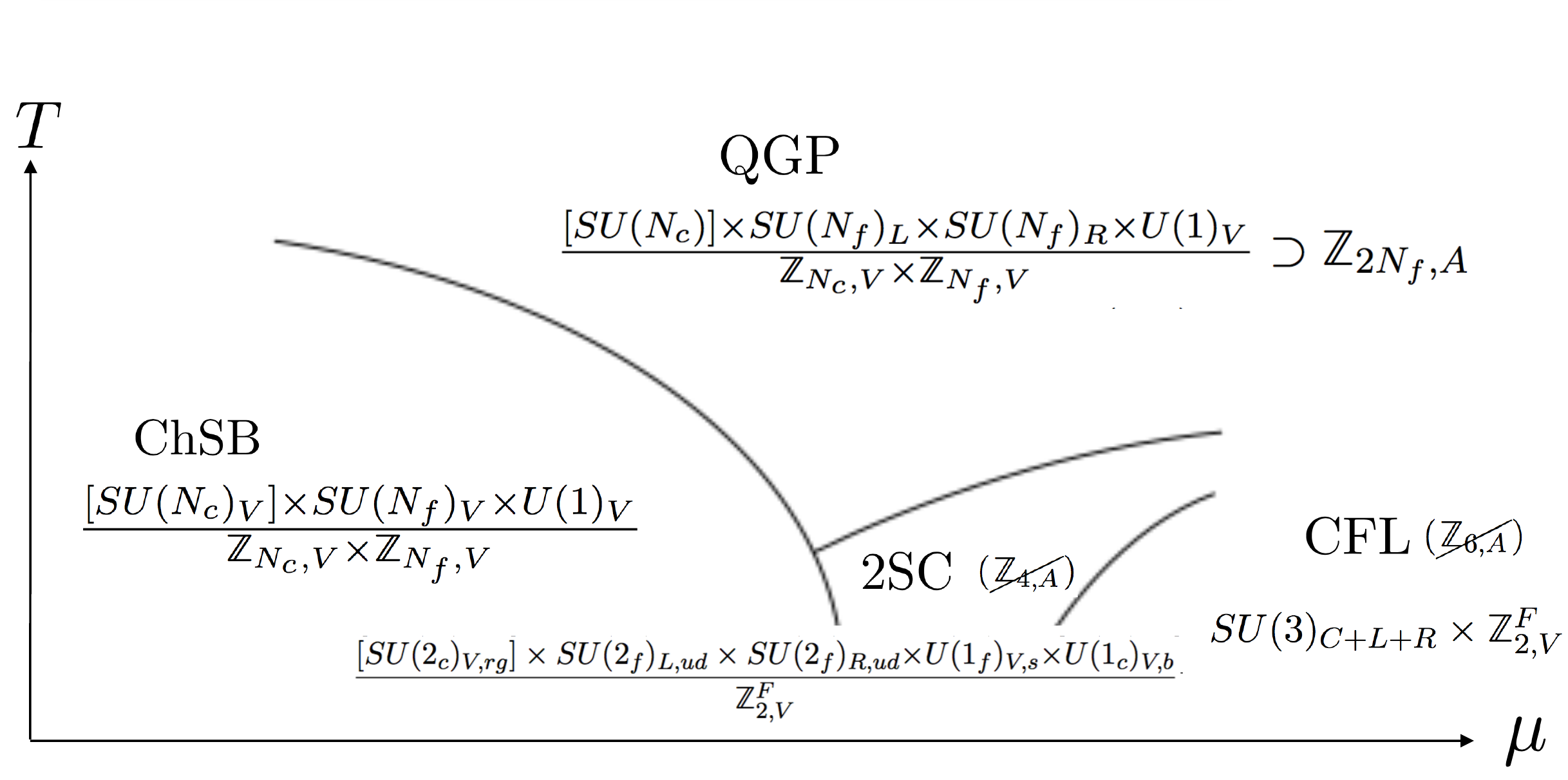}
\caption{We revisit the QCD$_4$ matter phase in the $\mu$-T (chemical potential-temperature) phase diagram:
The high T QGP (quark-gluon plasma/liquid),
the low T ChSB (chiral symmetry breaking), 2SC (2-color superconductivity) and CFL (3-color-flavor locking superconductivity) at high density.
We do not attempt to address the nature of phase transitions in this figure, thus we make some of the phase boundaries blur.
The $\cancel{\Z_{6,A}}$ or $\cancel{\Z_{4,A}}$ means that part of the discrete axial symmetry $(A)$ is broken: $\cancel{\Z_{2N_f,A}}$.
In general,  if the $G_{\text{sym}}$ group is broken, we denote it as $\cancel{G_{\text{sym}}}$.}
\label{fig:qcd-phase-1}
\end{figure}

\begin{figure}[!h]
\centering
\includegraphics[scale=.64]{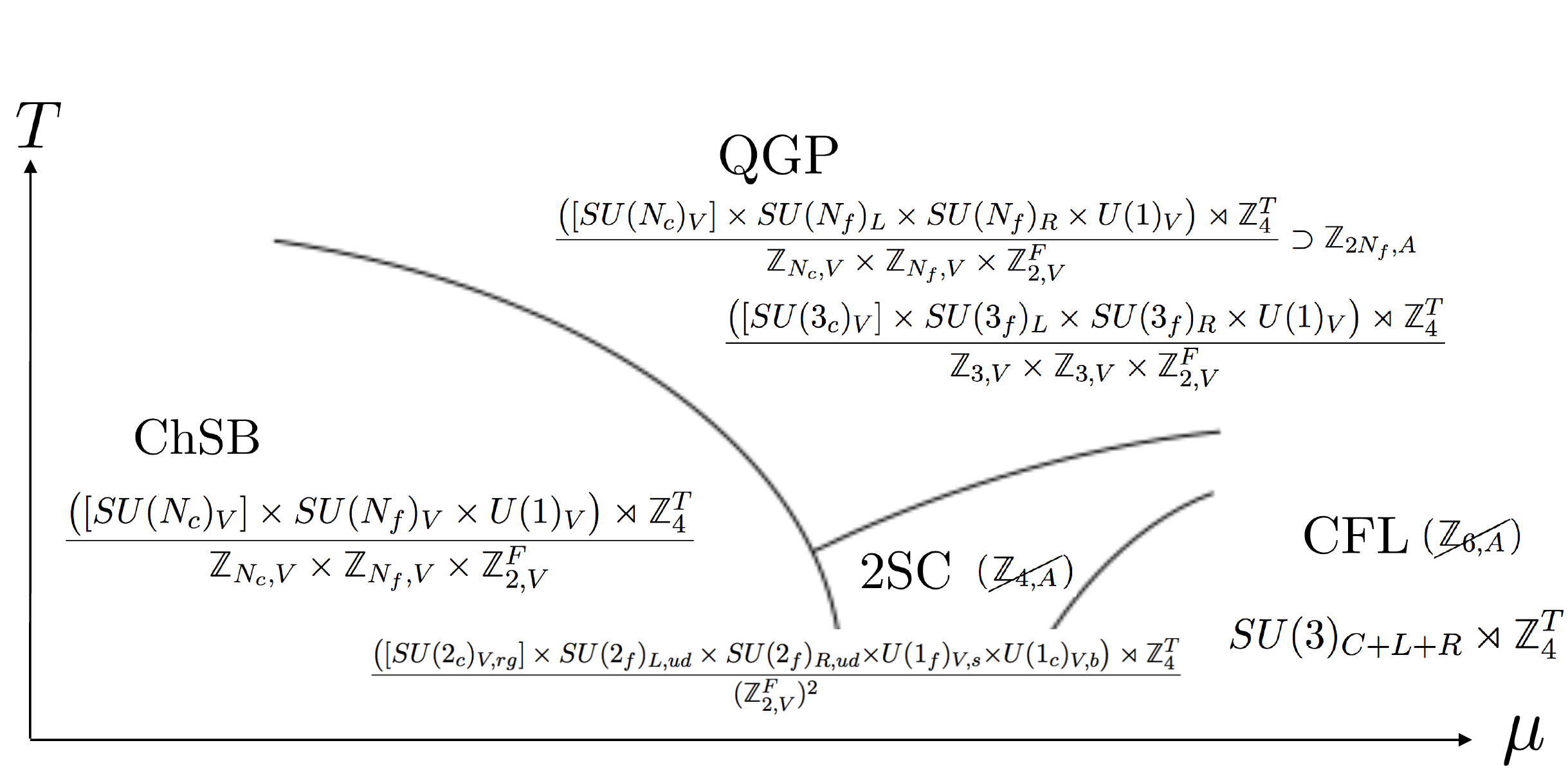}
\caption{Follow \Fig{fig:qcd-phase-1}, but now we include possible time-reversal symmetries, which can be any reasonable $\Z_2$-reflection symmetry by putting the Euclidean QCD$_4$ path integral on an unorientable spacetime. Here we choose a semi-direct product of $\rtimes \Z_4^T$, where $\Z_4^T \supset \Z_2^F$. 
The semi-direct product is more general, which also includes the direct product case of $\times \Z_4^{CT}$ as the $CT$ symmetry.
}
\label{fig:qcd-phase-T}
\end{figure}

By keeping in minds  and utilizing the above modern concepts of quantum systems and QFTs beyond Ginzburg-Landau
symmetry-breaking paradigm, 
here we revisit the standard lore of four QCD quark matter phases \cite{0011333Rajagopal, 0709.4635RMPQCD} and list down their global symmetries\footnote{It is worthwhile
to emphasize the old literature may happen to pay less attention to the finite group and discrete sectors. However,
the finite group and discrete sectors are important for topological terms and non-perturbative global anomalies later we compute from the cobordism group. Thus,
we aim to be as precise as possible writing down the global symmetries, see also \cite{1711.11587GPW}.} in \Fig{fig:qcd-phase-1} (without time-reversal symmetry)
and \Fig{fig:qcd-phase-T} (with certain time-reversal symmetries):
\begin{enumerate}[leftmargin=2.0mm, label=\textcolor{blue}{\arabic*}., ref=\textcolor{blue}{\arabic*}]
\item 
{QGP (quark-gluon plasma/liquid) at high T}:\\
For general $N_c$ and $N_f$, we have the symmetry:
\bea \label{eq:QGP-sym}
\frac{[\SU(N_c)_V] \times \SU(N_f)_L \times  \SU(N_f)_R \times \U(1)_V}{\Z_{N_c, V} \times \Z_{N_f, V}},
\eea
where the $[\SU(N_c)_V]$ color is gauged as
 a gauge group.\footnote{The $[G_g]$ specifies that $G_g$ is dynamically gauged.}
The $\U(1)_V/\Z_{N_c, V}$ is the vector symmetry associated to the baryon number $B$ conservation.
The $\SU(N_f)_L$ and $\SU(N_f)_R$
are the left/right-handed Weyl spinor $\SU(N_f)$ flavor symmetries under the projection of $P_{L/R}=\frac{1\mp\gamma^5}{2}$.\footnote{It
is worthwhile mentioning that the discrete axial symmetry $\Z_{2 N_f, A}$ that are not broken via the (ABJ) effect, still remains and sits inside:
$
\frac{[\SU(N_c)_V] \times \SU(N_f)_L \times  \SU(N_f)_R \times \U(1)_V}{\Z_{N_c, V} \times \Z_{N_f, V}} \supset \Z_{2 N_f, A}.
$}
We will mainly focus on $N_c=N_f=3$.

\item 
{ChSB (chiral symmetry breaking) at low T and at lower densities and low $\mu$}:\\
For general $N_c$ and $N_f$, we have the symmetry:
\bea
{\frac{ \big( [\SU(N_c)_V] \times  \SU(N_f)_V \times \U(1)_V\big) }{ \Z_{N_c, V} \times \Z_{N_f, V} }},
\eea
where the $[\SU(N_c)_V]$ color is gauged.
Due to the chiral condensate $\langle \bar \psi \psi \rangle \neq 0$ in this vacum, the
$\SU(N_f)_L$ and $\SU(N_f)_R$  flavor symmetries
of the left/right-handed Weyl spinor are broken down to
the diagonal vector subgroup $\SU(N_f)_V$.
We will mainly focus on $N_c=N_f=3$.
In ChSB,
by the spontaneously symmetry breaking (SSB) from QGP,
we gain 8 pseudo-Goldstone bosons (as mesons: $\pi^0, \pi^{\pm}, K^0, \bar{K}^0, \eta, K^{\pm}$), while there is one massive $\eta'$ meson.
If the ChSB further forms the nucleon superfluid, the $\U(1)_V$ is SSB, and we gain 1 more Goldstone boson.

\begin{table}[h!]
\footnotesize
\begin{center}
    \begin{tabular}{| c | c|c | c | c|c|c|  }
    \hline
  SC &  spin $s$ & 
  \begin{minipage}[c]{0.14\textwidth}  momentum\\
  (angular/orbital) \\
 $\vec{L}$,\;$\vec{k}$\\
  L/R (Parity,$J^P$) 
  \end{minipage}
 & color (SU(2)/SU(3))  & Flavor (SU(2)/SU(3))  & Condensate $\langle q q \rangle$ 
 \\ \hline\hline
\begin{minipage}[c]{0.06\textwidth}
 \quad 2SC \\
($ud,rg$) 
\end{minipage} 
& \begin{minipage}[c]{0.15\textwidth} singlet $(\up \down-\down \up)$ \\
$1$ in $2 \otimes 2= 1 \oplus 3$ \end{minipage} &
 \begin{minipage}[c]{0.115\textwidth} 
  0, 0\\
{LL $\pm$ RR} 
even/odd; $0^{\pm}$
\end{minipage} & 
\begin{minipage}[c]{0.17\textwidth}  $(rg-gr)$ singlet: \\
$1$ in $2 \otimes 2= 1 \oplus 3$\\
color-anti-triplet:\\
$A$ in $3 \otimes 3= \bar{3}_A \oplus 6_S$ \end{minipage}
 & 
\begin{minipage}[c]{0.17\textwidth} 
$(ud-du)$\\
$1$ in $2 \otimes 2= 1 \oplus 3$\\
$A$ in $3 \otimes 3= \bar{3}_A \oplus 6_S$
\end{minipage}
&  
\begin{minipage}[c]{0.15\textwidth} 
$ \langle \psi^{\rm T}_{\alpha i} \,C \gamma^5\,
  \psi_{\beta j} \rangle_{\text{2SC}}$ 
  $\propto \epsilon_{\alpha \beta A} \epsilon_{i j B} \delta^A_3 \delta^3_B$\\
 $=\epsilon_{\alpha \beta 3} \epsilon_{i j 3} $  
  \end{minipage} 
  \\ \hline   
CFL & \begin{minipage}[c]{0.15\textwidth} singlet $(\up \down-\down \up)$\\ $1$ in $2 \otimes 2= 1 \oplus 3$   \end{minipage} & 
\begin{minipage}[c]{0.115\textwidth} 
  0, 0\\
{LL $\pm$ RR} 
even/odd; $0^{\pm}$
\end{minipage} 
& 
\begin{minipage}[c]{0.17\textwidth} 
$\bar{3}_A$ in $3 \otimes 3= \bar{3}_A \oplus 6_S$\\
not-$SU(3)$-singlet
\end{minipage} 
 & $\bar{3}_A$ in $3 \otimes 3= \bar{3}_A \oplus 6_S$ 
&  
\begin{minipage}[c]{0.15\textwidth} 
$ \langle \psi^{\rm T}_{\alpha i} \,C \gamma^5\,
  \psi_{\beta j} \rangle_{\text{CFL}} $\\
  $\propto \epsilon_{\alpha \beta A} \epsilon_{i j B} \delta^A_B$
\end{minipage} 
 \\ \hline
 \hline   
    \end{tabular}
    \end{center}
\caption{Pairing of quark-quark condensate for
2SC (2-color superconductivity) and CFL (3-color-flavor locking superconductivity).
The L and R are for left/right-handed spinors.}
    \label{table:quarkPairing1}
\end{table}

\begin{table}[h!]
\begin{center}
    \begin{tabular}{ c  c  c  c  c   }
    \hline\\[-1em]
   &  Pairing function &   Wavefunction & Parity $P$    & Spin or orbital  \\ \hline\hline\\[-1em]
$\psi^\dagger \Delta C \gamma^5 \psi^*$ &  $\Delta_k=\Delta_{-k}$ &${(|L, L \rangle +|R,R  \rangle)} {| k, -k\rangle} (|\up \down \rangle -|\down \up \rangle)$     &  Even &  s wave/singlet   \\ \hline\\[-1em]
$\psi^\dagger \Delta C \gamma^5 \psi^*$ &  $\Delta_k=-\Delta_{-k}$ & $ {(|L, L \rangle -|R,R  \rangle)} {| k, -k\rangle} (|\up \down \rangle +|\down \up \rangle)$  & Even  &  p wave/triplet   \\ \hline
$\psi^\dagger \Delta C  \psi^*$   & $\Delta_k=\Delta_{-k}$ &  ${(|L, L \rangle -|R,R  \rangle)} {| k, -k\rangle} (|\up \down \rangle -|\down \up \rangle)$    & Odd   & s wave/singlet\\ \hline
$\psi^\dagger \Delta C  \psi^*$  & $\Delta_k=-\Delta_{-k}$  &  ${(|L, L \rangle +|R,R  \rangle)} {| k, -k\rangle} (|\up \down \rangle +|\down \up \rangle)$ &   Odd &  p wave/triplet \\ \hline
 \hline   
    \end{tabular}
    \end{center}
\caption{charge conjugate matrix $C=\rm i\gamma^2\gamma^0$ is unitary.
All pairings are $T$-symmetry invariant.
Under Parity $P$:
$L \leftrightarrow R$ and $k \leftrightarrow -k$, but {spin up/down} $(\up/\down) \leftrightarrow (\up/\down)$ is invariant.
Under Time Reversal $T$: $L/R \leftrightarrow L/R$ invariant, but $k \leftrightarrow -k$, and the spin up/down flips 
$(\up/\down) \leftrightarrow (\down/-\up)$.
}
    \label{table:quarkPairing2}
\end{table}

\item 
{2SC (2-color superconductivity) at low T and at intermediate densities and $\mu$}:
\bea
\frac{[\SU(2_c)_{V,rg}] \times \SU(2_f)_{L,ud} \times  \SU(2_f)_{R,ud}   {\times \U(1_f)_{V,s}} {\times \U(1_c)_{V,b}}}{{\Z_{2,V}^F}},
\eea
where the $[\SU(2_c)_V]$ color is gauged thus there is an SU(2) gauge theory.
The 2SC pairing is shown in Table \ref{table:quarkPairing1}.
The ${\Z_{2,V}^F}$ is the fermion parity symmetry, which is a vector symmetry.
The 2SC 
pairs 2-flavor $u$-$d$ and 2-color $r$-$g$ both into SU(2) singlets as 
a color superconductor. Since the $[\SU(3)_c]$ is broken to $[\SU(2_c)]$, this results in 5 massive gluons. 
There are 4 gapped quasiparticle fermions in the 2SC Bogoliubov basis. 
There are 5 un-paired and gapless quasiparticle fermion in the 2SC Bogoliubov basis. Thus many symmetries are still intact.

\item 
{CFL (3-color-flavor locking superconductivity) at low T and at high density and high $\mu$}:\\
The $\SU(3)_{C+L+R}$ means that the 
$\SU(3_c)$ are locked and rotated in the opposite manner as the diagonal of
$\SU(3_f)_L \times \SU(3_f)_R$.
This is a CFL superconductor, with 
9 gapped ($8 \oplus 1$) quasiparticle fermion  in the CFL Bogoliubov basis. 
By comparing the QGP and CFL phases, we see that the
continuous Lie group generators broken down from 25 generators ($8 \times 3 +1$) in QGP
to the only 8 generators of $\SU(3)_{C+L+R}$.
The missing 17 generators can be accounted via:
7 massive gluons and 1 mixture of a photon+gluon gauge boson, 
there are (8+1) Goldstone bosons. 
The 8 unbroken $\SU(3)_{C+L+R} \supset [\U(1)]_{\tilde Q}$ actually contains another mixture of the photon+gluon gauge boson.
Including the fermion parity symmetry ${\Z_{2,V}^F}$, we have the global symmetry:
\bea
\SU(3)_{C+L+R} \times \Z_{2,V}^F,
\eea
although there is a part of the global symmetry containing the electromagnetic $[\U(1)]_{\tilde Q}$ which is 
gauged not global.\footnote{In fact, since the strong forces are much dominant than the electromagnetism, we will focus on the 
$[\SU(3_c)]$ gauge sectors from the strong forces first, and treat the gauge sectors of electromagnetism $\U(1)_{EM}$ separately later.}
\end{enumerate}

By including time reversal symmetries into the QCD system, we can choose any suitable outer automorphism of 
the color gauge or flavor global symmetry group
as a $\Z_2$-time reversal symmetry,
which is a $\Z_2$-reflection symmetry by putting the Euclidean QCD$_4$ path integral on an unorientable spacetime.

In fact, several recent works have attempted to study QCD matter phases based on the languages of higher symmetries and anomalies above.
We should quickly overview some of these pioneer works:

\begin{enumerate}[leftmargin=2.0mm, label=\textcolor{blue}{\underline{(\arabic*)}}., ref=\textcolor{blue}{\underline{(\arabic*)}}]
\item 
{Whether the color superconductivity can be topological in some way was questioned in \Ref{1001.2555Nishida}. 
What \Ref{1001.2555Nishida} concerns is the topological insulators (TI) / topological superconductors (TSC)
in the free non-interacting quadratic mean-field Hamiltonian systems. Thus the classifications in \Ref{1001.2555Nishida} are only either 0 or $\Z$ classes for mean-field 
free fermion systems.
The new input in our context is that we consider fully interacting systems and enlist possible SPTs for these QCD matter phases by cobordism group classifications.}

\item \Ref{Anber2018tcjPoppitz1805.12290, Cordova2018acb1806.09592, Bi2018xvrSenthil1808.07465, Wan2018djlQCD1812.11955}
explores a related system of 4d adjoint quantum chromodynamics (QCD$_4$) with an SU(2) gauge group and two massless adjoint Weyl fermions.
Higher symmetries and higher anomalies play an important role. Depend on the complex mass parameters of fermions, we can land onto different phases,
and there are interesting quantum phase transitions between bulk phases.
There are implications and constraints for 3+1D deconfined quantum critical points (dQCP), quantum spin liquids (QSL) or fermionic liquids in condensed matter, 
and constraints on  3+1D ultraviolet-infrared (UV-IR) duality. See \Ref{Wan2018djlQCD1812.11955} for an overview of the proposed phases
at quantum critical points.

\item \Ref{Bi2019ers1910.12856, Wang2019obe1910.14664} explores quantum phase transitions between Landau ordering phase transitions but beyond the Landau
paradigm, for example, due to the effects of topological $\theta$-terms.
 \Ref{Wang2019obe1910.14664} suggests that
 SU(2) QCD$_4$ with large odd number of flavors of quarks 
 could be a direct second order phase transition between two phases of U(1) gauge theories as well as between a U(1) gauge theory and a trivial vacuum (e.g. a Landau symmetry-breaking gapped paramagnet). The gauge group is enhanced to be non-Abelian at and only at the transition. 
 It is characterized as Gauge Enhanced Quantum Critical Points.

\item \Ref{1706.05385Unsal, 1706.06104Yonekura, 1711.10487Tanizaki, Tanizaki2018wtg1807.07666,  Yonekura2019vyz1901.08188, Anber2019nzeBCF1909.09027} employs
global anomalies or global inconsistency of QCD$_4$ matter to constrain its QCD phase diagram, including either QCD zero-T phase or thermal phase.
 \Ref{1706.05385Unsal, 1706.06104Yonekura, 1711.10487Tanizaki, Tanizaki2018wtg1807.07666, Yonekura2019vyz1901.08188, Anber2019nzeBCF1909.09027}, implicitly or 
 explicitly, suggests that there are obstructions or anomalies to simultaneously 
 gauge or preserve both (1) a discrete axial symmetry and 
(2) a discrete 1-form color or flavor center symmetry (e.g. twisted discrete flavor-symmetry background boundary conditions or chemical potential $\mu$),
and/or (3) under certain twisted boundary conditions
 --- so preferably and naively either symmetries must be broken.
 In fact, there are still possibilities that symmetric gapped TQFTs can be constructed for some of these anomalies, for example, based on the 
symmetry extension method \cite{1705.06728WWW} or higher-symmetry extension method \cite{Wan2018djlQCD1812.11955}. 
In any case, \Ref{1706.05385Unsal, 1706.06104Yonekura, 1711.10487Tanizaki, Tanizaki2018wtg1807.07666, Yonekura2019vyz1901.08188, Anber2019nzeBCF1909.09027} can rule out trivial gapped phases under certain circumstances.

\item \emph{Quark-hadron continuity outside the Ginzburg-Landau paradigm}:
Quark-hadron continuity  \cite{Schafer1998efWilczek-Quark-hadron-continuity} asserts 
that hadronic matter superfluid phase is continuously connected to color-superconductor without phase transitions when the $\mu$ increases. 
This proposal is based on Ginzburg-Landau theory where two sides of phases have the similar symmetry breaking patterns and the similar gapless and gapped energetic spectrum. 
\Ref{Cherman2018jir1808.04827} questions the quark-hadron continuity to be invalid, by suggesting there must be a phase transition due to the topological fractionalization of excitations are different. \Ref{Hirono2018fjr1811.10608} re-analyzes the scenario based on higher-symmetry is not spontaneously broken, and found that
the quark-hadron continuity is still plausible. 

\end{enumerate}

In the remained of this article, 
we like to point out a higher anomaly involving discrete 0-form axial chiral and 1-form mixed chiral-flavor-locked center symmetries
that can be matched by the above four QCD phases. 
Then we will give a quick mathematical introduction of tools we used in \Sec{sec:math}.
After then, we will list down various data and tables computed from cobordism theory, with a view toward the applications of QCD matter phases, to be studied in the future \cite{toappear}.
The {QCD$_d$ matter symmetries, anomalies, and topological terms without time-reversal symmetry}, classified by 
the cobordism theory, are studied in \Sec{sec:WithoutTime-Reversal}
via a cobordism theory.
The {QCD$_d$ matter symmetries, anomalies, and topological terms with time-reversal symmetry}, classified by 
the cobordism theory, are studied in \Sec{sec:WithTime-Reversal}, in general in any spacetime dimensions $d=2,3,4,5$
via a cobordism theory.

\subsection{Approximate higher anomaly constraint on the QCD phase diagram}

In this section, we 
point out a higher 't Hooft anomaly involving discrete 0-form axial and 1-form mixed chiral-flavor-locked center symmetries
can be matched by the above four QCD phases.
Our approach is related but still somehow different from \Ref{1706.05385Unsal, 1706.06104Yonekura, 1711.10487Tanizaki, Tanizaki2018wtg1807.07666,  Yonekura2019vyz1901.08188, Anber2019nzeBCF1909.09027}.

For simplicity, we will set $N_c=N_f=N$ below. If $N_c \neq N_f$, we just need to replace the $N$ below to their greatest common divisor $\gcd(N_c,N_f)$
and make some moderate but a straightforward generalization. 

First, under some \emph{assumptions} that we will comment later, 
we hope to point out that there is an \emph{approximate} {1-form electric-magnetic (e-m)  global symmetry $\Z^{}_{N_{cf},[1]}$}
that mixed between 1-form color and flavor ({CF}) center symmetry: $\Z^{}_{N_{cf},[1]}$.
To recall, focus on the kinematics of the UV path integral of QFT,

\noindent
($\bullet$1) If we do not have quarks in the fundamental Rep of the color gauge group $[\SU(N_c)_V]$, 
then we have the 1-form electric symmetry $\Z^{}_{N,c,[1]}$ whose charged objects are the {color} 
SU$(N_c)$ fundamental Wilson line.
If we have quarks in the fundamental Rep of  $[\SU(N_c)_V]$, then the $\Z^{}_{N,c,[1]}$ is broken explicitly. 

\noindent
($\bullet$2) By looking at the largest symmetry at QCD phases --- \Eqn{eq:QGP-sym}'s QGP symmetry:
$\frac{[\SU(N_c)_V] \times \SU(N_f)_L \times  \SU(N_f)_R \times \U(1)_V}{\Z_{N_c, V} \times \Z_{N_f, V}}$
where $[\SU(N_c)_V]$ is gauged, we still are left with part of the projective special unitary symmetry $\frac{\SU(N_f)_V}{\Z_{N_f, V}}=\PSU(N_f)_V$ 
from the flavor symmetry.
We can regarded it as a 1-form magnetic symmetry that can be coupled to the {flavor} 
symmetry background probed fields $\frac{\SU(N_f)_V}{\Z_{N_f, V}}=\PSU(N_f)_V$,
by using the Stiefel-Whitney $w_2(\PSU(N_f)_V)$.
This idea is implemented already in \cite{1706.06104Yonekura, 1711.10487Tanizaki, Tanizaki2018wtg1807.07666,  Yonekura2019vyz1901.08188, Anber2019nzeBCF1909.09027}.

\noindent
($\bullet$3) Now we combine the above two 1-form e and m symmetries into a diagonal 1-form symmetry, 
and name it as {1-form electric-magnetic (e-m)  global symmetry $\Z^{}_{N_{cf},[1]}$}.
We define this 1-form e-m color-flavor-locked global symmetry $\Z^{}_{N_{cf},[1]}$
  as the diagonal symmetry that rotates oppositely 
the color-fundamental ($W_c$) and the flavor-fundamental ($W_f$)
Wilson line along the 1d curve ${\gamma^1}$, called 
\bea
W_c W_f = \exp(q_c \oint_{\gamma^1} a_c)  \exp(q_f \oint_{\gamma^1} a_f),
\eea
where the $a_c$ is.
The $(W_c W_f)$ have ends that can be opened by having bi-fundamental quark and anti-quark at each of two ends.
Let us call the 2-dimensional surface operator that is the 1-form $\Z^{}_{N_{cf},[1]}$ symmetry generator on 
2d area ${\Sigma^2}$ as: 
\bea
U\sim  (U_{{\rm e},c})  (U_{{\rm m},f})^{-1} \sim \exp(\oiint_{\Sigma^2} \Lambda) \exp(- \oiint_{\Sigma^2} w_2(\PSU(N_f)_V)).
\eea  
Here we write down the electric and magnetic 2-surface operators based on the conventions and notations in \cite{Wan2019oyr1904.00994}.
So that even if the bi-fundamental quarks can open up the the color-fundamental and the flavor-fundamental
Wilson line, it will \emph{not} be charged under the 1-form e-m $\Z^{}_{N_{cf},[1]}$ symmetry, because the obtained phase is exactly cancelled:
\bea
\langle (U) (W_c W_f) \rangle |_{\#_{U,(W_c W_f))}=1} 
&=&\exp(\ii \frac{2 \pi}{N})\exp(-\ii \frac{2 \pi}{N})
\langle (U) (W_c W_f) \rangle |_{\#_{U,(W_c W_f))}=0} \nn\\
&=&1 \cdot \langle (U) (W_c W_f) \rangle |_{\#_{U,(W_c W_f))}=0}
\eea
when the 2d $U$ surface and the 1d combined $W_c W_f$ line have a nontrivial linking number 
$$\#_{U,(W_c W_f))}=1$$ in a 4d spacetime.\footnote{The illustration and TQFT calculations
of such linking number, link invariants and link configurations in the spacetime picture, e.g. $\#_{({\Sigma^2},{\gamma^1})}=1$ on a 4-sphere $S^4$, can be found in \cite{1612.09298PutrovWang, Wang1901.11537WWY}.
}
Importantly, although bi-fundamental quarks can open up the $(W_c W_f)$ line,
physically this means that this (approximate) 1-form e-m $\Z^{}_{N_{cf},[1]}$ symmetry is not broken (explicitly and kinematically) by the quark which must sit as
bi-fundamental Rep in QCD$_4$!

However the 1-form e-m $\Z^{}_{N_{cf},[1]}$ symmetry still acts nontrivially on the 
color-fundamental Wilson line because this linking
\bea
\langle (U) (W_c ) \rangle |_{\#_{U,(W_c W_f))}=1} 
&=&\exp(\ii \frac{2 \pi}{N})
\langle (U) (W_c ) \rangle |_{\#_{U,(W_c W_f))}=0} 
\eea
shows $W_c$ is charged under 1-form $\Z^{}_{N_{cf},[1]}$ with a charge $\exp(\ii \frac{2 \pi}{N})$.
Similarly,  the 1-form e-m $\Z^{}_{N_{cf},[1]}$ symmetry still acts nontrivially on the 
flavor-fundamental Wilson line because this linking
\bea
\langle (U) (W_f ) \rangle |_{\#_{U,(W_c W_f))}=1} 
&=&\exp(-\ii \frac{2 \pi}{N})
\langle (U) (W_f ) \rangle |_{\#_{U,(W_c W_f))}=0} 
\eea
shows $W_f$ is charged under 1-form $\Z^{}_{N_{cf},[1]}$ symmetry with a charge $\exp(-\ii \frac{2 \pi}{N})$.
We propose that there is an \emph{approximate} anomaly mixing between
the discrete $\Z_{2N_f,A}$ axial (chiral) symmetry and the
 1-form e-m color-flavor-locked ${\Z^{}_{N_{cf},[1]}}$ global symmetry.
 We call the 2-form $\Z_N$ valued background field of 1-form $\Z^{}_{N_{cf},[1]}$ as $B_{cf}^{(2)}$.
 We find that there is an anomaly captured by the $\Z_{2N_f,A}$ axial (chiral) symmetry transformation,
 such that the partition function $\bZ$ gains a fractionalized term
\bea \label{eq:approximateanomalyBcf}
\bZ \xrightarrow[\text{}]{\Z_{2N_f,A} \text{transformation}} \bZ \cdot
{\exp\big( 
\frac{  { -\ii   N   }}{2 \pi }   \int\limits_{M_4} (
 B_{cf}^{(2)} \wedge  B_{cf}^{(2)} )\big)}.
\eea
The fractionalized term ${\exp\big( 
\frac{  { -\ii   N   }}{2 \pi }   \int\limits_{M_4} (
 B_{cf}^{(2)} \wedge  B_{cf}^{(2)} )\big)}$ is not a 4d SPTs but a fractionalized 4d SPTs, not a counter term, 
 thus cannot be absorbed into 4d, and should be regarded as a 5d higher-SPTs/higher-iTQFT -- which in fact is an indicator of the 4d
higher 't Hooft anomaly of the QCD$_4$.

The caveat however is that this 4d higher 't Hooft anomaly of the  QCD$_4$ is only \emph{approximate}.   
The subtlety is that it is only a precise anomaly if we also ``gauge'' the $\frac{\SU(N_f)_V}{\Z_{N_f, V}}=\PSU(N_f)_V$  sector
into the $[\frac{\SU(N_c)_V \times \SU(N_f)_V}{\Z_{\gcd(N_c,N_f), V} }]$
gauge theory.
The disadvantages of our \emph{approximate} anomaly \Eq{eq:approximateanomalyBcf}
are that the flavor Wilson lines are only probed but not dynamical objects, so we do not really have the 
1-form symmetry unless we at least \emph{weakly} gauge the flavor symmetry.\footnote{\emph{Weakly} gauge the flavor symmetry is thus related to 
a certain twisted flavor boundary condition studied in, for examples, \Ref{1706.06104Yonekura, 1711.10487Tanizaki}.}
Thus another interpretation of our gauge theory and anomaly 
is indeed the \emph{bi-fundamental gauge theory} in 3+1 dimensions studied in \cite{Tanizaki2017bam-BiFundamental-1705.01949, Karasik2019bxn1904.09551}.

In comparison, other approaches in
\Ref{1706.05385Unsal, 1706.06104Yonekura, 1711.10487Tanizaki, Tanizaki2018wtg1807.07666,  Yonekura2019vyz1901.08188, Anber2019nzeBCF1909.09027}
also have their own disadvantages.
For example, \Ref{1711.10487Tanizaki} derives a different kind of anomaly \emph{only under a certain twisted flavor boundary condition} in 4d:
\bea \label{eq:approximateanomalyBcBf4d}
\bZ \xrightarrow[\text{}]{\Z_{2N_f,A} \text{transformation}} \bZ \cdot
{\exp\big( 
\frac{  { -\ii   N   }}{2 \pi }   \int\limits_{M_4} (
 B_{c}^{(2)} \wedge  B_{f}^{(2)} )\big)}
\eea
and its dimensional reduction in 3d
\bea \label{eq:approximateanomalyBcBf3d}
\bZ \xrightarrow[\text{}]{\Z_{2N_f,A} \text{transformation}} \bZ \cdot
{\exp\big( 
\frac{  { -\ii   N   }}{2 \pi }   \int\limits_{M_4} (
 B_{c}^{(1)} \wedge  B_{f}^{(2)})\big)},
\eea
where $B_{c}$ and  $B_{f}$ are color/flavor background fields respectively.
\Ref{1706.06104Yonekura} derives constraints that have limitations on the chemical potential and boundary conditions as well.

\begin{figure}[!h]
\centering
\includegraphics[scale=.64]{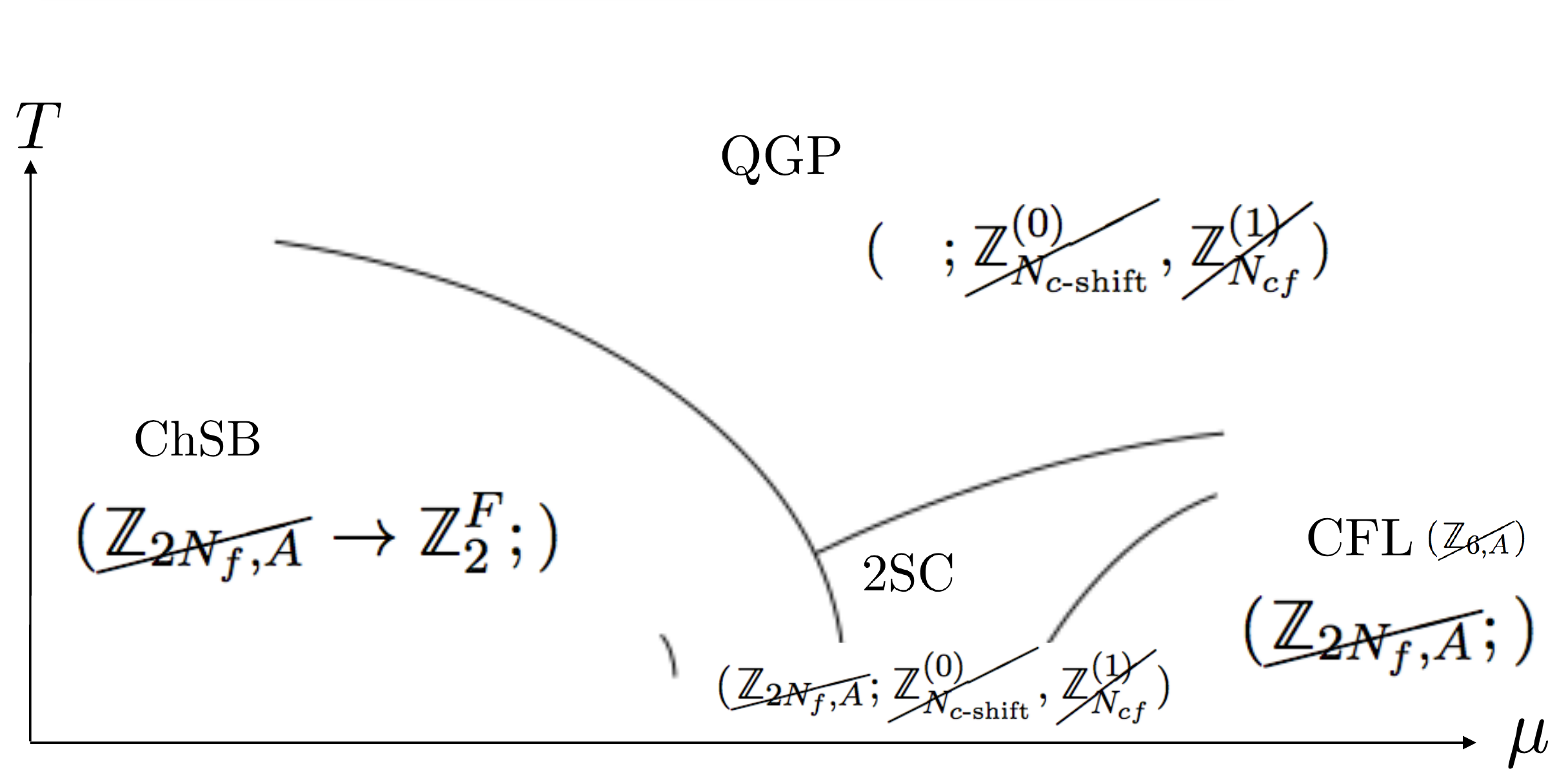}
\caption{Follow \Fig{fig:qcd-phase-1} and \Fig{fig:qcd-phase-T}, 
QCD$_4$ matter phases shown here include the high T QGP (quark-gluon plasma/liquid),
the low T ChSB (chiral symmetry breaking), 2SC (2-color superconductivity) and CFL (3-color-flavor locking superconductivity) at high density.
The four phases can be matched and cancelled by our approximate anomaly 
\Eq{eq:approximateanomalyBcf},
$\bZ \xrightarrow[\text{}]{\Z_{2N_f,A} \text{transformation}} \bZ \cdot
{\exp\big( 
\frac{  { -\ii   N   }}{2 \pi }   \int\limits_{M_4} (
 B_{cf}^{(2)} \wedge  B_{cf}^{(2)} )\big)}$.
The four phases can also be matched and cancelled by the anomaly 
\Eq{eq:approximateanomalyBcBf3d},
$\bZ \xrightarrow[\text{}]{\Z_{2N_f,A} \text{transformation}} \bZ \cdot
{\exp\big( 
\frac{  { -\ii   N   }}{2 \pi }   \int\limits_{M_4} (
 B_{c}^{(1)} \wedge  B_{f}^{(2)})\big)}$ introduced in \Ref{1706.06104Yonekura}. 
 We use the 
triple data $(\Z_{2N_f,A}; \Z_{N_{c\text{-shift}}}^{(0)}, {\Z^{{(1)}}_{N_{cf}}})$
to label
the discrete 0-form axial symmetry $\Z_{2N_f,A}$, 
a dimensionally reduced color-shift $\Z_{N_{c\text{-shift}}}^{(0)}$ symmetry introduced in \Ref{1706.06104Yonekura} , 
and the 1-form mixed chiral-flavor-locked center symmetry ${\Z^{{(1)}}_{N_{cf}}}$ that we introduce.
}
\label{fig:qcd-phase-break}
\end{figure}

In Figure \ref{fig:qcd-phase-break}, we show how our approximate anomaly in \Eq{eq:approximateanomalyBcf}
is still required and can be matched by the four QCD$_4$ matter phases.
We use the triple data with three imputs
\bea
(\Z_{2N_f,A}; \Z_{N_{c\text{-shift}}}^{}, {\Z^{}_{N_{cf},[1]}})
\eea
which in \Fig{fig:qcd-phase-break} can be also denoted as
\bea
(\Z_{2N_f,A}; \Z_{N_{c\text{-shift}}}^{(0)}, {\Z^{{(1)}}_{N_{cf}}}),
\eea
where the upper index ${{(1)}}$ in ${\Z^{{(1)}}_{N_{cf}}}$ indicates it is truly a 1-form symmetry in \Eq{eq:approximateanomalyBcf},
while the upper index ${{(0)}}$ in $\Z_{N_{c\text{-shift}}}^{(0)}$ indicates it is dimensionally reduced from 1-form symmetry  $\Z_{N_{c\text{-shift}},[1]}$ 
to a 0-form symmetry. The 1-form color-shift
 symmetry  $\Z_{N_{c\text{-shift}},[1]}$  is introduced in \Ref{1706.06104Yonekura}.
 
Let us indicate how the higher 't Hooft anomalies in \Eq{eq:approximateanomalyBcf} and in
\Eq{eq:approximateanomalyBcBf3d} can be matched by breaking some of the global symmetries 
in the triple data $(\Z_{2N_f,A}; {\Z^{{(1)}}_{N_{cf}}},\Z_{N_{c\text{-shift}}}^{(0)})$.
Our notations are that if the $G_{\text{sym}}$ is broken, we denote it as $\cancel{G_{\text{sym}}}$,
if the $G_{\text{sym}}$ is preserved, we would either indicate $G_{\text{sym}}$ remained, or simply omit the symbol as we did in the \Fig{fig:qcd-phase-break}.

\begin{enumerate}[leftmargin=2.0mm, label=\textcolor{blue}{\arabic*}., ref=\textcolor{blue}{\arabic*}]
\item 
{QGP (quark-gluon plasma/liquid) at high T}:
\bea\hspace{-5mm}
(\Z_{2N_f,A}; \Z_{N_{c\text{-shift}}}^{(0)}, {\Z^{{(1)}}_{N_{cf}}}) \xrightarrow[]{\text{SSB}}  
(\Z_{2N_f,A};   \cancel{\Z_{N_{c\text{-shift}}}^{(0)}} , \cancel{\Z^{{(1)}}_{N_{cf}}})
\sim
(\quad;  \cancel{\Z_{N_{c\text{-shift}}}^{(0)}} ,  \cancel{\Z^{{(1)}}_{N_{cf}}} )
 \text{ shown in \Fig{fig:qcd-phase-break} QGP}.
\eea
\item 
{ChSB (chiral symmetry breaking) at low T and at lower densities and low $\mu$}:
\bea \hspace{-5mm}
(\Z_{2N_f,A}; \Z_{N_{c\text{-shift}}}^{(0)}, {\Z^{{(1)}}_{N_{cf}}}) \xrightarrow[]{\text{SSB}} 
(\cancel{\Z_{2 N_f,A}}\to \Z_{2}^F;\Z_{N_{c\text{-shift}}}^{(0)}, {\Z^{{(1)}}_{N_{cf}}})  \sim
(\cancel{\Z_{2 N_f,A}}\to \Z_{2}^F; )
 \text{ shown in \Fig{fig:qcd-phase-break} ChSB}. \quad
\eea
\item 
{2SC (2-color superconductivity) at low T and at intermediate densities and $\mu$}:
\bea \hspace{-5mm}
(\Z_{2N_f,A}; \Z_{N_{c\text{-shift}}}^{(0)}, {\Z^{{(1)}}_{N_{cf}}}) \xrightarrow[]{\text{SSB}} 
(\cancel{\Z_{2 N_f,A}} ; \cancel{\Z_{N_{c\text{-shift}}}^{(0)}}, \cancel{\Z^{{(1)}}_{N_{cf}}})  
 \text{ shown in \Fig{fig:qcd-phase-break} 2SC}. \quad
\eea
\item \emph{CFL (3-color-flavor locking superconductivity) at low T and at high density and high $\mu$}:
\bea \hspace{-5mm}
(\Z_{2N_f,A}; \Z_{N_{c\text{-shift}}}^{(0)}, {\Z^{{(1)}}_{N_{cf}}}) \xrightarrow[]{\text{SSB}} 
(\cancel{\Z_{2 N_f,A}} ; {\Z_{N_{c\text{-shift}}}^{(0)}}, {\Z^{{(1)}}_{N_{cf}}})  
\sim
(\cancel{\Z_{2 N_f,A}} ; )
 \text{ shown in \Fig{fig:qcd-phase-break} CFL}. \quad
\eea
\end{enumerate}
What we have shown above is that 
higher 't Hooft anomalies in \Eq{eq:approximateanomalyBcf} and in
\Eq{eq:approximateanomalyBcBf3d} can indeed be matched by four phases via breaking some of the global symmetries.
We thus can constrain other possible QCD phases via the proposed approximate anomaly \Eq{eq:approximateanomalyBcf},
based on higher 't Hooft anomaly matching and cancellation.



\subsection{Mathematical Primer}
\label{sec:math}

In this article, we use spectral sequences (Adams spectral sequence, Atiyah-Hirzebruch spectral sequence, and Serre spectral sequence) to compute several cobordism groups which appear in QCD matter phases (QGP, ChSB, 2SC and CFL in \Sec{sec:QCD}). See \cite{1711.11587GPW, Wan2018bns1812.11967,2019arXiv191014668W} for a primer.

We aim to compute the cobordism group $\Omega_d^G$ for $d\le5$ where $G$ is the gauge group of QCD matter phases (QGP, ChSB, 2SC and CFL).

By the generalized Pontryagin-Thom isomorphism,
\bea
\Omega_d^{G}=\pi_d(MTG)
\eea
which identifies the cobordism group $\Omega_d^{G}$ with the homotopy group of the Madsen-Tillmann spectrum $MTG$.

We have the Adams spectral sequence
\bea
\Ext_{\A_p}^{s,t}(\H^*(Y,\Z_p),\Z_p)\Rightarrow\pi_{t-s}(Y)_p^{\wedge}.
\eea
Here $\A_p$ is the mod $p$ Steenrod algebra, $Y$ is any spectrum.
For any finitely generated abelian group $G$, $G_p^{\wedge}=\lim_{n\to\infty}G/p^nG$ is the $p$-completion of $G$.
In particular, $\A_2$ is generated by Steenrod squares $\Sq^i$.

In our cases, the cobordism groups only have 2-torsion and 3-torsion. 

We will use Adams spectral sequence to compute the 2-torsion part of the cobordism groups (we consider $Y=MTG$ in Adams spectral sequence, and we focus on $p=2$).

We will use Serre spectral sequence and Atiyah-Hirzebruch spectral sequence to compute the 3-torsion part of the cobordism groups. In our cases, in order to compute the 3-torsion part of the cobordism groups, we need only to compute the cobordism group $\Omega_d^{\SO}(\B G')$ for some group $G'$.

We have the Atiyah-Hirzebruch spectral sequence
\bea
\H_p(\B G',\Omega_q^{\SO})\Rightarrow\Omega_{p+q}^{\SO}(\B G').
\eea
Since 
\bea
\Omega_d^{\SO}=\left\{\begin{array}{ll}\Z&d=0\\0&d=1\\0&d=2\\0&d=3\\\Z&d=4\\\Z_2&d=5\end{array}\right.,
\eea
we need to know the integral homology groups $\H_p(\B G',\Z)$. In order to obtain this data, we compute the integral cohomology groups $\H^p(\B G',\Z)$ using Serre spectral sequence (we find a fibration of which $\B G'$ is the total space). 


For example, if $MTG=M\Spin\wedge X$ where $X$ is any spectrum,
by Corollary 5.1.2 of \cite{2018arXiv180107530B}, we have
\bea
\Ext_{\A_2}^{s,t}(\H^*(M\Spin\wedge X,\Z_2),\Z_2)=\Ext_{\A_2(1)}^{s,t}(\H^*(X,\Z_2),\Z_2)
\eea
for $t-s<8$.
Here $\A_2(1)$ is the subalgebra of $\A_2$ generated by $\Sq^1$ and $\Sq^2$.

So for the dimension $d=t-s<8$, we have
\bea\label{eq:ExtA_2(1)}
\Ext_{\A_2(1)}^{s,t}(\H^*(X,\Z_2),\Z_2)\Rightarrow(\Omega_{t-s}^{G})_2^{\wedge}.
\eea

The $\H^*(X,\Z_2)$ is an $\A_2(1)$-module whose internal degree $t$ is given by the $*$.

Our computation of $E_2$ pages of $\A_2(1)$-modules is based on 
Lemma 11 of \cite{Wan2018bns1812.11967}.
More precisely, we find a short exact sequence of $\A_2(1)$-modules $0\to L_1\to L_2\to L_3\to 0$, then apply Lemma 11 of \cite{Wan2018bns1812.11967} to compute $\Ext_{\A_2(1)}^{s,t}(L_2,\Z_2)$ by the data of $\Ext_{\A_2(1)}^{s,t}(L_1,\Z_2)$ and $\Ext_{\A_2(1)}^{s,t}(L_3,\Z_2)$. 
Our strategy is choosing $L_1$ to be the direct sum of suspensions of $\Z_2$ on which $\Sq^1$ and $\Sq^2$ act trivially, then we take $L_3$ to be the quotient of $L_2$ by $L_1$.
We can use this procedure again and again until $\Ext_{\A_2(1)}^{s,t}(L_3,\Z_2)$ is determined.

\section{QCD Symmetries, Anomalies and Topological Terms Without Time-Reversal}
\label{sec:WithoutTime-Reversal}

Before we consider the cobordism theory and co/bordism group of the following four QCD matter phases,
we need to convert our notations to involve both the internal symmetry and the spacetime symmetry.
Here are the results of conversions for the high T QGP (quark-gluon plasma/liquid),
the low T ChSB (chiral symmetry breaking), 2SC (2-color superconductivity) and CFL (3-color-flavor locking superconductivity) at high density.\\
QGP:
\bea
&&
\frac{[\SU(N_c)_V] \times \SU(N_f)_L \times  \SU(N_f)_R \times \U(1)_V}{\Z_{N_c, V} \times \Z_{N_f, V}} 
:=
\Big(\Spin(d) \times_{\Z_2} \frac{\SU(3)_L \times \SU(3)_R \times \U(1)_V}{(\Z_3)^2}
\Big)\nn\\
&&
=
\Big(\Spin(d) \times_{\Z_2} \frac{\U(3)_L \times \U(3)_R }{(\Z_3)^2 \times \U(1)_A}
\Big)
\times \Z_{3,L}
\times \Z_{3,R}
=
\Big(\Spin(d) \times_{\Z_2} \frac{\U(3)_L \times \U(3)_R }{(\Z_{3,V} \times \Z_{3,A} \times \U(1)_A)}
\Big).
\eea
Here $\U(1)_V=\frac{\U(1)_L \times \U(1)_R}{\U(1)_A}$, $\U(3)_L=\frac{\SU(3)_L \times \U(1)_L}{\Z_{3,L}}$,
$\U(3)_R=\frac{\SU(3)_R \times \U(1)_R}{\Z_{3,R}}$ where $L/R$ for $P_{L/R}=\frac{1 \pm \gamma^5}{2}$
ChSB:
\bea
{\frac{ \big( [\SU(N_c)_V] \times  \SU(N_f)_V \times \U(1)_V\big) }{ \Z_{N_c, V} \times \Z_{N_f, V} }} :
=
\Big(\Spin(d) \times_{\Z_2} \frac{\SU(3) \times \U(1)}{(\Z_3)^2}
\Big)
=
\Big(\Spin(d) \times_{\Z_2} \frac{\U(3)}{\Z_3}
\Big).
\eea
2SC:
\be
\frac{[\SU(2_c)_{V,rg}] \times \SU(2_f)_{L,ud} \times  \SU(2_f)_{R,ud}   {\times \U(1_f)_{V,s}} {\times \U(1_c)_{V,b}}}{{\Z_{2,V}^F}}:=
\Big(\Spin(d) \times_{\Z_2} (\SU(2) \times \SU(2))\Big) \times (\U(1) \times \U(1) ).
\ee
CFL:
\bea
\SU(3)_{C+L+R} \times \Z_{2,V}^F := \Spin(d) \times \SU(3)_{C+L+R}.
\eea

\subsection{Chiral symmetry breaking 
${\frac{ \big( [\SU(3)_V] \times  \SU(3)_V \times \U(1)_V\big) }{ \Z_{3_c, V} \times \Z_{3_f, V} }}$
as $(\Spin(d) \times_{\Z_2} \frac{\U(3)}{\Z_3})$}
For ChSB with a global symmetry:
$$
{\frac{ \big( [\SU(N_c)_V] \times  \SU(N_f)_V \times \U(1)_V\big) }{ \Z_{N_c, V} \times \Z_{N_f, V} }} :
=
\Big(\Spin(d) \times_{\Z_2} \frac{\SU(3) \times \U(1)}{(\Z_3)^2}
\Big)
=
\Big(\Spin(d) \times_{\Z_2} \frac{\U(3)}{\Z_3}
\Big).
$$

We have a fibration 
\bea
\xymatrix{\B\U(3)\ar[d]&\\
\B(\frac{\U(3)}{\Z_3})\ar[r]&\B^2\Z_3.}
\eea
Hence we have the Serre spectral sequence, see Figure \ref{fig:SSS}.
\bea
\H^p(\B^2\Z_3,\H^q(\B\U(3),\Z))\Rightarrow\H^{p+q}(\B(\frac{\U(3)}{\Z_3}),\Z).
\eea

We have
\bea
\H^*(\B\U(3),\Z)=\Z[c_1,c_2,c_3]
\eea
and
\bea
\H^p(\B^2\Z_3,\Z)=\left\{\begin{array}{llllllll}\Z&p=0\\0&p=1\\0&p=2\\\Z_3&p=3\\0&p=4\\\Z_3&p=5\\0&p=6\\\Z_9&p=7\end{array}\right..
\eea

\begin{figure}[H]
\center
\begin{sseq}[grid=none,labelstep=1,entrysize=1.5cm]{0...7}{0...6}
\ssdrop{\Z}
\ssmoveto 1 0 
\ssdrop{0}
\ssmoveto 2 0
\ssdrop{0}
\ssmoveto 3 0
\ssdrop{\Z_3}
\ssmoveto 4 0
\ssdrop{0}
\ssmoveto 5 0
\ssdrop{\Z_3}
\ssmoveto 6 0
\ssdrop{0}
\ssmoveto 7 0
\ssdrop{\Z_9}
\ssmoveto 0 1
\ssdrop{0}
\ssmoveto 1 1
\ssdrop{0}
\ssmoveto 2 1
\ssdrop{0}
\ssmoveto 3 1
\ssdrop{0}
\ssmoveto 4 1
\ssdrop{0}
\ssmoveto 5 1
\ssdrop{0}
\ssmoveto 6 1
\ssdrop{0}
\ssmoveto 7 1
\ssdrop{0}
\ssmoveto 0 2
\ssdrop{\Z}
\ssmoveto 1 2
\ssdrop{0}
\ssmoveto 2 2
\ssdrop{0}
\ssmoveto 3 2
\ssdrop{\Z_3}
\ssmoveto 4 2
\ssdrop{0}
\ssmoveto 5 2
\ssdrop{\Z_3}
\ssmoveto 6 2
\ssdrop{0}
\ssmoveto 7 2
\ssdrop{\Z_9}
\ssmoveto 0 3
\ssdrop{0}
\ssmoveto 1 3
\ssdrop{0}
\ssmoveto 2 3
\ssdrop{0}
\ssmoveto 3 3
\ssdrop{0}
\ssmoveto 4 3
\ssdrop{0}
\ssmoveto 5 3
\ssdrop{0}
\ssmoveto 6 3
\ssdrop{0}
\ssmoveto 7 3
\ssdrop{0}
\ssmoveto 0 4
\ssdrop{\Z^2}
\ssmoveto 1 4
\ssdrop{0}
\ssmoveto 2 4
\ssdrop{0}
\ssmoveto 3 4
\ssdrop{\Z_3^2}
\ssmoveto 4 4
\ssdrop{0}
\ssmoveto 5 4
\ssdrop{\Z_3^2}
\ssmoveto 6 4
\ssdrop{0}
\ssmoveto 7 4
\ssdrop{\Z_9^2}
\ssmoveto 0 5
\ssdrop{0}
\ssmoveto 1 5
\ssdrop{0}
\ssmoveto 2 5
\ssdrop{0}
\ssmoveto 3 5
\ssdrop{0}
\ssmoveto 4 5
\ssdrop{0}
\ssmoveto 5 5
\ssdrop{0}
\ssmoveto 6 5
\ssdrop{0}
\ssmoveto 7 5
\ssdrop{0}
\ssmoveto 0 6
\ssdrop{\Z^3}
\ssmoveto 1 6
\ssdrop{0}
\ssmoveto 2 6
\ssdrop{0}
\ssmoveto 3 6
\ssdrop{\Z_3^3}
\ssmoveto 4 6
\ssdrop{0}
\ssmoveto 5 6
\ssdrop{\Z_3^3}
\ssmoveto 6 6
\ssdrop{0}
\ssmoveto 7 6
\ssdrop{\Z_9^3}

\ssmoveto 0 4
\ssarrow[color=red] 5 {-4}

\ssmoveto 0 4
\ssarrow[color=red,dashed] 3 {-2}

\end{sseq}
\center
\caption{Serre spectral sequence for the fibration $\B\U(3)\to\B(\frac{\U(3)}{\Z_3})\to\B^2\Z_3$. The arrow from (0,4) to (5,0) is a nontrivial differential by comparison with the Serre spectral sequence of the fibration $\B\SU(3)\to\B\PSU(3)\to\B^2\Z_3$.
There is no differential from (0,2) to (3,0) since the 3d $\Z_3$ survives the spectral sequence in Figure \ref{fig:SSS-1}.
}
\label{fig:SSS}
\end{figure}

There is another approach: we have a fibration
\bea
\xymatrix{\B\PSU(3)\ar[d]&\\
\B(\frac{\U(3)}{\Z_3})\ar[r]&\B\U(1).}
\eea
Hence we have the Serre spectral sequence, see Figure \ref{fig:SSS-1}.
\bea
\H^p(\B\U(1),\H^q(\B\PSU(3),\Z))\Rightarrow\H^{p+q}(\B(\frac{\U(3)}{\Z_3}),\Z).
\eea

We have
\bea
\H^*(\B\U(1),\Z)=\Z[c_1]
\eea
and
\bea
\H^p(\B\PSU(3),\Z)=\left\{\begin{array}{llllllll}\Z&p=0\\0&p=1\\0&p=2\\\Z_3&p=3\\\Z&p=4\\0&p=5\\\Z&p=6\end{array}\right..
\eea

\begin{figure}[H]
\center
\begin{sseq}[grid=none,labelstep=1,entrysize=1.5cm]{0...6}{0...6}
\ssdrop{\Z}
\ssmoveto 1 0 
\ssdrop{0}
\ssmoveto 2 0
\ssdrop{\Z}
\ssmoveto 3 0
\ssdrop{0}
\ssmoveto 4 0
\ssdrop{\Z}
\ssmoveto 5 0
\ssdrop{0}
\ssmoveto 6 0
\ssdrop{\Z}

\ssmoveto 0 1
\ssdrop{0}
\ssmoveto 1 1
\ssdrop{0}
\ssmoveto 2 1
\ssdrop{0}
\ssmoveto 3 1
\ssdrop{0}
\ssmoveto 4 1
\ssdrop{0}
\ssmoveto 5 1
\ssdrop{0}
\ssmoveto 6 1
\ssdrop{0}

\ssmoveto 0 2
\ssdrop{0}
\ssmoveto 1 2
\ssdrop{0}
\ssmoveto 2 2
\ssdrop{0}
\ssmoveto 3 2
\ssdrop{0}
\ssmoveto 4 2
\ssdrop{0}
\ssmoveto 5 2
\ssdrop{0}
\ssmoveto 6 2
\ssdrop{0}

\ssmoveto 0 3
\ssdrop{\Z_3}
\ssmoveto 1 3
\ssdrop{0}
\ssmoveto 2 3
\ssdrop{\Z_3}
\ssmoveto 3 3
\ssdrop{0}
\ssmoveto 4 3
\ssdrop{\Z_3}
\ssmoveto 5 3
\ssdrop{0}
\ssmoveto 6 3
\ssdrop{\Z_3}

\ssmoveto 0 4
\ssdrop{\Z}
\ssmoveto 1 4
\ssdrop{0}
\ssmoveto 2 4
\ssdrop{\Z}
\ssmoveto 3 4
\ssdrop{0}
\ssmoveto 4 4
\ssdrop{\Z}
\ssmoveto 5 4
\ssdrop{0}
\ssmoveto 6 4
\ssdrop{\Z}

\ssmoveto 0 5
\ssdrop{0}
\ssmoveto 1 5
\ssdrop{0}
\ssmoveto 2 5
\ssdrop{0}
\ssmoveto 3 5
\ssdrop{0}
\ssmoveto 4 5
\ssdrop{0}
\ssmoveto 5 5
\ssdrop{0}
\ssmoveto 6 5
\ssdrop{0}

\ssmoveto 0 6
\ssdrop{\Z}
\ssmoveto 1 6
\ssdrop{0}
\ssmoveto 2 6
\ssdrop{\Z}
\ssmoveto 3 6
\ssdrop{0}
\ssmoveto 4 6
\ssdrop{\Z}
\ssmoveto 5 6
\ssdrop{0}
\ssmoveto 6 6
\ssdrop{\Z}

\ssmoveto 0 4
\ssarrow[color=red,dashed] 2 {-1}

\end{sseq}
\center
\caption{Serre spectral sequence for the fibration $\B\PSU(3)\to\B(\frac{\U(3)}{\Z_3})\to\B\U(1)$. 
}
\label{fig:SSS-1}
\end{figure}

In Figure \ref{fig:SSS-1}, we find that the 3d $\Z_3$ survives the spectral sequence, so in Figure \ref{fig:SSS}, there is no nontrivial differential from (0,2) to (3,0). Since the differential is a derivation, we conclude that in Figure \ref{fig:SSS} the dashed arrow from (0,4) to (3,2) does not actually exist. So there is a $\Z_3$ in 5d survives the spectral sequence, thus in Figure \ref{fig:SSS-1}, the dashed arrow from (0,4) to (2,3) also does not actually exist.

So 
\bea
\H^p(\B(\frac{\U(3)}{\Z_3}),\Z)=\left\{\begin{array}{lllllll}\Z&p=0\\0&p=1\\\Z&p=2\\\Z_3&p=3\\\Z^2&p=4\\\Z_3&p=5\\\Z^3&p=6\end{array}\right.
\eea
and
\bea
\H_p(\B(\frac{\U(3)}{\Z_3}),\Z)=\left\{\begin{array}{lllllll}\Z&p=0\\0&p=1\\\Z\times\Z_3&p=2\\0&p=3\\\Z^2\times\Z_3&p=4\\0&p=5\\\Z^3\times?&p=6\end{array}\right..
\eea
Here ? is an undetermined 3-torsion group.

By the Atiyah-Hirzebruch spectral sequence, we have
\bea
\H_p(\B(\frac{\U(3)}{\Z_3}),\Omega_q^{\SO})\Rightarrow\Omega_{p+q}^{\SO}(\B(\frac{\U(3)}{\Z_3})).
\eea

See Figure \ref{fig:AHSS}.

\begin{figure}[H]
\center
\begin{sseq}[grid=none,labelstep=1,entrysize=1.5cm]{0...6}{0...5}
\ssdrop{\Z}
\ssmoveto 1 0 
\ssdrop{0}
\ssmoveto 2 0
\ssdrop{\Z\times\Z_3}
\ssmoveto 3 0
\ssdrop{0}
\ssmoveto 4 0
\ssdrop{\Z^2\times\Z_3}
\ssmoveto 5 0
\ssdrop{0}
\ssmoveto 6 0
\ssdrop{\Z^3\times?}
\ssmoveto 0 1
\ssdrop{0}
\ssmoveto 1 1
\ssdrop{0}
\ssmoveto 2 1
\ssdrop{0}
\ssmoveto 3 1
\ssdrop{0}
\ssmoveto 4 1
\ssdrop{0}
\ssmoveto 5 1
\ssdrop{0}

\ssmoveto 0 2
\ssdrop{0}
\ssmoveto 1 2
\ssdrop{0}
\ssmoveto 2 2
\ssdrop{0}
\ssmoveto 3 2
\ssdrop{0}
\ssmoveto 4 2
\ssdrop{0}

\ssmoveto 0 3
\ssdrop{0}
\ssmoveto 1 3
\ssdrop{0}
\ssmoveto 2 3
\ssdrop{0}
\ssmoveto 3 3
\ssdrop{0}
\ssmoveto 0 4
\ssdrop{\Z}
\ssmoveto 1 4
\ssdrop{0}
\ssmoveto 2 4
\ssdrop{\Z\times\Z_3}
\ssmoveto 0 5
\ssdrop{\Z_2}

\end{sseq}
\center
\caption{Atiyah-Hirzebruch spectral sequence for $\Omega_d^{\SO \times \frac{\U(3)}{\Z_3}}$. Here ? is an undetermined 3-torsion group.}
\label{fig:AHSS}
\end{figure}

Since the localization of $\Spin\times_{\Z_2}\frac{\U(3)}{\Z_3}$ and $\SO\times\frac{\U(3)}{\Z_3}$ at the prime 3 are the same, the 3-torsion part of $\Omega_d^{\Spin\times_{\Z_2}\frac{\U(3)}{\Z_3}}$ and $\Omega_d^{\SO\times\frac{\U(3)}{\Z_3}}$ are the same.

Since the localization of $\Spin\times_{\Z_2}\frac{\U(3)}{\Z_3}$ and $\Spin\times_{\Z_2}\U(3)$ at the prime 2 are the same, the 2-torsion part of $\Omega_d^{\Spin\times_{\Z_2}\frac{\U(3)}{\Z_3}}$ and $\Omega_d^{\Spin\times_{\Z_2}\U(3)}$ are the same. 

Also since $\U(3)=\SU(3)\times_{\Z_3}\U(1)$, the localization of $\Spin\times_{\Z_2}\U(3)$ and $\Spin\times_{\Z_2}(\U(1)\times\SU(3))$ at the prime 2 are the same, the 2-torsion part of $\Omega_d^{\Spin\times_{\Z_2}\U(3)}$ and $\Omega_d^{\Spin\times_{\Z_2}(\U(1)\times\SU(3))}=\Omega_d^{\Spin^c\times\SU(3)}$ are the same.

We have $MT(\Spin^c \times \SU(3))=M\Spin\wedge\Sigma^{-2}M\U(1)\wedge (\B \SU(3))_+$.

For $t-s<8$, since there is no odd torsion, we have the Adams spectral sequence
\bea
\Ext_{\A_2(1)}^{s,t}(\H^{*+2}(M\U(1),\Z_2)\otimes\H^*(\B \SU(3),\Z_2),\Z_2)\Rightarrow\Omega_{t-s}^{\Spin^c \times \SU(3)}.
\eea

We have
\bea
\H^*(\B \SU(3),\Z_2)=\Z_2[c_2,c_3]
\eea
where $c_i$ is the Chern class of the $\SU(3)$ bundle.

By Thom isomorphism,
\bea
\H^{*+2}(M\U(1),\Z_2)=\Z_2[c_1]U
\eea
where $c_1$ is the Chern class of the $\U(1)$ bundle and $U$ is the Thom class.

\begin{figure}[H]
\begin{center}
\begin{tikzpicture}
\node at (0,-1) {0};
\node at (1,-1) {1};
\node at (2,-1) {2};
\node at (3,-1) {3};
\node at (4,-1) {4};
\node at (5,-1) {5};
\node at (6,-1) {$t-s$};
\node at (-1,0) {0};
\node at (-1,1) {1};
\node at (-1,2) {2};
\node at (-1,3) {3};
\node at (-1,4) {4};
\node at (-1,5) {5};
\node at (-1,6) {$s$};

\draw[->] (-0.5,-0.5) -- (-0.5,6);
\draw[->] (-0.5,-0.5) -- (6,-0.5);

\draw (0,0) -- (0,5);

\draw (2,1) -- (2,5);
\draw (4,2) -- (4,5);
\draw (4.1,0) -- (4.1,5);
\draw (4.2,0) -- (4.2,5);

\end{tikzpicture}
\end{center}
\caption{$\Omega_*^{\Spin^c \times \SU(3)}$.}
\label{fig:E_2SpincSU3}
\end{figure}
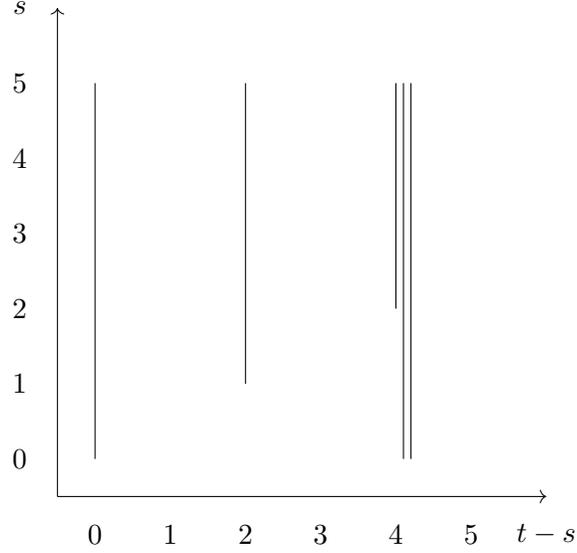

The $\A_2(1)$-module structure of $\H^{*+2}(M\U(1),\Z_2)\otimes\H^*(\B \SU(3),\Z_2)$ below degree 6 and the $E_2$ page are shown in Figure \ref{fig:A_2(1)MU1SU3}, \ref{fig:E_2SpincSU3}.

\begin{figure}[H]
\begin{center}
\begin{tikzpicture}[scale=0.5]

\node[below] at (0,0) {$U$};
\draw[fill] (0,0) circle(.1);

\node[right] at (0,2) {$c_1U$};
\node[right] at (0,4) {$c_1^2U$};

\node[right] at (0,6) {$c_1^3U$};
\draw[fill] (0,2) circle(.1);
\draw[fill] (0,4) circle(.1);
\draw[fill] (0,6) circle(.1);
\draw (0,0) to [out=150,in=150] (0,2);
\draw (0,4) to [out=150,in=150] (0,6);

\node[right] at (2,4) {$c_2U$};
\node[right] at (2,6) {$c_3U$};
\draw[fill] (2,4) circle(.1);
\draw[fill] (2,6) circle(.1);

\draw (2,4) to [out=150,in=150] (2,6);

\end{tikzpicture}
\end{center}
\caption{The $\A_2(1)$-module structure of $\H^{*+2}(M\U(1),\Z_2)\otimes\H^*(\B \SU(3),\Z_2)$ below degree 6.}
\label{fig:A_2(1)MU1SU3}
\end{figure}

So there is no 2-torsion in $\Omega^{\Spin \times_{\Z_2} \frac{\U(3)}{\Z_3}}_d$.

Combine the 2-torsion and 3-torsion results, we have
\begin{table}[H]
\centering
\begin{tabular}{ c c c}
\hline
\multicolumn{3}{c}{Bordism group}\\
\hline
$d$ & 
$\Omega^{\Spin \times_{\Z_2} \frac{\U(3)}{\Z_3}}_d$
& generators \\
\hline
0& $\Z$\\
1& 0\\
2& $\Z\times\Z_3$ \\
3 & $0$\\
4 & $e(\Z^2\times\Z_3,\Z)$ \\
5 & $0$ \\
\hline
\end{tabular}
\caption{Bordism group. The notation $e(A,B)$ denotes a group extension of A by B, that is, a group that fits into the following short exact sequence
$0\to B\to e(A,B)\to A\to0$. 
}
\label{table:U3Z3Z2Bordism}
\end{table}

\subsection{3-Color-Flavor locking superconductivity $\SU(3)_{C+L+R} \times \Z_{2,V}^F$ as $\Spin\times\SU(3)$}

We have $MT(\Spin \times \SU(3))=M\Spin\wedge (\B \SU(3))_+$.

For $t-s<8$, since there is no odd torsion, we have the Adams spectral sequence
\bea
\Ext_{\A_2(1)}^{s,t}(\H^*(\B \SU(3),\Z_2),\Z_2)\Rightarrow\Omega_{t-s}^{\Spin \times \SU(3)}.
\eea

We have
\bea
\H^*(\B \SU(3),\Z_2)=\Z_2[c_2,c_3]
\eea
where $c_i$ is the Chern class of the $\SU(3)$ bundle.

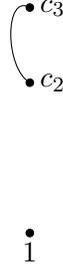
\begin{figure}[H]
\begin{center}
\begin{tikzpicture}[scale=0.5]

\node[below] at (0,0) {$1$};
\draw[fill] (0,0) circle(.1);

\node[right] at (0,4) {$c_2$};
\node[right] at (0,6) {$c_3$};
\draw[fill] (0,4) circle(.1);
\draw[fill] (0,6) circle(.1);

\draw (0,4) to [out=150,in=150] (0,6);

\end{tikzpicture}
\end{center}
\caption{The $\A_2(1)$-module structure of $\H^*(\B \SU(3),\Z_2)$ below degree 6.}
\label{fig:A_2(1)SU3}
\end{figure}

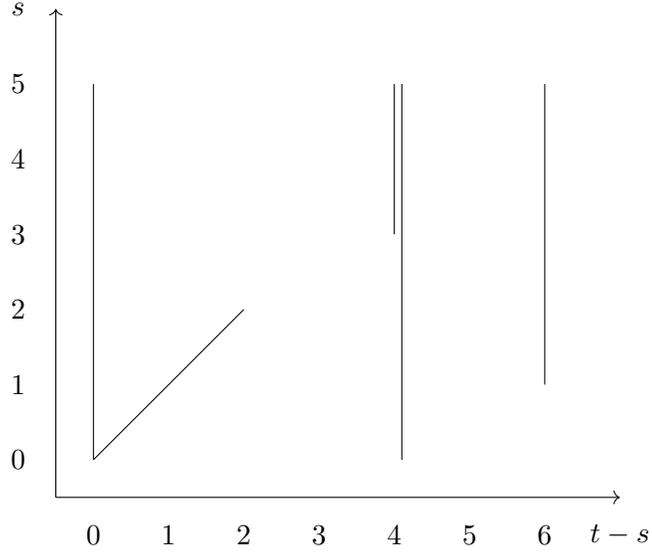
\begin{figure}[H]
\begin{center}
\begin{tikzpicture}
\node at (0,-1) {0};
\node at (1,-1) {1};
\node at (2,-1) {2};
\node at (3,-1) {3};
\node at (4,-1) {4};
\node at (5,-1) {5};
\node at (6,-1) {6};
\node at (7,-1) {$t-s$};
\node at (-1,0) {0};
\node at (-1,1) {1};
\node at (-1,2) {2};
\node at (-1,3) {3};
\node at (-1,4) {4};
\node at (-1,5) {5};
\node at (-1,6) {$s$};

\draw[->] (-0.5,-0.5) -- (-0.5,6);
\draw[->] (-0.5,-0.5) -- (7,-0.5);

\draw (0,0) -- (0,5);
\draw (0,0) -- (2,2);
\draw (4,3) -- (4,5);
\draw (4.1,0) -- (4.1,5);
\draw (6,1) -- (6,5);

\end{tikzpicture}
\end{center}
\caption{$\Omega_*^{\Spin \times \SU(3)}$.}
\label{fig:E_2SU3}
\end{figure}

The $\A_2(1)$-module structure of $\H^*(\B \SU(3),\Z_2)$ below degree 6 and the $E_2$ page are shown in Figure \ref{fig:A_2(1)SU3}, \ref{fig:E_2SU3}.

\begin{table}[H]
\centering
\begin{tabular}{ c c c}
\hline
\multicolumn{3}{c}{Bordism group}\\
\hline
$d$ & 
$\Omega^{\Spin \times \SU(3)}_d$
& generators \\
\hline
0& $\Z$\\
1& $\Z_2$ & $\tilde\eta$\\
2& $\Z_2$ & Arf\\
3 & $0$\\
4 & $\Z^2$ & $\frac{\sigma}{16},c_2$\\
5 & $0$ \\
6 & $\Z$ & $\frac{c_3}{2}$ \\
\hline
\end{tabular}
\caption{Bordism group. 
$\tilde\eta$ is a mod 2 index of 1d Dirac operator.
Arf is a 2d Arf invariant.
$\sigma$ is the signature of manifold.
Note that $c_3=\Sq^2c_2=(w_2+w_1^2)c_2=0\mod2$ on Spin 6-manifolds.
}
\label{table:SU3Bordism}
\end{table}

\subsection{2-Color Superconductivity: $\frac{[\SU(2_c)_{V,rg}] \times \SU(2_f)_{L,ud} \times  \SU(2_f)_{R,ud}   {\times \U(1_f)_{V,s}} {\times \U(1_c)_{V,b}}}{{\Z_{2,V}^F}}$ as $\Spin\times_{\Z_2}\Spin(4)\times\U(1)\times\U(1)$}

We have $MT(\Spin\times_{\Z_2}\Spin(4)\times\U(1)\times\U(1))=M\Spin\wedge\Sigma^{-4}M\SO(4)\wedge(\B\U(1)\times\B\U(1))_+$.

For $t-s<8$, since there is no odd torsion, we have the Adams spectral sequence
\bea
\Ext_{\A_2(1)}^{s,t}(\H^{*+4}(M\SO(4),\Z_2)\otimes\H^*(\B\U(1),\Z_2)\otimes\H^*(\B\U(1),\Z_2),\Z_2)\Rightarrow \Omega_{t-s}^{\Spin\times_{\Z_2}\Spin(4)\times\U(1)\times\U(1)}.
\eea
By Thom isomorphism, we have 
\bea
\H^{*+4}(M\SO(4),\Z_2)=\Z_2[w_2',w_3',w_4']U
\eea
where $w_i'$ is the Stiefel-Whitney class of the $\SO(4)$ bundle and $U$ is the Thom class.

We also have 
\bea
\H^*(\B\U(1),\Z_2)=\Z_2[c_1]
\eea
where $c_1$ is the first Chern class of the $\U(1)$ bundle.

\begin{figure}[H]
\begin{center}
\begin{tikzpicture}
\node at (0,-1) {0};
\node at (1,-1) {1};
\node at (2,-1) {2};
\node at (3,-1) {3};
\node at (4,-1) {4};
\node at (5,-1) {5};
\node at (6,-1) {$t-s$};
\node at (-1,0) {0};
\node at (-1,1) {1};
\node at (-1,2) {2};
\node at (-1,3) {3};
\node at (-1,4) {4};
\node at (-1,5) {5};
\node at (-1,6) {$s$};

\draw[->] (-0.5,-0.5) -- (-0.5,6);
\draw[->] (-0.5,-0.5) -- (6,-0.5);

\draw (0,0) -- (0,5);

\draw (2,0) -- (2,5);

\draw (2.1,0) -- (2.1,5);
\draw (3.7,0) -- (3.7,5);
\draw (3.8,0) -- (3.8,5);

\draw (3.9,0) -- (3.9,5);

\draw (4,1) -- (4,5);
\draw (4,1) -- (5,2);
\draw (4.1,0) -- (4.1,5);
\draw (4.2,0) -- (4.2,5);
\draw (4.2,0) -- (5.2,1);
\draw[fill] (5,0) circle(0.05);

\end{tikzpicture}
\end{center}
\caption{$\Omega_*^{\Spin\times_{\Z_2}\Spin(4)\times\U(1)\times\U(1)}$.}
\label{fig:E_2Spin4Z2U1U1}
\end{figure}
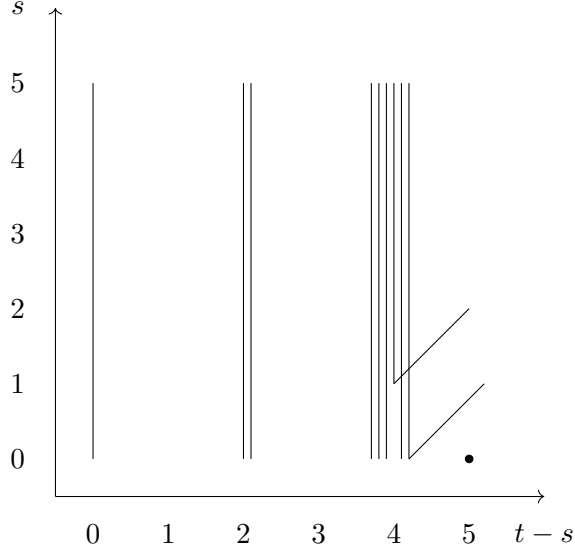

The $\A_2(1)$-module structure of $\H^{*+4}(M\SO(4),\Z_2)\otimes\H^*(\B\U(1),\Z_2)\otimes\H^*(\B\U(1),\Z_2)$ below degree 5 and the $E_2$ page are shown in Figure \ref{fig:A_2(1)MSO4BU1BU1}, \ref{fig:E_2Spin4Z2U1U1}.

\begin{figure}[H]
\begin{center}
\begin{tikzpicture}[scale=0.5]

\node[below] at (0,0) {$1$};

\draw[fill] (0,0) circle(.1);

\node[right] at (0,2) {$c_1$};

\draw[fill] (0,2) circle(.1);

\node[right] at (0,4) {$c_1^2$};

\draw[fill] (0,4) circle(.1);

\draw (0,2) to [out=150,in=150] (0,4);

\node[right] at (2,2) {$c_1'$};

\draw[fill] (2,2) circle(.1);

\node[right] at (2,4) {$c_1'^2$};

\draw[fill] (2,4) circle(.1);

\draw (2,2) to [out=150,in=150] (2,4);

\node[right] at (4,4) {$c_1c_1'$};

\draw[fill] (4,4) circle(.1);

\draw[fill] (4,6) circle(.1);

\draw (4,4) to [out=150,in=150] (4,6);

\node at (6,3) {$\bigotimes$};

\node[below] at (8,0) {$U$};

\draw[fill] (8,0) circle(.1);
\draw[fill] (8,2) circle(.1);
\draw (8,0) to [out=150,in=150] (8,2);
\draw[fill] (8,3) circle(.1);
\draw (8,2) -- (8,3);

\node[left] at (8,4) {$w_4'U$};

\draw[fill] (8,4) circle(.1);

\node[below] at (10,4) {$w_2'^2U$};

\draw[fill] (10,4) circle(.1);
\draw[fill] (10,5) circle(.1);
\draw[fill] (10,6) circle(.1);
\draw[fill] (11,6) circle(.1);
\draw (10,4) to [out=30,in=150] (11,6);
\draw (10,5) -- (10,6);
\draw[fill] (11,7) circle(.1);
\draw (11,6) -- (11,7);
\draw (10,5) to [out=30,in=150] (11,7);
\draw[fill] (11,8) circle(.1);
\draw[fill] (11,9) circle(.1);
\draw (10,6) to [out=30,in=150] (11,8);
\draw (11,7) to [out=30,in=30] (11,9);
\draw (11,8) -- (11,9);

\node at (-2,-10) {$=$};

\node[below] at (0,-15) {$U$};

\draw[fill] (0,-15) circle(.1);
\draw[fill] (0,-13) circle(.1);
\draw (0,-15) to [out=150,in=150] (0,-13);
\draw[fill] (0,-12) circle(.1);
\draw (0,-13) -- (0,-12);

\node[left] at (0,-11) {$w_4'U$};

\draw[fill] (0,-11) circle(.1);

\node[below] at (2,-11) {$w_2'^2U$};

\draw[fill] (2,-11) circle(.1);
\draw[fill] (2,-10) circle(.1);
\draw[fill] (2,-9) circle(.1);
\draw[fill] (3,-9) circle(.1);
\draw (2,-11) to [out=30,in=150] (3,-9);
\draw (2,-10) -- (2,-9);
\draw[fill] (3,-8) circle(.1);
\draw (3,-9) -- (3,-8);
\draw (2,-10) to [out=30,in=150] (3,-8);
\draw[fill] (3,-7) circle(.1);
\draw[fill] (3,-6) circle(.1);
\draw (2,-9) to [out=30,in=150] (3,-7);
\draw (3,-8) to [out=30,in=30] (3,-6);
\draw (3,-7) -- (3,-6);

\node[below] at (4,-13) {$c_1U$};

\draw[fill] (4,-13) circle(.1);
\draw[fill] (4,-11) circle(.1);
\draw (4,-13) to [out=150,in=150] (4,-11);
\draw[fill] (4,-10) circle(.1);
\draw (4,-11) -- (4,-10);
\draw[fill] (4,-9) circle(.1);
\draw[fill] (4,-8) circle(.1);
\draw (4,-10) to [out=150,in=150] (4,-8);
\draw (4,-9) -- (4,-8);
\draw[fill] (5,-11) circle(.1);
\draw (5,-11) to [out=150,in=30] (4,-9);
\node[below] at (5,-11) {$c_1^2U$};

\node[below] at (6,-13) {$c_1'U$};

\draw[fill] (6,-13) circle(.1);
\draw[fill] (6,-11) circle(.1);
\draw (6,-13) to [out=150,in=150] (6,-11);
\draw[fill] (6,-10) circle(.1);
\draw (6,-11) -- (6,-10);
\draw[fill] (6,-9) circle(.1);
\draw[fill] (6,-8) circle(.1);
\draw (6,-10) to [out=150,in=150] (6,-8);
\draw (6,-9) -- (6,-8);
\draw[fill] (7,-11) circle(.1);
\draw (7,-11) to [out=150,in=30] (6,-9);
\node[below] at (7,-11) {$c_1'^2U$};

\node[below] at (9,-11) {$c_1c_1'U$};

\draw[fill] (9,-11) circle(.1);
\draw[fill] (9,-9) circle(.1);
\draw (9,-11) to [out=150,in=150] (9,-9);
\draw[fill] (9,-8) circle(.1);
\draw (9,-9) -- (9,-8);
\draw[fill] (9,-7) circle(.1);
\draw[fill] (9,-6) circle(.1);
\draw (9,-8) to [out=150,in=150] (9,-6);
\draw (9,-7) -- (9,-6);
\draw[fill] (10,-9) circle(.1);
\draw (10,-9) to [out=150,in=30] (9,-7);

\end{tikzpicture}
\end{center}
\caption{The $\A_2(1)$-module structure of $\H^{*+4}(M\SO(4),\Z_2)\otimes\H^*(\B\U(1),\Z_2)\otimes\H^*(\B\U(1),\Z_2)$ below degree 5.}
\label{fig:A_2(1)MSO4BU1BU1}
\end{figure}

\begin{table}[H]
\centering
\begin{tabular}{ c c c}
\hline
\multicolumn{3}{c}{Bordism group}\\
\hline
$d$ & 
$\Omega^{\Spin\times_{\Z_2}\Spin(4)\times\U(1)\times\U(1)}_d$
& generators \\
\hline
0& $\Z$\\
1& $0$\\
2& $\Z^2$  & $c_1,c_1'$\\
3 & $0$\\
4 & $\Z^6$ & $p_1',e_4',c_1^2,c_1'^2,c_1c_1',?$ \\
5 & $\Z_2^3$  & $w_2'w_3',w_4'\tilde\eta,w_3'\text{Arf}(?)$ \\
\hline
\end{tabular}
\caption{Bordism group. Here $e_i$ is the Euler class, $p_i$ is the Pontryagin class, $c_i$ is the Chern class. 
Here $w_i'=w_i(\SO(4))$, $p_1'=p_1(\SO(4))$, $e_i'=e_i(\SO(4))$.
$\tilde\eta$ is the mod 2 index of 1d Dirac operator, Arf is the Arf invariant.
The ? is an undetermined cobordism invariant which also appears in \cite{2019arXiv191014668W}.
}
\label{table:Spin4Z2U1U1Bordism}
\end{table}

\subsection{Quark Gluon Plasma/Liquid
$\frac{[\SU(3)_V] \times \SU(3)_L \times  \SU(3)_R \times \U(1)_V}{\Z_{3, V} \times \Z_{3, V}} $ as
$\Spin(d)\times_{\Z_2}\frac{\U(3)_L \times \U(3)_R }{(\Z_{3,V} \times \Z_{3,A} \times \U(1)_A)}$ 
}

Since the localization of $\Spin\times_{\Z_2}\frac{\U(3)\times\U(3)}{\Z_3\times\Z_3\times\U(1)}$ and $\SO\times \frac{\U(3)\times\U(3)}{\Z_3\times\Z_3\times\U(1)}$ at the prime 3 are the same, the 3-torsion of $\Omega_d^{\Spin\times_{\Z_2}\frac{\U(3)\times\U(3)}{\Z_3\times\Z_3\times\U(1)}}$ and $\Omega_d^{\SO\times \frac{\U(3)\times\U(3)}{\Z_3\times\Z_3\times\U(1)}}$ are the same.

Since the localization of $\Spin\times_{\Z_2}\frac{\U(3)\times\U(3)}{\Z_3\times\Z_3\times\U(1)}$ and $\Spin\times_{\Z_2}(\U(1)\times\SU(3)\times\SU(3))$ at the prime 2 are the same, the 2-torsion of $\Omega_d^{\Spin\times_{\Z_2}\frac{\U(3)\times\U(3)}{\Z_3\times\Z_3\times\U(1)}}$ and $\Omega_d^{\Spin^c\times\SU(3)\times\SU(3)}$ are the same.

We have $MT(\Spin^c  \times \SU(3)\times \SU(3))=M\Spin\wedge\Sigma^{-2}M\U(1)\wedge (\B\SU(3)\times\B \SU(3))_+$.

For $t-s<8$, since there is no odd torsion, we have the Adams spectral sequence
\bea
\Ext_{\A_2(1)}^{s,t}(\H^{*+2}(M\U(1),\Z_2)\otimes\H^*(\B \SU(3),\Z_2)\otimes\H^*(\B \SU(3),\Z_2),\Z_2)\Rightarrow\Omega_{t-s}^{\Spin^c  \times \SU(3)\times \SU(3)}.
\eea

We have
\bea
\H^*(\B \SU(3)\times\B\SU(3),\Z_2)=\Z_2[c_2,c_3,c_2',c_3']
\eea
where $c_i$ is the Chern class of the $\SU(3)$ bundle.

By Thom isomorphism,
\bea
\H^{*+2}(M\U(1),\Z_2)=\Z_2[c_1]U
\eea
where $c_1$ is the Chern class of the $\U(1)$ bundle and $U$ is the Thom class.

\begin{figure}[H]
\begin{center}
\begin{tikzpicture}
\node at (0,-1) {0};
\node at (1,-1) {1};
\node at (2,-1) {2};
\node at (3,-1) {3};
\node at (4,-1) {4};
\node at (5,-1) {5};
\node at (6,-1) {$t-s$};
\node at (-1,0) {0};
\node at (-1,1) {1};
\node at (-1,2) {2};
\node at (-1,3) {3};
\node at (-1,4) {4};
\node at (-1,5) {5};
\node at (-1,6) {$s$};

\draw[->] (-0.5,-0.5) -- (-0.5,6);
\draw[->] (-0.5,-0.5) -- (6,-0.5);

\draw (0,0) -- (0,5);

\draw (2,1) -- (2,5);
\draw (4,2) -- (4,5);
\draw (4.1,0) -- (4.1,5);
\draw (4.2,0) -- (4.2,5);

\draw (4.3,0) -- (4.3,5);

\end{tikzpicture}
\end{center}
\caption{$\Omega_*^{\Spin^c \times\SU(3)\times \SU(3)}$.}
\label{fig:E_2SpincSU3SU3}
\end{figure}
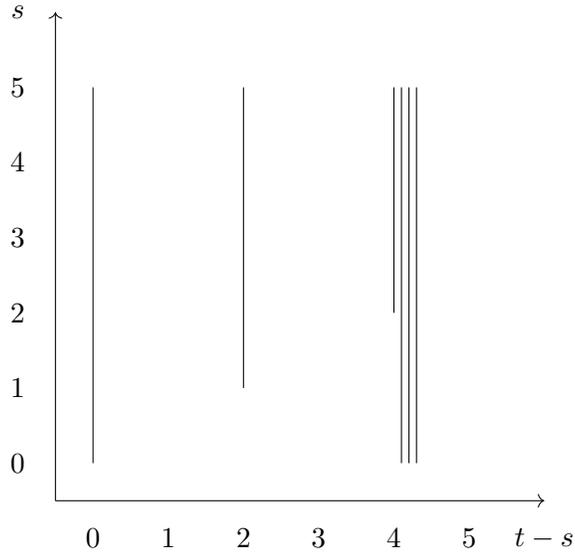

The $\A_2(1)$-module structure of $\H^{*+2}(M\U(1),\Z_2)\otimes\H^*(\B \SU(3),\Z_2)\otimes\H^*(\B \SU(3),\Z_2)$ below degree 6 and the $E_2$ page are shown in Figure \ref{fig:A_2(1)MU1SU3SU3}, \ref{fig:E_2SpincSU3SU3}.

\begin{figure}[H]
\begin{center}
\begin{tikzpicture}[scale=0.5]

\node[below] at (0,0) {$U$};
\draw[fill] (0,0) circle(.1);

\node[right] at (0,2) {$c_1U$};
\node[right] at (0,4) {$c_1^2U$};

\node[right] at (0,6) {$c_1^3U$};
\draw[fill] (0,2) circle(.1);
\draw[fill] (0,4) circle(.1);
\draw[fill] (0,6) circle(.1);
\draw (0,0) to [out=150,in=150] (0,2);
\draw (0,4) to [out=150,in=150] (0,6);

\node[right] at (2,4) {$c_2U$};
\node[right] at (2,6) {$c_3U$};
\draw[fill] (2,4) circle(.1);
\draw[fill] (2,6) circle(.1);

\draw (2,4) to [out=150,in=150] (2,6);

\node[right] at (4,4) {$c_2'U$};
\node[right] at (4,6) {$c_3'U$};
\draw[fill] (4,4) circle(.1);
\draw[fill] (4,6) circle(.1);

\draw (4,4) to [out=150,in=150] (4,6);

\end{tikzpicture}
\end{center}
\caption{The $\A_2(1)$-module structure of $\H^{*+2}(M\U(1),\Z_2)\otimes\H^*(\B \SU(3),\Z_2)\otimes\H^*(\B \SU(3),\Z_2)$ below degree 6.}
\label{fig:A_2(1)MU1SU3SU3}
\end{figure}

So there is no 2-torsion in $\Omega^{\Spin\times_{\Z_2}\frac{\U(3)\times\U(3)}{\Z_3\times\Z_3\times\U(1)}}_d$.

\begin{figure}[H]
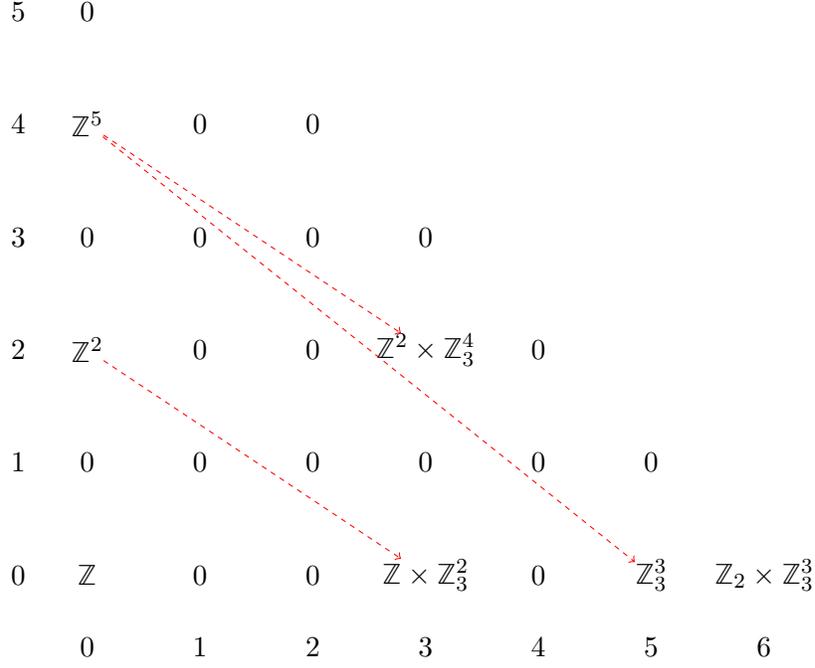

\center
\begin{sseq}[grid=none,labelstep=1,entrysize=1.5cm]{0...6}{0...5}
\ssdrop{\Z}
\ssmoveto 1 0 
\ssdrop{0}
\ssmoveto 2 0
\ssdrop{0}
\ssmoveto 3 0
\ssdrop{\Z\times\Z_3^2}
\ssmoveto 4 0
\ssdrop{0}
\ssmoveto 5 0
\ssdrop{\Z_3^3}
\ssmoveto 6 0
\ssdrop{\Z_2\times\Z_3^3}

\ssmoveto 0 1
\ssdrop{0}
\ssmoveto 1 1
\ssdrop{0}
\ssmoveto 2 1
\ssdrop{0}
\ssmoveto 3 1
\ssdrop{0}
\ssmoveto 4 1
\ssdrop{0}
\ssmoveto 5 1
\ssdrop{0}

\ssmoveto 0 2
\ssdrop{\Z^2}
\ssmoveto 1 2
\ssdrop{0}
\ssmoveto 2 2
\ssdrop{0}
\ssmoveto 3 2
\ssdrop{\Z^2\times\Z_3^4}
\ssmoveto 4 2
\ssdrop{0}

\ssmoveto 0 3
\ssdrop{0}
\ssmoveto 1 3
\ssdrop{0}
\ssmoveto 2 3
\ssdrop{0}
\ssmoveto 3 3
\ssdrop{0}

\ssmoveto 0 4
\ssdrop{\Z^5}
\ssmoveto 1 4
\ssdrop{0}
\ssmoveto 2 4
\ssdrop{0}

\ssmoveto 0 5
\ssdrop{0}

\ssmoveto 0 2
\ssarrow[color=red,dashed] 3 {-2}
\ssmoveto 0 4
\ssarrow[color=red,dashed] 3 {-2}
\ssmoveto 0 4
\ssarrow[color=red,dashed] 5 {-4}
\end{sseq}
\center
\caption{Serre spectral sequence for the fibration $\B\U(3)\times\B\U(3)\to\B(\frac{\U(3)\times\U(3)}{\Z_3\times\Z_3\times\U(1)})\to\B^2\Z_3\times\B^2\Z_3\times\B^2\U(1)$. The dashed arrows are possible differentials.
}
\label{fig:SSS-2}
\end{figure}

We have a fibration 
\bea
\xymatrix{\B\U(3)\times\B\U(3)\ar[d]&\\
\B(\frac{\U(3)\times\U(3)}{\Z_3\times\Z_3\times\U(1)})\ar[r]&\B^2\Z_3\times\B^2\Z_3\times\B^2\U(1).}
\eea
Hence we have the Serre spectral sequence, see Figure \ref{fig:SSS-2}.
\bea
\H^p(\B^2\Z_3\times\B^2\Z_3\times\B^2\U(1),\H^q(\B\U(3)\times\B\U(3),\Z))\Rightarrow\H^{p+q}(\B(\frac{\U(3)\times\U(3)}{\Z_3\times\Z_3\times\U(1)}),\Z).
\eea

We have
\bea
\H^*(\B\U(3)\times\B\U(3),\Z)=\Z[c_1,c_2,c_3,c_1',c_2',c_3']
\eea
and
\bea
\H^p(\B^2\Z_3,\Z)=\left\{\begin{array}{llllllll}\Z&p=0\\0&p=1\\0&p=2\\\Z_3&p=3\\0&p=4\\\Z_3&p=5\\0&p=6\\\Z_9&p=7\end{array}\right.
\eea
and
\bea
\H^p(\B^2\U(1),\Z)=\left\{\begin{array}{llllllll}\Z&p=0\\0&p=1\\0&p=2\\\Z&p=3\\0&p=4\\0&p=5\\\Z_2&p=6\end{array}\right..
\eea

By K\"unneth formula,
\bea
\H^p(\B^2\Z_3\times\B^2\Z_3\times\B^2\U(1),\Z)=\left\{\begin{array}{llllllll}\Z&p=0\\0&p=1\\0&p=2\\\Z\times\Z_3^2&p=3\\0&p=4\\\Z_3^3&p=5\\\Z_2\times\Z_3^3&p=6\end{array}\right..
\eea

Compared with the 2-torsion result, we find that the differential from (0,2) to (3,0) kills one $\Z$ since the 2d cobordism group contains only one $\Z$, and the differential from (0,4) to (3,2) kills two $\Z$ since the 4d cobordism group contains four $\Z$ while one $\Z$ is from $\Omega_4^{\SO}$.

So 
\bea
\H^p(\B(\frac{\U(3)\times\U(3)}{\Z_3\times\Z_3\times\U(1)}),\Z)=\left\{\begin{array}{lllllll}\Z&p=0\\0&p=1\\\Z&p=2\\\Z_3 \text{ or }\Z_3^2&p=3\\\Z^3&p=4\\?&p=5\end{array}\right.
\eea
and
\bea
\H_p(\B(\frac{\U(3)\times\U(3)}{\Z_3\times\Z_3\times\U(1)}),\Z)=\left\{\begin{array}{lllllll}\Z&p=0\\0&p=1\\\Z\times\Z_3\text{ or }\Z\times\Z_3^2&p=2\\0&p=3\\\Z^3\times?&p=4\\0&p=5\end{array}\right..
\eea
Here ? is an undetermined 3-torsion group.

By the Atiyah-Hirzebruch spectral sequence, we have
\bea
\H_p(\B(\frac{\U(3)\times\U(3)}{\Z_3\times\Z_3\times\U(1)}),\Omega_q^{\SO})\Rightarrow\Omega_{p+q}^{\SO}(\B(\frac{\U(3)\times\U(3)}{\Z_3\times\Z_3\times\U(1)})).
\eea

See Figure \ref{fig:AHSS-2}.

\begin{figure}[H]
\center
\begin{sseq}[grid=none,labelstep=1,entrysize=1.5cm]{0...5}{0...5}
\ssdrop{\Z}
\ssmoveto 1 0 
\ssdrop{0}
\ssmoveto 2 0
\ssdrop{\Z\times\Z_3^n}
\ssmoveto 3 0
\ssdrop{0}
\ssmoveto 4 0
\ssdrop{\Z^3\times?}
\ssmoveto 5 0
\ssdrop{0}

\ssmoveto 0 1
\ssdrop{0}
\ssmoveto 1 1
\ssdrop{0}
\ssmoveto 2 1
\ssdrop{0}
\ssmoveto 3 1
\ssdrop{0}
\ssmoveto 4 1
\ssdrop{0}

\ssmoveto 0 2
\ssdrop{0}
\ssmoveto 1 2
\ssdrop{0}
\ssmoveto 2 2
\ssdrop{0}
\ssmoveto 3 2
\ssdrop{0}

\ssmoveto 0 3
\ssdrop{0}
\ssmoveto 1 3
\ssdrop{0}
\ssmoveto 2 3
\ssdrop{0}

\ssmoveto 0 4
\ssdrop{\Z}
\ssmoveto 1 4
\ssdrop{0}

\ssmoveto 0 5
\ssdrop{\Z_2}

\end{sseq}
\center
\caption{Atiyah-Hirzebruch spectral sequence for $\Omega_d^{\SO \times \frac{\U(3)\times\U(3)}{\Z_3\times\Z_3\times\U(1)}}$. 
Here $n=1$ or 2, while ? is an undetermined 3-torsion group.}
\label{fig:AHSS-2}
\end{figure}

Combine the 2-torsion and 3-torsion results, we have
\begin{table}[H]
\centering
\begin{tabular}{ c c c}
\hline
\multicolumn{3}{c}{Bordism group}\\
\hline
$d$ & 
$\Omega^{\Spin \times_{\Z_2} \frac{\U(3)\times\U(3)}{\Z_3\times\Z_3\times\U(1)}}_d$
& generators \\
\hline
0& $\Z$\\
1& 0\\
2& $\Z\times\Z_3^n$ \\
3 & $0$\\
4 & $e(\Z^3\times?,\Z)$ \\
5 & $0$ \\
\hline
\end{tabular}
\caption{Bordism group. The notation $e(A,B)$ denotes a group extension of A by B, that is, a group that fits into the following short exact sequence
$0\to B\to e(A,B)\to A\to0$. Here $n=1$ or 2, while ? is an undetermined 3-torsion group.
}
\label{table:U3U3Z3Z3U1Z2Bordism}
\end{table}

\section{QCD Symmetries, Anomalies and Topological Terms With Time-Reversal}
\label{sec:WithTime-Reversal}


Now we consider putting the QCD matters on the smooth differentiable and unorientable spacetime manifolds  -- if the fermions/spinor can live
on them, we require Spin structure; if we require time-reversal $T=CP$, or $CT$ or other reflection symmetries, we 
require Pin$^+$, Pin$^-$ or other semi-direct ($\ltimes$) product or twisted structures between the spacetime tangent bundle $TM$ and the gauge bundle $E_G$ of the gauge group $G$. See more in the main text and see an overview of our setting in \cite{1711.11587GPW}.
{Follow \Fig{fig:qcd-phase-1}
and \Fig{fig:qcd-phase-T},  
 we can choose any suitable outer automorphism of the color gauge or flavor global symmetry group as
possible time-reversal symmetries, which can be any reasonable $\Z_2$-reflection symmetry. 
This implies putting the Euclidean QCD$_4$ path integral on an unorientable spacetime.
The most general case is a semi-direct product $\rtimes \Z_4^T$,
which is all allowed total group made from the exact sequence:
$$
1 \to \Z_2^F \to \Z_4^T\to \Z_2^T \to 1.
$$
Here we will only focus on two cases:
The direct product $\times \Z_4^T$ which implies the Pin$^+$ structure, where $\Z_4^T \supset \Z_2^F$. 
We may also denote such a $\Z_4^T := \Z_4^{TF}$ to indicates it includes $\Z_2^F$ as a normal subgroup.
The direct product $\times \Z_2^T \times \Z_2^F$ which implies the Pin$^-$ structure. 
For other possible  time-reversal symmetries, we leave them in a future work \cite{toappear}.

\subsection{Chiral symmetry breaking 
${\frac{ \big( [\SU(3)_V] \times  \SU(3)_V \times \U(1)_V\big) \rtimes \Z_4^T  }{ \Z_{3_c, V} \times \Z_{3_f, V} \times \Z_{2,V}^F}}$}

\subsubsection{${\frac{ \big( [\SU(3)_V] \times  \SU(3)_V \times \U(1)_V\big) \times \Z_4^{TF}  }{ \Z_{3_c, V} \times \Z_{3_f, V} \times \Z_{2,V}^F}}$
as
$(\Pin^+(d) \times_{\Z_2} \frac{\U(3)}{\Z_3})$ and\\ 
${\frac{ \big( [\SU(3)_V] \times  \SU(3)_V \times \U(1)_V\big) \times \Z_2^T  }{ \Z_{3_c, V} \times \Z_{3_f, V} }}$
as $(\Pin^-(d) \times_{\Z_2} \frac{\U(3)}{\Z_3})$}

Since the localization of $\Pin^{\pm} \times_{\Z_2} \frac{\U(3)}{\Z_3}$ and $\O\times \frac{\U(3)}{\Z_3}$ at the prime 3 are the same, so the 3-torsion of $\Omega_d^{\Pin^{\pm} \times_{\Z_2} \frac{\U(3)}{\Z_3}}$ and $\Omega_d^{\O\times \frac{\U(3)}{\Z_3}}$ are the same, hence there is no 3-torsion in $\Omega_d^{\Pin^{\pm} \times_{\Z_2} \frac{\U(3)}{\Z_3}}$.

Since the localization of $\Pin^{\pm} \times_{\Z_2} \frac{\U(3)}{\Z_3}$ and $\Pin^{\pm}\times_{\Z_2}(\U(1)\times\SU(3))$ at the prime 2 are the same, so the 2-torsion of $\Omega_d^{\Pin^{\pm} \times_{\Z_2} \frac{\U(3)}{\Z_3}}$ and $\Omega_d^{\Pin^c\times\SU(3)}$ are the same.

We have $MT(\Pin^c \times \SU(3))=M\Spin\wedge\Sigma^{-2}M\U(1)\wedge\Sigma^{-1}M\O(1)\wedge (\B \SU(3))_+$.

For $t-s<8$, since there is no odd torsion, we have the Adams spectral sequence
\bea
\Ext_{\A_2(1)}^{s,t}(\H^{*+2}(M\U(1),\Z_2)\otimes\H^{*+1}(M\O(1),\Z_2)\otimes\H^*(\B \SU(3),\Z_2),\Z_2)\Rightarrow\Omega_{t-s}^{\Pin^c \times \SU(3)}.
\eea

We have
\bea
\H^*(\B \SU(3),\Z_2)=\Z_2[c_2,c_3]
\eea
where $c_i$ is the Chern class of the $\SU(3)$ bundle.

By Thom isomorphism,
\bea
\H^{*+2}(M\U(1),\Z_2)=\Z_2[c_1]U
\eea
where $c_1$ is the Chern class of the $\U(1)$ bundle and $U$ is the Thom class.

Also by Thom isomorphism,
\bea
\H^{*+1}(M\O(1),\Z_2)=\Z_2[w_1]V
\eea
where $w_1$ is the Stiefel-Whitney class of the $\O(1)$ bundle and $V$ is the Thom class.

The $\A_2(1)$-module structure of $\H^{*+2}(M\U(1),\Z_2)\otimes\H^{*+1}(M\O(1),\Z_2)\otimes\H^*(\B \SU(3),\Z_2)$ below degree 6 and the $E_2$ page are shown in Figure \ref{fig:A_2(1)MU1MO1SU3}, \ref{fig:E_2PincSU3}.

\begin{figure}[H]
\begin{center}
\begin{tikzpicture}[scale=0.5]

\node[below] at (0,0) {$UV$};

\draw[fill] (0,0) circle(.1);
\draw[fill] (0,1) circle(.1);
\draw (0,0) -- (0,1);
\draw[fill] (0,2) circle(.1);
\draw (0,0) to [out=150,in=150] (0,2);
\draw[fill] (1,2) circle(.1);
\draw[fill] (1,3) circle(.1);
\draw (1,2) -- (1,3);
\draw (0,1) to [out=30,in=150] (1,3);
\draw[fill] (0,3) circle(.1);
\draw (0,2) -- (0,3);
\draw[fill] (0,4) circle(.1);
\draw[fill] (0,5) circle(.1);
\draw (0,4) -- (0,5);
\draw (0,3) to [out=150,in=150] (0,5);
\draw[fill] (1,4) circle(.1);
\draw (1,2) to [out=30,in=30] (1,4);
\draw[fill] (1,5) circle(.1);
\draw (1,4) -- (1,5);
\draw[fill] (0,6) circle(.1);
\draw (0,4) to [out=30,in=30] (0,6);
\draw[fill] (1,6) circle(.1);
\draw[fill] (1,7) circle(.1);
\draw (1,6) -- (1,7);
\draw (1,5) to [out=150,in=150] (1,7);
\draw[fill] (0,7) circle(.1);
\draw (0,6) -- (0,7);
\draw[fill] (1,8) circle(.1);
\draw[fill] (1,9) circle(.1);
\draw (1,8) -- (1,9);
\draw (1,6) to [out=30,in=30] (1,8);

\node[below] at (3,4) {$c_1^2UV$};

\draw[fill] (3,4) circle(.1);
\draw[fill] (3,5) circle(.1);
\draw (3,4) -- (3,5);
\draw[fill] (3,6) circle(.1);
\draw (3,4) to [out=150,in=150] (3,6);
\draw[fill] (4,6) circle(.1);
\draw[fill] (4,7) circle(.1);
\draw (4,6) -- (4,7);
\draw (3,5) to [out=30,in=150] (4,7);
\draw[fill] (3,7) circle(.1);
\draw (3,6) -- (3,7);
\draw[fill] (3,8) circle(.1);
\draw[fill] (3,9) circle(.1);
\draw (3,8) -- (3,9);
\draw (3,7) to [out=150,in=150] (3,9);
\draw[fill] (4,8) circle(.1);
\draw (4,6) to [out=30,in=30] (4,8);
\draw[fill] (4,9) circle(.1);
\draw (4,8) -- (4,9);
\draw[fill] (3,10) circle(.1);
\draw (3,8) to [out=30,in=30] (3,10);
\draw[fill] (4,10) circle(.1);
\draw[fill] (4,11) circle(.1);
\draw (4,10) -- (4,11);
\draw (4,9) to [out=150,in=150] (4,11);
\draw[fill] (3,11) circle(.1);
\draw (3,10) -- (3,11);
\draw[fill] (4,12) circle(.1);
\draw[fill] (4,13) circle(.1);
\draw (4,12) -- (4,13);
\draw (4,10) to [out=30,in=30] (4,12);

\node[below] at (6,4) {$c_2UV$};

\draw[fill] (6,4) circle(.1);
\draw[fill] (6,5) circle(.1);
\draw (6,4) -- (6,5);
\draw[fill] (6,6) circle(.1);
\draw (6,4) to [out=150,in=150] (6,6);
\draw[fill] (7,6) circle(.1);
\draw[fill] (7,7) circle(.1);
\draw (7,6) -- (7,7);
\draw (6,5) to [out=30,in=150] (7,7);
\draw[fill] (6,7) circle(.1);
\draw (6,6) -- (6,7);
\draw[fill] (6,8) circle(.1);
\draw[fill] (6,9) circle(.1);
\draw (6,8) -- (6,9);
\draw (6,7) to [out=150,in=150] (6,9);
\draw[fill] (7,8) circle(.1);
\draw (7,6) to [out=30,in=30] (7,8);
\draw[fill] (7,9) circle(.1);
\draw (7,8) -- (7,9);
\draw[fill] (6,10) circle(.1);
\draw (6,8) to [out=30,in=30] (6,10);
\draw[fill] (7,10) circle(.1);
\draw[fill] (7,11) circle(.1);
\draw (7,10) -- (7,11);
\draw (7,9) to [out=150,in=150] (7,11);
\draw[fill] (6,11) circle(.1);
\draw (6,10) -- (6,11);
\draw[fill] (7,12) circle(.1);
\draw[fill] (7,13) circle(.1);
\draw (7,12) -- (7,13);
\draw (7,10) to [out=30,in=30] (7,12);

\end{tikzpicture}
\end{center}
\caption{The $\A_2(1)$-module structure of $\H^{*+2}(M\U(1),\Z_2)\otimes\H^{*+1}(M\O(1),\Z_2)\otimes\H^*(\B \SU(3),\Z_2)$ below degree 6.}
\label{fig:A_2(1)MU1MO1SU3}
\end{figure}

\begin{figure}[H]
\begin{center}
\begin{tikzpicture}
\node at (0,-1) {0};
\node at (1,-1) {1};
\node at (2,-1) {2};
\node at (3,-1) {3};
\node at (4,-1) {4};
\node at (5,-1) {5};
\node at (6,-1) {$t-s$};
\node at (-1,0) {0};
\node at (-1,1) {1};
\node at (-1,2) {2};
\node at (-1,3) {3};
\node at (-1,4) {4};
\node at (-1,5) {5};
\node at (-1,6) {$s$};

\draw[->] (-0.5,-0.5) -- (-0.5,6);
\draw[->] (-0.5,-0.5) -- (6,-0.5);

\draw[fill] (0,0) circle(.05);

\draw (2,0) -- (2,1);
\draw (4,0) -- (4,2);
\draw[fill] (4.1,0) circle(.05);

\draw[fill] (4.2,0) circle(.05);

\end{tikzpicture}
\end{center}
\caption{$\Omega_*^{\Pin^c \times \SU(3)}$.}
\label{fig:E_2PincSU3}
\end{figure}
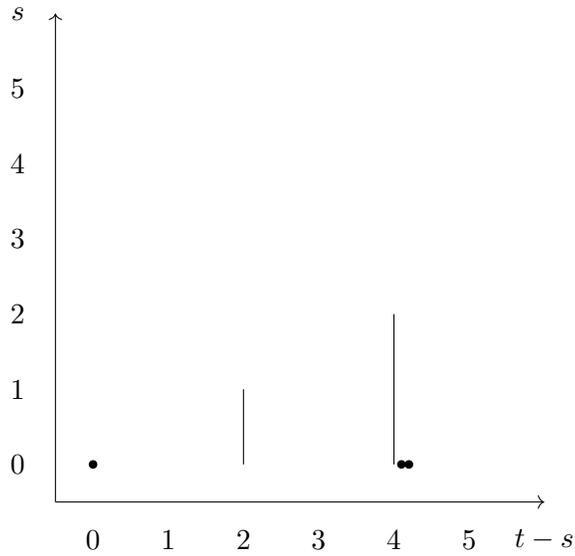

Combine the 2-torsion and 3-torsion results, we have
\begin{table}[H]
\centering
\begin{tabular}{ c c c}
\hline
\multicolumn{3}{c}{Bordism group}\\
\hline
$d$ & 
$\Omega^{\Pin^c\times\SU(3) }_d=\Omega_d^{\Pin^+ \times_{\Z_2} \frac{\U(3)}{\Z_3}}=\Omega_d^{\Pin^- \times_{\Z_2} \frac{\U(3)}{\Z_3}}$
& generators \\
\hline
0& $\Z_2$\\
1& 0\\
2& $\Z_4$ & ABK mod 4\\
3 & 0\\
4 & $\Z_2^2\times\Z_8$ & $c_2\mod2,c_1^2\mod2,(c_1\mod2)\text{ABK}$\\
5 & 0\\
\hline
\end{tabular}
\caption{Bordism group. Here $c_1$ is the Chern class of the $\U(1)$ bundle, $c_2$ is the Chern class of the $\SU(3)$ bundle, ABK is the Arf-Brown-Kervaire invariant.
}
\label{table:PincSU3Bordism}
\end{table}

\subsection{3-Color-Flavor locking superconductivity $\SU(3)_{C+L+R} \rtimes \Z_{4}^{T}$} 

\subsubsection{$\SU(3)_{C+L+R} \times \Z_{4}^{TF}$ as $\Pin^+\times\SU(3)$}

We have $MT(\Pin^+\times\SU(3))=MT\Pin^+\wedge (\B\SU(3))_+$.

$MT\Pin^+=M\Spin\wedge\Sigma^1MT\O(1)$.

By K\"unneth formula,
\bea
\H^*(\Sigma^1MT\O(1)\wedge (\B\SU(3))_+,\Z_2)=\H^{*-1}(MT\O(1),\Z_2)\otimes\H^*(\B\SU(3),\Z_2).
\eea

For $t-s<8$, since there is no odd torsion, we have the Adams spectral sequence 
\bea
\Ext_{\A_2(1)}^{s,t}(\H^{*-1}(MT\O(1),\Z_2)\otimes\H^*(\B\SU(3),\Z_2),\Z_2)\Rightarrow\Omega_{t-s}^{\Pin^+\times\SU(3)}.
\eea

The $\A_2(1)$-module structure of $\H^{*-1}(MT\O(1),\Z_2)\otimes\H^*(\B\SU(3),\Z_2)$ below degree 5 and the $E_2$ page are shown in Figure \ref{fig:A_2(1)MTO1BSU3}, \ref{fig:E_2Pin+SU3}.

\begin{figure}[H]
\begin{center}
\begin{tikzpicture}[scale=0.5]

\draw[fill] (0,0) circle(.1);
\draw[fill] (0,1) circle(.1);
\draw (0,0) -- (0,1);
\draw[fill] (0,2) circle(.1);
\draw (0,0) to [out=150,in=150] (0,2);
\draw[fill] (0,3) circle(.1);
\draw (0,2) -- (0,3);
\draw[fill] (0,4) circle(.1);
\draw[fill] (0,5) circle(.1);
\draw (0,3) to [out=150,in=150] (0,5);
\draw (0,4) -- (0,5);
\draw[fill] (0,6) circle(.1);
\draw (0,4) to [out=30,in=30] (0,6);
\draw[fill] (0,7) circle(.1);
\draw (0,6) -- (0,7);

\node at (2,2) {$\bigotimes$};

\node[below] at (4,0) {$1$};

\draw[fill] (4,0) circle(.1);

\node[right] at (4,4) {$c_2$};

\draw[fill] (4,4) circle(.1);

\node[right] at (4,6) {$c_3$};

\draw[fill] (4,6) circle(.1);

\draw (4,4) to [out=150,in=150] (4,6);

\node at (6,2) {$=$};

\draw[fill] (8,0) circle(.1);
\draw[fill] (8,1) circle(.1);
\draw (8,0) -- (8,1);
\draw[fill] (8,2) circle(.1);
\draw (8,0) to [out=150,in=150] (8,2);
\draw[fill] (8,3) circle(.1);
\draw (8,2) -- (8,3);
\draw[fill] (8,4) circle(.1);
\draw[fill] (8,5) circle(.1);
\draw (8,3) to [out=150,in=150] (8,5);
\draw (8,4) -- (8,5);
\draw[fill] (8,6) circle(.1);
\draw (8,4) to [out=30,in=30] (8,6);
\draw[fill] (8,7) circle(.1);
\draw (8,6) -- (8,7);

\node[below] at (10,4) {$c_2$}; 

\draw[fill] (10,4) circle(.1);
\draw[fill] (10,5) circle(.1);
\draw (10,4) -- (10,5);
\draw[fill] (10,6) circle(.1);
\draw (10,4) to [out=150,in=150] (10,6);
\draw[fill] (11,6) circle(.1);
\draw[fill] (11,7) circle(.1);
\draw (11,6) -- (11,7);
\draw (10,5) to [out=30,in=150] (11,7);
\draw[fill] (10,7) circle(.1);
\draw (10,6) -- (10,7);
\draw[fill] (10,8) circle(.1);
\draw[fill] (10,9) circle(.1);
\draw (10,8) -- (10,9);
\draw (10,7) to [out=150,in=150] (10,9);
\draw[fill] (11,8) circle(.1);
\draw (11,6) to [out=30,in=30] (11,8);
\draw[fill] (11,9) circle(.1);
\draw (11,8) -- (11,9);
\draw[fill] (10,10) circle(.1);
\draw (10,8) to [out=30,in=30] (10,10);
\draw[fill] (11,10) circle(.1);
\draw[fill] (11,11) circle(.1);
\draw (11,10) -- (11,11);
\draw (11,9) to [out=150,in=150] (11,11);
\draw[fill] (10,11) circle(.1);
\draw (10,10) -- (10,11);
\draw[fill] (11,12) circle(.1);
\draw[fill] (11,13) circle(.1);
\draw (11,12) -- (11,13);
\draw (11,10) to [out=30,in=30] (11,12);

\end{tikzpicture}
\end{center}
\caption{The $\A_2(1)$-module structure of $\H^{*-1}(MT\O(1),\Z_2)\otimes\H^*(\B\SU(3),\Z_2)$ below degree 5.}
\label{fig:A_2(1)MTO1BSU3}
\end{figure}

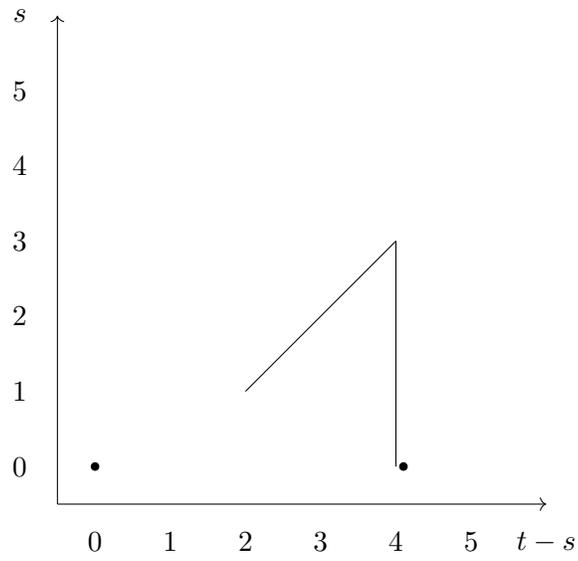
\begin{figure}[H]
\begin{center}
\begin{tikzpicture}
\node at (0,-1) {0};
\node at (1,-1) {1};
\node at (2,-1) {2};
\node at (3,-1) {3};
\node at (4,-1) {4};
\node at (5,-1) {5};
\node at (6,-1) {$t-s$};
\node at (-1,0) {0};
\node at (-1,1) {1};
\node at (-1,2) {2};
\node at (-1,3) {3};
\node at (-1,4) {4};
\node at (-1,5) {5};
\node at (-1,6) {$s$};

\draw[->] (-0.5,-0.5) -- (-0.5,6);
\draw[->] (-0.5,-0.5) -- (6,-0.5);

\draw[fill] (0,0) circle(.05);
\draw (2,1) -- (4,3);
\draw (4,0) -- (4,3);
\draw[fill] (4.1,0) circle(.05);

\end{tikzpicture}
\end{center}
\caption{$\Omega_*^{\Pin^+ \times \SU(3)}$}
\label{fig:E_2Pin+SU3}
\end{figure}

\begin{table}[H]
\centering
\begin{tabular}{ c c c}
\hline
\multicolumn{3}{c}{Bordism group}\\
\hline
$d$ & 
$\Omega^{\Pin^+\times \SU(3) }_d$
& generators \\
\hline
0& $\Z_2$\\
1& 0\\
2& $\Z_2$ & $w_1\tilde\eta$\\
3 & $\Z_2$ & $w_1\text{Arf}$\\
4 & $\Z_2\times\Z_{16}$ & $c_2\mod2,\eta$\\
5 & 0\\
\hline
\end{tabular}
\caption{Bordism group. Here $w_1$ is the Stiefel-Whitney class of the tangent bundle, $\tilde\eta$ is the mod 2 index of 1d Dirac operator, Arf is the Arf invariant, $\eta$ is the 4d eta invariant, $c_2$ is the Chern class of the $\SU(3)$ bundle.
}
\label{table:Pin+SU3Bordism}
\end{table}

\subsubsection{$\SU(3)_{C+L+R} \times \Z_{2,V}^F \times \Z_{2}^T$ as $\Pin^- \times\SU(3)$}

We have $MT(\Pin^-\times\SU(3))=MT\Pin^-\wedge (\B\SU(3))_+$.

$MT\Pin^-=M\Spin\wedge\Sigma^{-1}M\O(1)$.

By K\"unneth formula,
\bea
\H^*(\Sigma^{-1}M\O(1)\wedge (\B\SU(3))_+,\Z_2)=\H^{*+1}(M\O(1),\Z_2)\otimes\H^*(\B\SU(3),\Z_2).
\eea

For $t-s<8$, since there is no odd torsion, we have the Adams spectral sequence
\bea
\Ext_{\A_2(1)}^{s,t}(\H^{*+1}(M\O(1),\Z_2)\otimes\H^*(\B\SU(3),\Z_2),\Z_2)\Rightarrow\Omega_{t-s}^{\Pin^-\times\SU(3)}.
\eea

The $\A_2(1)$-module structure of $\H^{*+1}(M\O(1),\Z_2)\otimes\H^*(\B\SU(3),\Z_2)$ below degree 5 and the $E_2$ page are shown in Figure \ref{fig:A_2(1)MO1BSU3}, \ref{fig:E_2Pin-SU3}.

\begin{figure}[H]
\begin{center}
\begin{tikzpicture}[scale=0.5]

\draw[fill] (0,0) circle(.1);
\draw[fill] (0,1) circle(.1);
\draw (0,0) -- (0,1);
\draw[fill] (0,2) circle(.1);
\draw[fill] (0,3) circle(.1);
\draw (0,2) -- (0,3);
\draw (0,1) to [out=150,in=150] (0,3);
\draw[fill] (0,4) circle(.1);
\draw[fill] (0,5) circle(.1);
\draw (0,4) -- (0,5);
\draw (0,2) to [out=30,in=30] (0,4);

\node at (2,2) {$\bigotimes$};

\node[below] at (4,0) {$1$};

\draw[fill] (4,0) circle(.1);

\node[right] at (4,4) {$c_2$};

\draw[fill] (4,4) circle(.1);

\node[right] at (4,6) {$c_3$};

\draw[fill] (4,6) circle(.1);

\draw (4,4) to [out=150,in=150] (4,6);

\node at (6,2) {$=$};

\draw[fill] (8,0) circle(.1);
\draw[fill] (8,1) circle(.1);
\draw (8,0) -- (8,1);
\draw[fill] (8,2) circle(.1);
\draw[fill] (8,3) circle(.1);
\draw (8,2) -- (8,3);
\draw (8,1) to [out=150,in=150] (8,3);
\draw[fill] (8,4) circle(.1);
\draw[fill] (8,5) circle(.1);
\draw (8,4) -- (8,5);
\draw (8,2) to [out=30,in=30] (8,4);

\node[below] at (10,4) {$c_2$}; 

\draw[fill] (10,4) circle(.1);
\draw[fill] (10,5) circle(.1);
\draw (10,4) -- (10,5);
\draw[fill] (10,6) circle(.1);
\draw (10,4) to [out=150,in=150] (10,6);
\draw[fill] (11,6) circle(.1);
\draw[fill] (11,7) circle(.1);
\draw (11,6) -- (11,7);
\draw (10,5) to [out=30,in=150] (11,7);
\draw[fill] (10,7) circle(.1);
\draw (10,6) -- (10,7);
\draw[fill] (10,8) circle(.1);
\draw[fill] (10,9) circle(.1);
\draw (10,8) -- (10,9);
\draw (10,7) to [out=150,in=150] (10,9);
\draw[fill] (11,8) circle(.1);
\draw (11,6) to [out=30,in=30] (11,8);
\draw[fill] (11,9) circle(.1);
\draw (11,8) -- (11,9);
\draw[fill] (10,10) circle(.1);
\draw (10,8) to [out=30,in=30] (10,10);
\draw[fill] (11,10) circle(.1);
\draw[fill] (11,11) circle(.1);
\draw (11,10) -- (11,11);
\draw (11,9) to [out=150,in=150] (11,11);
\draw[fill] (10,11) circle(.1);
\draw (10,10) -- (10,11);
\draw[fill] (11,12) circle(.1);
\draw[fill] (11,13) circle(.1);
\draw (11,12) -- (11,13);
\draw (11,10) to [out=30,in=30] (11,12);

\end{tikzpicture}
\end{center}
\caption{The $\A_2(1)$-module structure of $\H^{*+1}(M\O(1),\Z_2)\otimes\H^*(\B\SU(3),\Z_2)$ below degree 5.}
\label{fig:A_2(1)MO1BSU3}
\end{figure}

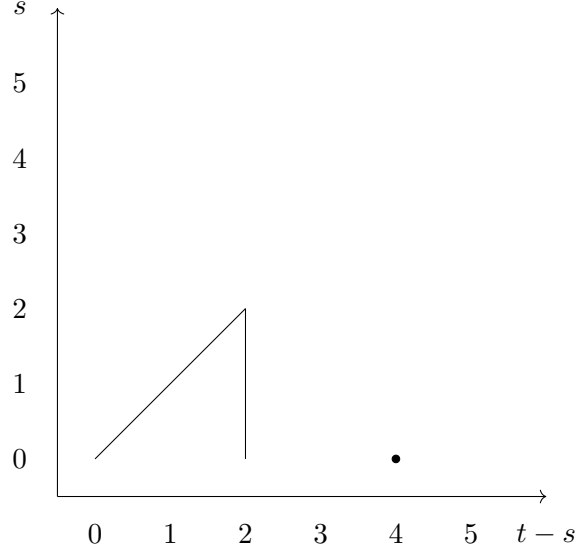
\begin{figure}[H]
\begin{center}
\begin{tikzpicture}
\node at (0,-1) {0};
\node at (1,-1) {1};
\node at (2,-1) {2};
\node at (3,-1) {3};
\node at (4,-1) {4};
\node at (5,-1) {5};
\node at (6,-1) {$t-s$};
\node at (-1,0) {0};
\node at (-1,1) {1};
\node at (-1,2) {2};
\node at (-1,3) {3};
\node at (-1,4) {4};
\node at (-1,5) {5};
\node at (-1,6) {$s$};

\draw[->] (-0.5,-0.5) -- (-0.5,6);
\draw[->] (-0.5,-0.5) -- (6,-0.5);

\draw (0,0) -- (2,2);

\draw (2,0) -- (2,2);

\draw[fill] (4,0) circle(.05);

\end{tikzpicture}
\end{center}
\caption{$\Omega_*^{\Pin^- \times \SU(3)}$}
\label{fig:E_2Pin-SU3}
\end{figure}

\begin{table}[H]
\centering
\begin{tabular}{ c c c}
\hline
\multicolumn{3}{c}{Bordism group}\\
\hline
$d$ & 
$\Omega^{\Pin^-\times \SU(3) }_d$
& generators \\
\hline
0& $\Z_2$\\
1& $\Z_2$ & $\tilde\eta$\\
2& $\Z_8$ & ABK\\
3 & 0 \\
4 & $\Z_2$ & $c_2\mod2$\\
5 & 0\\
\hline
\end{tabular}
\caption{Bordism group. Here $\tilde\eta$ is the mod 2 index of 1d Dirac operator, ABK is the Arf-Brown-Kervaire invariant, $c_2$ is the Chern class of the $\SU(3)$ bundle.
}
\label{table:Pin-SU3Bordism}
\end{table}

\subsection{2-Color Superconductivity: 
$\frac{( [\SU(2_c)_{V,rg}] \times \SU(2_f)_{L,ud} \times  \SU(2_f)_{R,ud}   {\times \U(1_f)_{V,s}} {\times \U(1_c)_{V,b}}) 
 \rtimes \Z_4^T }{{(\Z_{2,V}^F)^2}}$
} 

\subsubsection{$\frac{( [\SU(2_c)_{V,rg}] \times \SU(2_f)_{L,ud} \times  \SU(2_f)_{R,ud}   {\times \U(1_f)_{V,s}} {\times \U(1_c)_{V,b}}) 
 \times \Z_4^{TF} }{{(\Z_{2,V}^F)^2}}$
 as 
$\Pin^+\times_{\Z_2}\Spin(4)\times\U(1)\times\U(1)$}

We have $MT(\Pin^+\times_{\Z_2}\Spin(4)\times\U(1)\times\U(1))=MT\Pin^+\wedge\Sigma^{-4}M\SO(4)\wedge(\B\U(1)\times\B\U(1))_+=M\Spin\wedge\Sigma^1MT\O(1)\wedge\Sigma^{-4}M\SO(4)\wedge(\B\U(1)\times\B\U(1))_+$.

We have the constraint $w_2=w_2'$ where $w_i$ is the Stiefel-Whitney class of the tangent bundle, $w_i'$ is the Stiefel-Whitney class of the $\SO(4)$ bundle.

For $t-s<8$, since there is no odd torsion, we have the Adams spectral sequence
\bea
&&\Ext_{\A_2(1)}^{s,t}(\H^{*-1}(MT\O(1),\Z_2)\otimes\H^{*+4}(M\SO(4),\Z_2)\otimes\H^*(\B\U(1),\Z_2)\otimes\H^*(\B\U(1),\Z_2),\Z_2)\nn\\
&&\Rightarrow \Omega_{t-s}^{\Pin^+\times_{\Z_2}\Spin(4)\times\U(1)\times\U(1)}.
\eea
By Thom isomorphism, we have 
\bea
\H^{*+4}(M\SO(4),\Z_2)=\Z_2[w_2',w_3',w_4']U
\eea
where $w_i'$ is the Stiefel-Whitney class of the $\SO(4)$ bundle and $U$ is the Thom class.

Also by Thom isomorphism, we have 
\bea
\H^{*-1}(MT\O(1),\Z_2)=\Z_2[w_1]V
\eea
where $w_1$ is the Stiefel-Whitney class of the $\O(1)$ bundle $V_1$, and $V$ is the Thom class of $-V_1$.

We also have 
\bea
\H^*(\B\U(1),\Z_2)=\Z_2[c_1]
\eea
where $c_1$ is the first Chern class of the $\U(1)$ bundle.

The $\A_2(1)$-module structure of $\H^{*-1}(MT\O(1),\Z_2)\otimes\H^{*+4}(M\SO(4),\Z_2)\otimes\H^*(\B\U(1),\Z_2)\otimes\H^*(\B\U(1),\Z_2)$ below degree 5 and the $E_2$ page are shown in Figure \ref{fig:A_2(1)MTO1MSO4BU1BU1}, \ref{fig:E_2Pin+Spin4Z2U1U1}.

\begin{figure}[H]
\begin{center}
\begin{tikzpicture}[scale=0.5]

\node[below] at (0,0) {$V$};

\draw[fill] (0,0) circle(.1);
\draw[fill] (0,1) circle(.1);
\draw (0,0) -- (0,1);
\draw[fill] (0,2) circle(.1);
\draw (0,0) to [out=150,in=150] (0,2);
\draw[fill] (0,3) circle(.1);
\draw (0,2) -- (0,3);
\draw[fill] (0,4) circle(.1);
\draw[fill] (0,5) circle(.1);
\draw (0,3) to [out=150,in=150] (0,5);
\draw (0,4) -- (0,5);
\draw[fill] (0,6) circle(.1);
\draw (0,4) to [out=30,in=30] (0,6);
\draw[fill] (0,7) circle(.1);
\draw (0,6) -- (0,7);

\node at (2,2) {$\bigotimes$};

\node[below] at (4,0) {$U$};

\draw[fill] (4,0) circle(.1);
\draw[fill] (4,2) circle(.1);
\draw (4,0) to [out=150,in=150] (4,2);
\draw[fill] (4,3) circle(.1);
\draw (4,2) -- (4,3);

\node[left] at (4,4) {$w_4'U$};

\draw[fill] (4,4) circle(.1);

\node[below] at (6,4) {$w_2'^2U$};

\draw[fill] (6,4) circle(.1);
\draw[fill] (6,5) circle(.1);
\draw[fill] (6,6) circle(.1);
\draw[fill] (7,6) circle(.1);
\draw (6,4) to [out=30,in=150] (7,6);
\draw (6,5) -- (6,6);
\draw[fill] (7,7) circle(.1);
\draw (7,6) -- (7,7);
\draw (6,5) to [out=30,in=150] (7,7);
\draw[fill] (7,8) circle(.1);
\draw[fill] (7,9) circle(.1);
\draw (6,6) to [out=30,in=150] (7,8);
\draw (7,7) to [out=30,in=30] (7,9);
\draw (7,8) -- (7,9);

\node[below] at (8,2) {$c_1U$};

\draw[fill] (8,2) circle(.1);
\draw[fill] (8,4) circle(.1);
\draw (8,2) to [out=150,in=150] (8,4);
\draw[fill] (8,5) circle(.1);
\draw (8,4) -- (8,5);
\draw[fill] (8,6) circle(.1);
\draw[fill] (8,7) circle(.1);
\draw (8,5) to [out=150,in=150] (8,7);
\draw (8,6) -- (8,7);
\draw[fill] (9,4) circle(.1);
\draw (9,4) to [out=150,in=30] (8,6);
\node[below] at (9,4) {$c_1^2U$};

\node[below] at (10,2) {$c_1'U$};

\draw[fill] (10,2) circle(.1);
\draw[fill] (10,4) circle(.1);
\draw (10,2) to [out=150,in=150] (10,4);
\draw[fill] (10,5) circle(.1);
\draw (10,4) -- (10,5);
\draw[fill] (10,6) circle(.1);
\draw[fill] (10,7) circle(.1);
\draw (10,5) to [out=150,in=150] (10,7);
\draw (10,6) -- (10,7);
\draw[fill] (11,4) circle(.1);
\draw (11,4) to [out=150,in=30] (10,6);
\node[below] at (11,4) {$c_1'^2U$};

\node[below] at (13,4) {$c_1c_1'U$};

\draw[fill] (13,4) circle(.1);
\draw[fill] (13,6) circle(.1);
\draw (13,4) to [out=150,in=150] (13,6);
\draw[fill] (13,7) circle(.1);
\draw (13,6) -- (13,7);
\draw[fill] (13,8) circle(.1);
\draw[fill] (13,9) circle(.1);
\draw (13,7) to [out=150,in=150] (13,9);
\draw (13,8) -- (13,9);
\draw[fill] (14,6) circle(.1);
\draw (14,6) to [out=150,in=30] (13,8);

\node at (-2,-10) {$=$};

\node[below] at (0,-12) {$UV$};

\draw[fill] (0,-12) circle(.1);
\draw[fill] (0,-11) circle(.1);
\draw (0,-12) -- (0,-11);
\draw[fill] (0,-10) circle(.1);
\draw (0,-12) to [out=150,in=150] (0,-10);
\draw[fill] (0,-9) circle(.1);
\draw (0,-10) -- (0,-9);
\draw[fill] (0,-8) circle(.1);
\draw[fill] (0,-7) circle(.1);
\draw (0,-9) to [out=150,in=150] (0,-7);
\draw (0,-8) -- (0,-7);
\draw[fill] (0,-6) circle(.1);
\draw (0,-8) to [out=30,in=30] (0,-6);
\draw[fill] (0,-5) circle(.1);
\draw (0,-6) -- (0,-5);
\draw[fill] (1,-9) circle(.1);
\draw[fill] (1,-8) circle(.1);
\draw (1,-9) -- (1,-8);
\draw (0,-11) to [out=30,in=150] (1,-9);
\draw (0,-10) to [out=30,in=150] (1,-8);

\node[below] at (2,-10) {$w_1^2UV$};

\draw[fill] (2,-10) circle(.1);

\draw[fill] (2,-9) circle(.1);

\draw (2,-10) -- (2,-9);

\draw[fill] (2,-8) circle(.1);

\draw[fill] (3,-7) circle(.1);

\draw[fill] (3,-6) circle(.1);
\draw (3,-7) -- (3,-6);

\draw (2,-10) to [out=150,in=150] (2,-8);

\draw (2,-9) to [out=30,in=150] (3,-7);

\draw (2,-8) to [out=30,in=150] (3,-6);

\draw[fill] (2,-7) circle(.1);

\draw[fill] (3,-5) circle(.1);

\draw[fill] (3,-4) circle(.1);
\draw (3,-5) -- (3,-4);
\draw (2,-8) -- (2,-7);

\draw (2,-7) to [out=30,in=150] (3,-5);
\draw (3,-6) to [out=30,in=30] (3,-4);

\node[below] at (4,-8) {$w_4'UV$};

\draw[fill] (4,-8) circle(.1);
\draw[fill] (4,-7) circle(.1);
\draw (4,-8) -- (4,-7);
\draw[fill] (4,-6) circle(.1);
\draw (4,-8) to [out=150,in=150] (4,-6);
\draw[fill] (4,-5) circle(.1);
\draw (4,-6) -- (4,-5);
\draw[fill] (4,-4) circle(.1);
\draw[fill] (4,-3) circle(.1);
\draw (4,-5) to [out=150,in=150] (4,-3);
\draw (4,-4) -- (4,-3);
\draw[fill] (4,-2) circle(.1);
\draw (4,-4) to [out=30,in=30] (4,-2);
\draw[fill] (4,-1) circle(.1);
\draw (4,-2) -- (4,-1);

\node[below] at (6,-8) {$w_2'^2UV$};

\draw[fill] (6,-8) circle(.1);

\draw[fill] (6,-7) circle(.1);

\draw (6,-8) -- (6,-7);

\draw[fill] (6,-6) circle(.1);

\draw[fill] (7,-5) circle(.1);

\draw[fill] (7,-4) circle(.1);
\draw (7,-5) -- (7,-4);

\draw (6,-8) to [out=150,in=150] (6,-6);

\draw (6,-7) to [out=30,in=150] (7,-5);

\draw (6,-6) to [out=30,in=150] (7,-4);

\draw[fill] (6,-5) circle(.1);

\draw[fill] (7,-3) circle(.1);

\draw[fill] (7,-2) circle(.1);
\draw (7,-3) -- (7,-2);
\draw (6,-6) -- (6,-5);

\draw (6,-5) to [out=30,in=150] (7,-3);
\draw (7,-4) to [out=30,in=30] (7,-2);

\node[below] at (8,-7) {$w_2'w_3'UV$};

\draw[fill] (8,-7) circle(.1);

\draw[fill] (8,-6) circle(.1);

\draw (8,-7) -- (8,-6);

\draw[fill] (8,-5) circle(.1);

\draw[fill] (9,-4) circle(.1);

\draw[fill] (9,-3) circle(.1);
\draw (9,-4) -- (9,-3);

\draw (8,-7) to [out=150,in=150] (8,-5);

\draw (8,-6) to [out=30,in=150] (9,-4);

\draw (8,-5) to [out=30,in=150] (9,-3);

\draw[fill] (8,-4) circle(.1);

\draw[fill] (9,-2) circle(.1);

\draw[fill] (9,-1) circle(.1);
\draw (9,-2) -- (9,-1);
\draw (8,-5) -- (8,-4);

\draw (8,-4) to [out=30,in=150] (9,-2);
\draw (9,-3) to [out=30,in=30] (9,-1);

\node[below] at (10,-8) {$c_1c_1'UV$};

\draw[fill] (10,-8) circle(.1);

\draw[fill] (10,-7) circle(.1);

\draw (10,-8) -- (10,-7);

\draw[fill] (10,-6) circle(.1);

\draw[fill] (11,-5) circle(.1);

\draw[fill] (11,-4) circle(.1);
\draw (11,-5) -- (11,-4);

\draw (10,-8) to [out=150,in=150] (10,-6);

\draw (10,-7) to [out=30,in=150] (11,-5);

\draw (10,-6) to [out=30,in=150] (11,-4);

\draw[fill] (10,-5) circle(.1);

\draw[fill] (11,-3) circle(.1);

\draw[fill] (11,-2) circle(.1);
\draw (11,-3) -- (11,-2);
\draw (10,-6) -- (10,-5);

\draw (10,-5) to [out=30,in=150] (11,-3);
\draw (11,-4) to [out=30,in=30] (11,-2);

\node[below] at (0,-24) {$c_1UV$};

\draw[fill] (0,-24) circle(.1);

\draw[fill] (0,-23) circle(.1);

\draw (0,-24) -- (0,-23);

\draw[fill] (0,-22) circle(.1);

\draw[fill] (1,-21) circle(.1);

\draw[fill] (1,-20) circle(.1);
\draw (1,-21) -- (1,-20);

\draw (0,-24) to [out=150,in=150] (0,-22);

\draw (0,-23) to [out=30,in=150] (1,-21);

\draw (0,-22) to [out=30,in=150] (1,-20);

\draw[fill] (0,-21) circle(.1);

\draw[fill] (1,-19) circle(.1);

\draw[fill] (1,-18) circle(.1);
\draw (1,-19) -- (1,-18);
\draw (0,-22) -- (0,-21);

\draw (0,-21) to [out=30,in=150] (1,-19);
\draw (1,-20) to [out=30,in=30] (1,-18);

\node[below] at (2,-22) {$w_1^2c_1UV$};

\draw[fill] (2,-22) circle(.1);

\draw[fill] (2,-21) circle(.1);

\draw (2,-22) -- (2,-21);

\draw[fill] (2,-20) circle(.1);

\draw[fill] (3,-19) circle(.1);

\draw[fill] (3,-18) circle(.1);
\draw (3,-19) -- (3,-18);

\draw (2,-22) to [out=150,in=150] (2,-20);

\draw (2,-21) to [out=30,in=150] (3,-19);

\draw (2,-20) to [out=30,in=150] (3,-18);

\draw[fill] (2,-19) circle(.1);

\draw[fill] (3,-17) circle(.1);

\draw[fill] (3,-16) circle(.1);
\draw (3,-17) -- (3,-16);
\draw (2,-20) -- (2,-19);

\draw (2,-19) to [out=30,in=150] (3,-17);
\draw (3,-18) to [out=30,in=30] (3,-16);

\node[below] at (4,-22) {$c_1^2UV$};

\draw[fill] (4,-22) circle(.1);
\draw[fill] (4,-21) circle(.1);
\draw (4,-22) -- (4,-21);
\draw[fill] (4,-20) circle(.1);
\draw (4,-22) to [out=150,in=150] (4,-20);
\draw[fill] (5,-20) circle(.1);
\draw[fill] (5,-19) circle(.1);
\draw (5,-20) -- (5,-19);
\draw (4,-21) to [out=30,in=150] (5,-19);
\draw[fill] (4,-19) circle(.1);
\draw (4,-20) -- (4,-19);
\draw[fill] (4,-18) circle(.1);
\draw[fill] (4,-17) circle(.1);
\draw (4,-18) -- (4,-17);
\draw (4,-19) to [out=150,in=150] (4,-17);
\draw[fill] (5,-18) circle(.1);
\draw (5,-20) to [out=30,in=30] (5,-18);
\draw[fill] (5,-17) circle(.1);
\draw (5,-18) -- (5,-17);
\draw[fill] (4,-16) circle(.1);
\draw (4,-18) to [out=30,in=30] (4,-16);
\draw[fill] (5,-16) circle(.1);
\draw[fill] (5,-15) circle(.1);
\draw (5,-16) -- (5,-15);
\draw (5,-17) to [out=150,in=150] (5,-15);
\draw[fill] (4,-15) circle(.1);
\draw (4,-16) -- (4,-15);
\draw[fill] (5,-14) circle(.1);
\draw[fill] (5,-13) circle(.1);
\draw (5,-14) -- (5,-13);
\draw (5,-16) to [out=30,in=30] (5,-14);

\node[below] at (6,-24) {$c_1'UV$};

\draw[fill] (6,-24) circle(.1);

\draw[fill] (6,-23) circle(.1);

\draw (6,-24) -- (6,-23);

\draw[fill] (6,-22) circle(.1);

\draw[fill] (7,-21) circle(.1);

\draw[fill] (7,-20) circle(.1);
\draw (7,-21) -- (7,-20);

\draw (6,-24) to [out=150,in=150] (6,-22);

\draw (6,-23) to [out=30,in=150] (7,-21);

\draw (6,-22) to [out=30,in=150] (7,-20);

\draw[fill] (6,-21) circle(.1);

\draw[fill] (7,-19) circle(.1);

\draw[fill] (7,-18) circle(.1);
\draw (7,-19) -- (7,-18);
\draw (6,-22) -- (6,-21);

\draw (6,-21) to [out=30,in=150] (7,-19);
\draw (7,-20) to [out=30,in=30] (7,-18);

\node[below] at (8,-22) {$w_1^2c_1'UV$};

\draw[fill] (8,-22) circle(.1);

\draw[fill] (8,-21) circle(.1);

\draw (8,-22) -- (8,-21);

\draw[fill] (8,-20) circle(.1);

\draw[fill] (9,-19) circle(.1);

\draw[fill] (9,-18) circle(.1);
\draw (9,-19) -- (9,-18);

\draw (8,-22) to [out=150,in=150] (8,-20);

\draw (8,-21) to [out=30,in=150] (9,-19);

\draw (8,-20) to [out=30,in=150] (9,-18);

\draw[fill] (8,-19) circle(.1);

\draw[fill] (9,-17) circle(.1);

\draw[fill] (9,-16) circle(.1);
\draw (9,-17) -- (9,-16);
\draw (8,-20) -- (8,-19);

\draw (8,-19) to [out=30,in=150] (9,-17);
\draw (9,-18) to [out=30,in=30] (9,-16);

\node[below] at (10,-22) {$c_1'^2UV$};

\draw[fill] (10,-22) circle(.1);
\draw[fill] (10,-21) circle(.1);
\draw (10,-22) -- (10,-21);
\draw[fill] (10,-20) circle(.1);
\draw (10,-22) to [out=150,in=150] (10,-20);
\draw[fill] (11,-20) circle(.1);
\draw[fill] (11,-19) circle(.1);
\draw (11,-20) -- (11,-19);
\draw (10,-21) to [out=30,in=150] (11,-19);
\draw[fill] (10,-19) circle(.1);
\draw (10,-20) -- (10,-19);
\draw[fill] (10,-18) circle(.1);
\draw[fill] (10,-17) circle(.1);
\draw (10,-18) -- (10,-17);
\draw (10,-19) to [out=150,in=150] (10,-17);
\draw[fill] (11,-18) circle(.1);
\draw (11,-20) to [out=30,in=30] (11,-18);
\draw[fill] (11,-17) circle(.1);
\draw (11,-18) -- (11,-17);
\draw[fill] (10,-16) circle(.1);
\draw (10,-18) to [out=30,in=30] (10,-16);
\draw[fill] (11,-16) circle(.1);
\draw[fill] (11,-15) circle(.1);
\draw (11,-16) -- (11,-15);
\draw (11,-17) to [out=150,in=150] (11,-15);
\draw[fill] (10,-15) circle(.1);
\draw (10,-16) -- (10,-15);
\draw[fill] (11,-14) circle(.1);
\draw[fill] (11,-13) circle(.1);
\draw (11,-14) -- (11,-13);
\draw (11,-16) to [out=30,in=30] (11,-14);

\end{tikzpicture}
\end{center}
\caption{The $\A_2(1)$-module structure of $\H^{*-1}(MT\O(1),\Z_2)\otimes\H^{*+4}(M\SO(4),\Z_2)\otimes\H^*(\B\U(1),\Z_2)\otimes\H^*(\B\U(1),\Z_2)$ below degree 5.}
\label{fig:A_2(1)MTO1MSO4BU1BU1}
\end{figure}

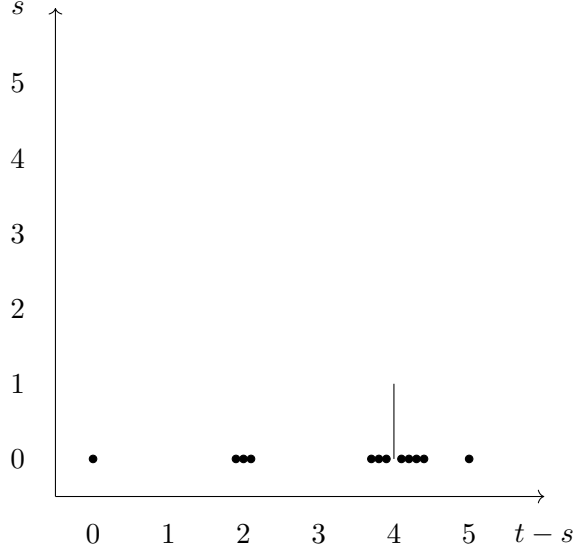
\begin{figure}[H]
\begin{center}
\begin{tikzpicture}
\node at (0,-1) {0};
\node at (1,-1) {1};
\node at (2,-1) {2};
\node at (3,-1) {3};
\node at (4,-1) {4};
\node at (5,-1) {5};
\node at (6,-1) {$t-s$};
\node at (-1,0) {0};
\node at (-1,1) {1};
\node at (-1,2) {2};
\node at (-1,3) {3};
\node at (-1,4) {4};
\node at (-1,5) {5};
\node at (-1,6) {$s$};

\draw[->] (-0.5,-0.5) -- (-0.5,6);
\draw[->] (-0.5,-0.5) -- (6,-0.5);

\draw[fill] (0,0) circle(0.05);

\draw[fill] (1.9,0) circle(0.05);
\draw[fill] (2,0) circle(0.05);
\draw[fill] (2.1,0) circle(0.05);

\draw (4,0) -- (4,1);

\draw[fill] (3.7,0) circle(0.05);
\draw[fill] (3.8,0) circle(0.05);
\draw[fill] (3.9,0) circle(0.05);
\draw[fill] (4.1,0) circle(0.05);
\draw[fill] (4.2,0) circle(0.05);
\draw[fill] (4.3,0) circle(0.05);
\draw[fill] (4.4,0) circle(0.05);

\draw[fill] (5,0) circle(0.05);

\end{tikzpicture}
\end{center}
\caption{$\Omega_*^{\Pin^+\times_{\Z_2}\Spin(4)\times\U(1)\times\U(1)}$.}
\label{fig:E_2Pin+Spin4Z2U1U1}
\end{figure}

\begin{table}[H]
\centering
\begin{tabular}{ c c c}
\hline
\multicolumn{3}{c}{Bordism group}\\
\hline
$d$ & 
$\Omega^{\Pin^+\times_{\Z_2}\Spin(4)\times\U(1)\times\U(1)}_d$
& generators \\
\hline
0& $\Z_2$\\
1& $0$\\
2& $\Z_2^3$  & $w_1^2,c_1\mod2,c_1'\mod2$\\
3 & $0$\\
4 & $\Z_2^7\times\Z_4$ & $w_4',w_2'^2,w_1^2c_1,w_1^2c_1',c_1^2\mod2,c_1'^2\mod2,c_1c_1'\mod2,\eta_{\Spin(4)}$ \\
5 & $\Z_2$  & $w_2'w_3'$ \\
\hline
\end{tabular}
\caption{Bordism group. Here $w_i$ is the Stiefel-Whitney class of the tangent bundle, $w_i'$ is the Stiefel-Whitney class of the $\SO(4)$ bundle, $c_1$ ($c_1'$) is the Chern class of the $\U(1)$ bundle, $\eta_{\Spin(4)}$ is similar to the $\eta_{\SU(2)}$ defined in  \cite{1711.11587GPW}.
}
\label{table:Pin+Spin4Z2U1U1Bordism}
\end{table}

\subsubsection{$\frac{( [\SU(2_c)_{V,rg}] \times \SU(2_f)_{L,ud} \times  \SU(2_f)_{R,ud}   {\times \U(1_f)_{V,s}} {\times \U(1_c)_{V,b}}) 
 \times \Z_2^T }{{\Z_{2,V}^F}}$
$\Pin^-\times_{\Z_2}\Spin(4)\times\U(1)\times\U(1)$}

We have $MT(\Pin^-\times_{\Z_2}\Spin(4)\times\U(1)\times\U(1))=MT\Pin^-\wedge\Sigma^{-4}M\SO(4)\wedge(\B\U(1)\times\B\U(1))_+=M\Spin\wedge\Sigma^{-1}M\O(1)\wedge\Sigma^{-4}M\SO(4)\wedge(\B\U(1)\times\B\U(1))_+$.

We have the constraint $w_2+w_1^2=w_2'$ where $w_i$ is the Stiefel-Whitney class of the tangent bundle, $w_i'$ is the Stiefel-Whitney class of the $\SO(4)$ bundle.

For $t-s<8$, since there is no odd torsion, we have the Adams spectral sequence
\bea
&&\Ext_{\A_2(1)}^{s,t}(\H^{*+1}(M\O(1),\Z_2)\otimes\H^{*+4}(M\SO(4),\Z_2)\otimes\H^*(\B\U(1),\Z_2)\otimes\H^*(\B\U(1),\Z_2),\Z_2)\nn\\
&&\Rightarrow \Omega_{t-s}^{\Pin^-\times_{\Z_2}\Spin(4)\times\U(1)\times\U(1)}.
\eea
By Thom isomorphism, we have 
\bea
\H^{*+4}(M\SO(4),\Z_2)=\Z_2[w_2',w_3',w_4']U
\eea
where $w_i'$ is the Stiefel-Whitney class of the $\SO(4)$ bundle and $U$ is the Thom class.

Also by Thom isomorphism, we have 
\bea
\H^{*+1}(M\O(1),\Z_2)=\Z_2[w_1]V
\eea
where $w_1$ is the Stiefel-Whitney class of the $\O(1)$ bundle $V_1$, and $V$ is the Thom class of $V_1$.

We also have 
\bea
\H^*(\B\U(1),\Z_2)=\Z_2[c_1]
\eea
where $c_1$ is the first Chern class of the $\U(1)$ bundle.

The $\A_2(1)$-module structure of $\H^{*+1}(M\O(1),\Z_2)\otimes\H^{*+4}(M\SO(4),\Z_2)\otimes\H^*(\B\U(1),\Z_2)\otimes\H^*(\B\U(1),\Z_2)$ below degree 5 and the $E_2$ page are shown in Figure \ref{fig:A_2(1)MO1MSO4BU1BU1}, \ref{fig:E_2Pin-Spin4Z2U1U1}.

\begin{figure}[H]
\begin{center}
\begin{tikzpicture}[scale=0.5]

\node[below] at (0,0) {$V$};

\draw[fill] (0,0) circle(.1);
\draw[fill] (0,1) circle(.1);
\draw (0,0) -- (0,1);
\draw[fill] (0,2) circle(.1);
\draw[fill] (0,3) circle(.1);
\draw (0,2) -- (0,3);
\draw (0,1) to [out=150,in=150] (0,3);
\draw[fill] (0,4) circle(.1);
\draw[fill] (0,5) circle(.1);
\draw (0,4) -- (0,5);
\draw (0,2) to [out=30,in=30] (0,4);

\node at (2,2) {$\bigotimes$};

\node[below] at (4,0) {$U$};

\draw[fill] (4,0) circle(.1);
\draw[fill] (4,2) circle(.1);
\draw (4,0) to [out=150,in=150] (4,2);
\draw[fill] (4,3) circle(.1);
\draw (4,2) -- (4,3);

\node[left] at (4,4) {$w_4'U$};

\draw[fill] (4,4) circle(.1);

\node[below] at (6,4) {$w_2'^2U$};

\draw[fill] (6,4) circle(.1);
\draw[fill] (6,5) circle(.1);
\draw[fill] (6,6) circle(.1);
\draw[fill] (7,6) circle(.1);
\draw (6,4) to [out=30,in=150] (7,6);
\draw (6,5) -- (6,6);
\draw[fill] (7,7) circle(.1);
\draw (7,6) -- (7,7);
\draw (6,5) to [out=30,in=150] (7,7);
\draw[fill] (7,8) circle(.1);
\draw[fill] (7,9) circle(.1);
\draw (6,6) to [out=30,in=150] (7,8);
\draw (7,7) to [out=30,in=30] (7,9);
\draw (7,8) -- (7,9);

\node[below] at (8,2) {$c_1U$};

\draw[fill] (8,2) circle(.1);
\draw[fill] (8,4) circle(.1);
\draw (8,2) to [out=150,in=150] (8,4);
\draw[fill] (8,5) circle(.1);
\draw (8,4) -- (8,5);
\draw[fill] (8,6) circle(.1);
\draw[fill] (8,7) circle(.1);
\draw (8,5) to [out=150,in=150] (8,7);
\draw (8,6) -- (8,7);
\draw[fill] (9,4) circle(.1);
\draw (9,4) to [out=150,in=30] (8,6);
\node[below] at (9,4) {$c_1^2U$};

\node[below] at (10,2) {$c_1'U$};

\draw[fill] (10,2) circle(.1);
\draw[fill] (10,4) circle(.1);
\draw (10,2) to [out=150,in=150] (10,4);
\draw[fill] (10,5) circle(.1);
\draw (10,4) -- (10,5);
\draw[fill] (10,6) circle(.1);
\draw[fill] (10,7) circle(.1);
\draw (10,5) to [out=150,in=150] (10,7);
\draw (10,6) -- (10,7);
\draw[fill] (11,4) circle(.1);
\draw (11,4) to [out=150,in=30] (10,6);
\node[below] at (11,4) {$c_1'^2U$};

\node[below] at (13,4) {$c_1c_1'U$};

\draw[fill] (13,4) circle(.1);
\draw[fill] (13,6) circle(.1);
\draw (13,4) to [out=150,in=150] (13,6);
\draw[fill] (13,7) circle(.1);
\draw (13,6) -- (13,7);
\draw[fill] (13,8) circle(.1);
\draw[fill] (13,9) circle(.1);
\draw (13,7) to [out=150,in=150] (13,9);
\draw (13,8) -- (13,9);
\draw[fill] (14,6) circle(.1);
\draw (14,6) to [out=150,in=30] (13,8);

\node at (-2,-10) {$=$};

\node[below] at (0,-12) {$UV$};

\draw[fill] (0,-12) circle(.1);

\draw[fill] (0,-11) circle(.1);

\draw (0,-12) -- (0,-11);

\draw[fill] (0,-10) circle(.1);

\draw[fill] (1,-9) circle(.1);

\draw[fill] (1,-8) circle(.1);
\draw (1,-9) -- (1,-8);

\draw (0,-12) to [out=150,in=150] (0,-10);

\draw (0,-11) to [out=30,in=150] (1,-9);

\draw (0,-10) to [out=30,in=150] (1,-8);

\draw[fill] (0,-9) circle(.1);

\draw[fill] (1,-7) circle(.1);

\draw[fill] (1,-6) circle(.1);
\draw (1,-7) -- (1,-6);
\draw (0,-10) -- (0,-9);

\draw (0,-9) to [out=30,in=150] (1,-7);
\draw (1,-8) to [out=30,in=30] (1,-6);

\node[below] at (2,-10) {$w_2UV$};

\draw[fill] (2,-10) circle(.1);
\draw[fill] (2,-9) circle(.1);
\draw (2,-10) -- (2,-9);
\draw[fill] (2,-8) circle(.1);
\draw (2,-10) to [out=150,in=150] (2,-8);
\draw[fill] (2,-7) circle(.1);
\draw (2,-8) -- (2,-7);
\draw[fill] (2,-6) circle(.1);
\draw[fill] (2,-5) circle(.1);
\draw (2,-7) to [out=150,in=150] (2,-5);
\draw (2,-6) -- (2,-5);
\draw[fill] (2,-4) circle(.1);
\draw (2,-6) to [out=30,in=30] (2,-4);
\draw[fill] (2,-3) circle(.1);
\draw (2,-4) -- (2,-3);

\node[below] at (4,-8) {$w_1^4UV$};

\draw[fill] (4,-8) circle(.1);

\draw[fill] (4,-7) circle(.1);

\draw (4,-8) -- (4,-7);

\draw[fill] (4,-6) circle(.1);

\draw[fill] (5,-5) circle(.1);

\draw[fill] (5,-4) circle(.1);
\draw (5,-5) -- (5,-4);

\draw (4,-8) to [out=150,in=150] (4,-6);

\draw (4,-7) to [out=30,in=150] (5,-5);

\draw (4,-6) to [out=30,in=150] (5,-4);

\draw[fill] (4,-5) circle(.1);

\draw[fill] (5,-3) circle(.1);

\draw[fill] (5,-2) circle(.1);
\draw (5,-3) -- (5,-2);
\draw (4,-6) -- (4,-5);

\draw (4,-5) to [out=30,in=150] (5,-3);
\draw (5,-4) to [out=30,in=30] (5,-2);

\node[below] at (6,-8) {$w_2'^2UV$};

\draw[fill] (6,-8) circle(.1);

\draw[fill] (6,-7) circle(.1);

\draw (6,-8) -- (6,-7);

\draw[fill] (6,-6) circle(.1);

\draw[fill] (7,-5) circle(.1);

\draw[fill] (7,-4) circle(.1);
\draw (7,-5) -- (7,-4);

\draw (6,-8) to [out=150,in=150] (6,-6);

\draw (6,-7) to [out=30,in=150] (7,-5);

\draw (6,-6) to [out=30,in=150] (7,-4);

\draw[fill] (6,-5) circle(.1);

\draw[fill] (7,-3) circle(.1);

\draw[fill] (7,-2) circle(.1);
\draw (7,-3) -- (7,-2);
\draw (6,-6) -- (6,-5);

\draw (6,-5) to [out=30,in=150] (7,-3);
\draw (7,-4) to [out=30,in=30] (7,-2);

\node[below] at (8,-7) {$w_2'w_3'UV$};

\draw[fill] (8,-7) circle(.1);

\draw[fill] (8,-6) circle(.1);

\draw (8,-7) -- (8,-6);

\draw[fill] (8,-5) circle(.1);

\draw[fill] (9,-4) circle(.1);

\draw[fill] (9,-3) circle(.1);
\draw (9,-4) -- (9,-3);

\draw (8,-7) to [out=150,in=150] (8,-5);

\draw (8,-6) to [out=30,in=150] (9,-4);

\draw (8,-5) to [out=30,in=150] (9,-3);

\draw[fill] (8,-4) circle(.1);

\draw[fill] (9,-2) circle(.1);

\draw[fill] (9,-1) circle(.1);
\draw (9,-2) -- (9,-1);
\draw (8,-5) -- (8,-4);

\draw (8,-4) to [out=30,in=150] (9,-2);
\draw (9,-3) to [out=30,in=30] (9,-1);

\node[below] at (10,-8) {$c_1c_1'UV$};

\draw[fill] (10,-8) circle(.1);

\draw[fill] (10,-7) circle(.1);

\draw (10,-8) -- (10,-7);

\draw[fill] (10,-6) circle(.1);

\draw[fill] (11,-5) circle(.1);

\draw[fill] (11,-4) circle(.1);
\draw (11,-5) -- (11,-4);

\draw (10,-8) to [out=150,in=150] (10,-6);

\draw (10,-7) to [out=30,in=150] (11,-5);

\draw (10,-6) to [out=30,in=150] (11,-4);

\draw[fill] (10,-5) circle(.1);

\draw[fill] (11,-3) circle(.1);

\draw[fill] (11,-2) circle(.1);
\draw (11,-3) -- (11,-2);
\draw (10,-6) -- (10,-5);

\draw (10,-5) to [out=30,in=150] (11,-3);
\draw (11,-4) to [out=30,in=30] (11,-2);

\node[below] at (12,-8) {$w_4'UV$};

\draw[fill] (12,-8) circle(.1);
\draw[fill] (12,-7) circle(.1);
\draw (12,-8) -- (12,-7);
\draw[fill] (12,-6) circle(.1);
\draw[fill] (12,-5) circle(.1);
\draw (12,-6) -- (12,-5);
\draw (12,-7) to [out=150,in=150] (12,-5);
\draw[fill] (12,-4) circle(.1);
\draw[fill] (12,-3) circle(.1);
\draw (12,-4) -- (12,-3);
\draw (12,-6) to [out=30,in=30] (12,-4);

\node[below] at (0,-24) {$c_1UV$};

\draw[fill] (0,-24) circle(.1);

\draw[fill] (0,-23) circle(.1);

\draw (0,-24) -- (0,-23);

\draw[fill] (0,-22) circle(.1);

\draw[fill] (1,-21) circle(.1);

\draw[fill] (1,-20) circle(.1);
\draw (1,-21) -- (1,-20);

\draw (0,-24) to [out=150,in=150] (0,-22);

\draw (0,-23) to [out=30,in=150] (1,-21);

\draw (0,-22) to [out=30,in=150] (1,-20);

\draw[fill] (0,-21) circle(.1);

\draw[fill] (1,-19) circle(.1);

\draw[fill] (1,-18) circle(.1);
\draw (1,-19) -- (1,-18);
\draw (0,-22) -- (0,-21);

\draw (0,-21) to [out=30,in=150] (1,-19);
\draw (1,-20) to [out=30,in=30] (1,-18);

\node[below] at (2,-22) {$w_1^2c_1UV$};

\draw[fill] (2,-22) circle(.1);

\draw[fill] (2,-21) circle(.1);

\draw (2,-22) -- (2,-21);

\draw[fill] (2,-20) circle(.1);

\draw[fill] (3,-19) circle(.1);

\draw[fill] (3,-18) circle(.1);
\draw (3,-19) -- (3,-18);

\draw (2,-22) to [out=150,in=150] (2,-20);

\draw (2,-21) to [out=30,in=150] (3,-19);

\draw (2,-20) to [out=30,in=150] (3,-18);

\draw[fill] (2,-19) circle(.1);

\draw[fill] (3,-17) circle(.1);

\draw[fill] (3,-16) circle(.1);
\draw (3,-17) -- (3,-16);
\draw (2,-20) -- (2,-19);

\draw (2,-19) to [out=30,in=150] (3,-17);
\draw (3,-18) to [out=30,in=30] (3,-16);

\node[below] at (4,-23) {$w_2c_1UV$};

\draw[color=red,->] (4,-23) -- (4,-22);

\draw[fill] (4,-22) circle(.1);
\draw[fill] (4,-21) circle(.1);
\draw (4,-22) -- (4,-21);
\draw[fill] (4,-20) circle(.1);
\draw (4,-22) to [out=150,in=150] (4,-20);
\draw[fill] (5,-20) circle(.1);
\draw[fill] (5,-19) circle(.1);
\draw (5,-20) -- (5,-19);
\draw (4,-21) to [out=30,in=150] (5,-19);
\draw[fill] (4,-19) circle(.1);
\draw (4,-20) -- (4,-19);
\draw[fill] (4,-18) circle(.1);
\draw[fill] (4,-17) circle(.1);
\draw (4,-18) -- (4,-17);
\draw (4,-19) to [out=150,in=150] (4,-17);
\draw[fill] (5,-18) circle(.1);
\draw (5,-20) to [out=30,in=30] (5,-18);
\draw[fill] (5,-17) circle(.1);
\draw (5,-18) -- (5,-17);
\draw[fill] (4,-16) circle(.1);
\draw (4,-18) to [out=30,in=30] (4,-16);
\draw[fill] (5,-16) circle(.1);
\draw[fill] (5,-15) circle(.1);
\draw (5,-16) -- (5,-15);
\draw (5,-17) to [out=150,in=150] (5,-15);
\draw[fill] (4,-15) circle(.1);
\draw (4,-16) -- (4,-15);
\draw[fill] (5,-14) circle(.1);
\draw[fill] (5,-13) circle(.1);
\draw (5,-14) -- (5,-13);
\draw (5,-16) to [out=30,in=30] (5,-14);

\node[below] at (6,-24) {$c_1'UV$};

\draw[fill] (6,-24) circle(.1);

\draw[fill] (6,-23) circle(.1);

\draw (6,-24) -- (6,-23);

\draw[fill] (6,-22) circle(.1);

\draw[fill] (7,-21) circle(.1);

\draw[fill] (7,-20) circle(.1);
\draw (7,-21) -- (7,-20);

\draw (6,-24) to [out=150,in=150] (6,-22);

\draw (6,-23) to [out=30,in=150] (7,-21);

\draw (6,-22) to [out=30,in=150] (7,-20);

\draw[fill] (6,-21) circle(.1);

\draw[fill] (7,-19) circle(.1);

\draw[fill] (7,-18) circle(.1);
\draw (7,-19) -- (7,-18);
\draw (6,-22) -- (6,-21);

\draw (6,-21) to [out=30,in=150] (7,-19);
\draw (7,-20) to [out=30,in=30] (7,-18);

\node[below] at (8,-22) {$w_1^2c_1'UV$};

\draw[fill] (8,-22) circle(.1);

\draw[fill] (8,-21) circle(.1);

\draw (8,-22) -- (8,-21);

\draw[fill] (8,-20) circle(.1);

\draw[fill] (9,-19) circle(.1);

\draw[fill] (9,-18) circle(.1);
\draw (9,-19) -- (9,-18);

\draw (8,-22) to [out=150,in=150] (8,-20);

\draw (8,-21) to [out=30,in=150] (9,-19);

\draw (8,-20) to [out=30,in=150] (9,-18);

\draw[fill] (8,-19) circle(.1);

\draw[fill] (9,-17) circle(.1);

\draw[fill] (9,-16) circle(.1);
\draw (9,-17) -- (9,-16);
\draw (8,-20) -- (8,-19);

\draw (8,-19) to [out=30,in=150] (9,-17);
\draw (9,-18) to [out=30,in=30] (9,-16);

\node[below] at (10,-23) {$w_2c_1'UV$};

\draw[color=red,->] (10,-23) -- (10,-22);

\draw[fill] (10,-22) circle(.1);
\draw[fill] (10,-21) circle(.1);
\draw (10,-22) -- (10,-21);
\draw[fill] (10,-20) circle(.1);
\draw (10,-22) to [out=150,in=150] (10,-20);
\draw[fill] (11,-20) circle(.1);
\draw[fill] (11,-19) circle(.1);
\draw (11,-20) -- (11,-19);
\draw (10,-21) to [out=30,in=150] (11,-19);
\draw[fill] (10,-19) circle(.1);
\draw (10,-20) -- (10,-19);
\draw[fill] (10,-18) circle(.1);
\draw[fill] (10,-17) circle(.1);
\draw (10,-18) -- (10,-17);
\draw (10,-19) to [out=150,in=150] (10,-17);
\draw[fill] (11,-18) circle(.1);
\draw (11,-20) to [out=30,in=30] (11,-18);
\draw[fill] (11,-17) circle(.1);
\draw (11,-18) -- (11,-17);
\draw[fill] (10,-16) circle(.1);
\draw (10,-18) to [out=30,in=30] (10,-16);
\draw[fill] (11,-16) circle(.1);
\draw[fill] (11,-15) circle(.1);
\draw (11,-16) -- (11,-15);
\draw (11,-17) to [out=150,in=150] (11,-15);
\draw[fill] (10,-15) circle(.1);
\draw (10,-16) -- (10,-15);
\draw[fill] (11,-14) circle(.1);
\draw[fill] (11,-13) circle(.1);
\draw (11,-14) -- (11,-13);
\draw (11,-16) to [out=30,in=30] (11,-14);

\end{tikzpicture}
\end{center}
\caption{The $\A_2(1)$-module structure of $\H^{*+1}(M\O(1),\Z_2)\otimes\H^{*+4}(M\SO(4),\Z_2)\otimes\H^*(\B\U(1),\Z_2)\otimes\H^*(\B\U(1),\Z_2)$ below degree 5.}
\label{fig:A_2(1)MO1MSO4BU1BU1}
\end{figure}

\begin{figure}[H]
\begin{center}
\begin{tikzpicture}
\node at (0,-1) {0};
\node at (1,-1) {1};
\node at (2,-1) {2};
\node at (3,-1) {3};
\node at (4,-1) {4};
\node at (5,-1) {5};
\node at (6,-1) {$t-s$};
\node at (-1,0) {0};
\node at (-1,1) {1};
\node at (-1,2) {2};
\node at (-1,3) {3};
\node at (-1,4) {4};
\node at (-1,5) {5};
\node at (-1,6) {$s$};

\draw[->] (-0.5,-0.5) -- (-0.5,6);
\draw[->] (-0.5,-0.5) -- (6,-0.5);

\draw[fill] (0,0) circle(0.05);

\draw[fill] (1.9,0) circle(0.05);
\draw[fill] (2,0) circle(0.05);
\draw[fill] (2.1,0) circle(0.05);

\draw (4,1) -- (5,2);
\draw (4,0) -- (5,1);
\draw[fill] (3.7,0) circle(0.05);
\draw[fill] (3.8,0) circle(0.05);
\draw[fill] (3.9,0) circle(0.05);
\draw[fill] (4.1,0) circle(0.05);
\draw[fill] (4.2,0) circle(0.05);
\draw[fill] (4.3,0) circle(0.05);
\draw[fill] (4.4,0) circle(0.05);

\draw[fill] (5,0) circle(0.05);

\end{tikzpicture}
\end{center}
\caption{$\Omega_*^{\Pin^-\times_{\Z_2}\Spin(4)\times\U(1)\times\U(1)}$.}
\label{fig:E_2Pin-Spin4Z2U1U1}
\end{figure}
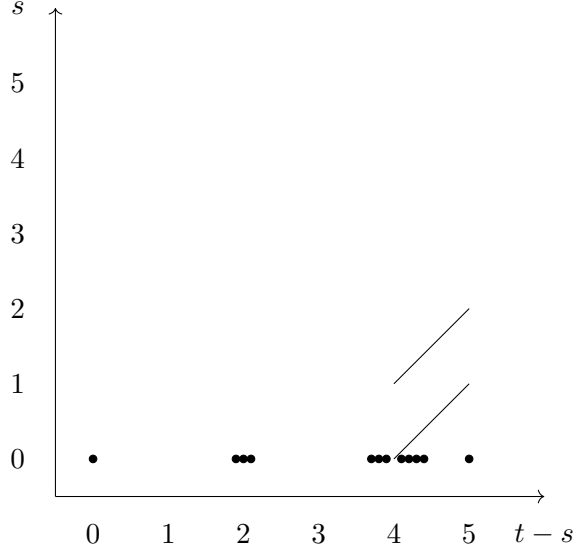

\begin{table}[H]
\centering
\begin{tabular}{ c c c}
\hline
\multicolumn{3}{c}{Bordism group}\\
\hline
$d$ & 
$\Omega^{\Pin^-\times_{\Z_2}\Spin(4)\times\U(1)\times\U(1)}_d$
& generators \\
\hline
0& $\Z_2$\\
1& $0$\\
2& $\Z_2^3$  & $w_1^2,c_1\mod2,c_1'\mod2$\\
3 & $0$\\
4 & $\Z_2^9$ & $w_1^4,w_4',w_2'^2,w_1^2c_1,w_1^2c_1',w_2c_1,w_2c_1',c_1c_1'\mod2,(w_1^3+w_1w_2'+w_3')\tilde\eta$ \\
5 & $\Z_2^3$  & $w_2'w_3',w_4'\tilde\eta, (w_1^3+w_1w_2'+w_3')\text{Arf}$ \\
\hline
\end{tabular}
\caption{Bordism group. Here $w_i$ is the Stiefel-Whitney class of the tangent bundle, $w_i'$ is the Stiefel-Whitney class of the $\SO(4)$ bundle, $c_1$ ($c_1'$) is the Chern class of the $\U(1)$ bundle, $\tilde\eta$ is the mod 2 index of 1d Dirac operator, Arf is the Arf invariant.
}
\label{table:Pin-Spin4Z2U1U1Bordism}
\end{table}

\subsection{Quark Gluon Plasma/Liquid
$\frac{([\SU(3)_V] \times \SU(3)_L \times  \SU(3)_R \times \U(1)_V)  \rtimes \Z_4^T
}{\Z_{3, V} \times \Z_{3, V} \times \Z_{2,V}^F} $
}

\subsubsection{$\frac{([\SU(3)_V] \times \SU(3)_L \times  \SU(3)_R \times \U(1)_V)  \times \Z_4^{TF}
}{\Z_{3, V} \times \Z_{3, V} \times \Z_{2,V}^F} $ as
$\Pin^+ \times_{\Z_2} \frac{\U(3)_L \times \U(3)_R }{(\Z_{3,V} \times \Z_{3,A} \times \U(1)_A)}$ and\\ 
$\frac{([\SU(3)_V] \times \SU(3)_L \times  \SU(3)_R \times \U(1)_V)  \times \Z_2^T
}{\Z_{3, V} \times \Z_{3, V} } $ as
$\Pin^- \times_{\Z_2} \frac{\U(3)_L \times \U(3)_R }{(\Z_{3,V} \times \Z_{3,A} \times \U(1)_A)}$}

Since the localization of $\Pin^{\pm} \times_{\Z_2} \frac{\U(3) \times \U(3) }{(\Z_3 \times \Z_3 \times \U(1))}$ and $\O\times\frac{\U(3) \times \U(3) }{(\Z_3 \times \Z_3 \times \U(1))}$ at the prime 3 are the same, the 3-torsion of $\Omega_d^{\Pin^{\pm} \times_{\Z_2} \frac{\U(3) \times \U(3) }{(\Z_3 \times \Z_3 \times \U(1))}}$ and $\Omega_d^{\O\times\frac{\U(3) \times \U(3) }{(\Z_3 \times \Z_3 \times \U(1))}}$ are the same, hence there is no 3-torsion in $\Omega_d^{\Pin^{\pm} \times_{\Z_2} \frac{\U(3) \times \U(3) }{(\Z_3 \times \Z_3 \times \U(1))}}$.

Since the localization of $\Pin^{\pm} \times_{\Z_2} \frac{\U(3) \times \U(3) }{(\Z_3 \times \Z_3 \times \U(1))}$ and $\Pin^{\pm} \times_{\Z_2} (\U(1)\times\SU(3)\times\SU(3))$ at the prime 2 are the same, the 2-torsion of $\Omega_d^{\Pin^{\pm} \times_{\Z_2} \frac{\U(3) \times \U(3) }{(\Z_3 \times \Z_3 \times \U(1))}}$ and $\Omega_d^{\Pin^c\times\SU(3)\times\SU(3)}$ are the same.

We have $MT(\Pin^c \times \SU(3)\times\SU(3))=M\Spin\wedge\Sigma^{-2}M\U(1)\wedge\Sigma^{-1}M\O(1)\wedge (\B\SU(3)\times\B \SU(3))_+$.

For $t-s<8$, since there is no odd torsion, we have the Adams spectral sequence
\bea
&&\Ext_{\A_2(1)}^{s,t}(\H^{*+2}(M\U(1),\Z_2)\otimes\H^{*+1}(M\O(1),\Z_2)\otimes\H^*(\B \SU(3),\Z_2)\otimes\H^*(\B \SU(3),\Z_2),\Z_2)\nn\\
&&\Rightarrow\Omega_{t-s}^{\Pin^c \times \SU(3)\times\SU(3)}.
\eea

We have
\bea
\H^*(\B\SU(3)\times\B \SU(3),\Z_2)=\Z_2[c_2,c_3,c_2',c_3']
\eea
where $c_i$ is the Chern class of the $\SU(3)$ bundle.

By Thom isomorphism,
\bea
\H^{*+2}(M\U(1),\Z_2)=\Z_2[c_1]U
\eea
where $c_1$ is the Chern class of the $\U(1)$ bundle and $U$ is the Thom class.

Also by Thom isomorphism,
\bea
\H^{*+1}(M\O(1),\Z_2)=\Z_2[w_1]V
\eea
where $w_1$ is the Stiefel-Whitney class of the $\O(1)$ bundle and $V$ is the Thom class.

The $\A_2(1)$-module structure of $\H^{*+2}(M\U(1),\Z_2)\otimes\H^{*+1}(M\O(1),\Z_2)\otimes\H^*(\B \SU(3),\Z_2)\otimes\H^*(\B \SU(3),\Z_2)$ below degree 6 and the $E_2$ page are shown in Figure \ref{fig:A_2(1)MU1MO1SU3SU3}, \ref{fig:E_2PincSU3SU3}.

\begin{figure}[H]
\begin{center}
\begin{tikzpicture}[scale=0.5]

\node[below] at (0,0) {$UV$};

\draw[fill] (0,0) circle(.1);
\draw[fill] (0,1) circle(.1);
\draw (0,0) -- (0,1);
\draw[fill] (0,2) circle(.1);
\draw (0,0) to [out=150,in=150] (0,2);
\draw[fill] (1,2) circle(.1);
\draw[fill] (1,3) circle(.1);
\draw (1,2) -- (1,3);
\draw (0,1) to [out=30,in=150] (1,3);
\draw[fill] (0,3) circle(.1);
\draw (0,2) -- (0,3);
\draw[fill] (0,4) circle(.1);
\draw[fill] (0,5) circle(.1);
\draw (0,4) -- (0,5);
\draw (0,3) to [out=150,in=150] (0,5);
\draw[fill] (1,4) circle(.1);
\draw (1,2) to [out=30,in=30] (1,4);
\draw[fill] (1,5) circle(.1);
\draw (1,4) -- (1,5);
\draw[fill] (0,6) circle(.1);
\draw (0,4) to [out=30,in=30] (0,6);
\draw[fill] (1,6) circle(.1);
\draw[fill] (1,7) circle(.1);
\draw (1,6) -- (1,7);
\draw (1,5) to [out=150,in=150] (1,7);
\draw[fill] (0,7) circle(.1);
\draw (0,6) -- (0,7);
\draw[fill] (1,8) circle(.1);
\draw[fill] (1,9) circle(.1);
\draw (1,8) -- (1,9);
\draw (1,6) to [out=30,in=30] (1,8);

\node[below] at (3,4) {$c_1^2UV$};

\draw[fill] (3,4) circle(.1);
\draw[fill] (3,5) circle(.1);
\draw (3,4) -- (3,5);
\draw[fill] (3,6) circle(.1);
\draw (3,4) to [out=150,in=150] (3,6);
\draw[fill] (4,6) circle(.1);
\draw[fill] (4,7) circle(.1);
\draw (4,6) -- (4,7);
\draw (3,5) to [out=30,in=150] (4,7);
\draw[fill] (3,7) circle(.1);
\draw (3,6) -- (3,7);
\draw[fill] (3,8) circle(.1);
\draw[fill] (3,9) circle(.1);
\draw (3,8) -- (3,9);
\draw (3,7) to [out=150,in=150] (3,9);
\draw[fill] (4,8) circle(.1);
\draw (4,6) to [out=30,in=30] (4,8);
\draw[fill] (4,9) circle(.1);
\draw (4,8) -- (4,9);
\draw[fill] (3,10) circle(.1);
\draw (3,8) to [out=30,in=30] (3,10);
\draw[fill] (4,10) circle(.1);
\draw[fill] (4,11) circle(.1);
\draw (4,10) -- (4,11);
\draw (4,9) to [out=150,in=150] (4,11);
\draw[fill] (3,11) circle(.1);
\draw (3,10) -- (3,11);
\draw[fill] (4,12) circle(.1);
\draw[fill] (4,13) circle(.1);
\draw (4,12) -- (4,13);
\draw (4,10) to [out=30,in=30] (4,12);

\node[below] at (6,4) {$c_2UV$};

\draw[fill] (6,4) circle(.1);
\draw[fill] (6,5) circle(.1);
\draw (6,4) -- (6,5);
\draw[fill] (6,6) circle(.1);
\draw (6,4) to [out=150,in=150] (6,6);
\draw[fill] (7,6) circle(.1);
\draw[fill] (7,7) circle(.1);
\draw (7,6) -- (7,7);
\draw (6,5) to [out=30,in=150] (7,7);
\draw[fill] (6,7) circle(.1);
\draw (6,6) -- (6,7);
\draw[fill] (6,8) circle(.1);
\draw[fill] (6,9) circle(.1);
\draw (6,8) -- (6,9);
\draw (6,7) to [out=150,in=150] (6,9);
\draw[fill] (7,8) circle(.1);
\draw (7,6) to [out=30,in=30] (7,8);
\draw[fill] (7,9) circle(.1);
\draw (7,8) -- (7,9);
\draw[fill] (6,10) circle(.1);
\draw (6,8) to [out=30,in=30] (6,10);
\draw[fill] (7,10) circle(.1);
\draw[fill] (7,11) circle(.1);
\draw (7,10) -- (7,11);
\draw (7,9) to [out=150,in=150] (7,11);
\draw[fill] (6,11) circle(.1);
\draw (6,10) -- (6,11);
\draw[fill] (7,12) circle(.1);
\draw[fill] (7,13) circle(.1);
\draw (7,12) -- (7,13);
\draw (7,10) to [out=30,in=30] (7,12);

\node[below] at (9,4) {$c_2'UV$};

\draw[fill] (9,4) circle(.1);
\draw[fill] (9,5) circle(.1);
\draw (9,4) -- (9,5);
\draw[fill] (9,6) circle(.1);
\draw (9,4) to [out=150,in=150] (9,6);
\draw[fill] (10,6) circle(.1);
\draw[fill] (10,7) circle(.1);
\draw (10,6) -- (10,7);
\draw (9,5) to [out=30,in=150] (10,7);
\draw[fill] (9,7) circle(.1);
\draw (9,6) -- (9,7);
\draw[fill] (9,8) circle(.1);
\draw[fill] (9,9) circle(.1);
\draw (9,8) -- (9,9);
\draw (9,7) to [out=150,in=150] (9,9);
\draw[fill] (10,8) circle(.1);
\draw (10,6) to [out=30,in=30] (10,8);
\draw[fill] (10,9) circle(.1);
\draw (10,8) -- (10,9);
\draw[fill] (9,10) circle(.1);
\draw (9,8) to [out=30,in=30] (9,10);
\draw[fill] (10,10) circle(.1);
\draw[fill] (10,11) circle(.1);
\draw (10,10) -- (10,11);
\draw (10,9) to [out=150,in=150] (10,11);
\draw[fill] (9,11) circle(.1);
\draw (9,10) -- (9,11);
\draw[fill] (10,12) circle(.1);
\draw[fill] (10,13) circle(.1);
\draw (10,12) -- (10,13);
\draw (10,10) to [out=30,in=30] (10,12);

\end{tikzpicture}
\end{center}
\caption{The $\A_2(1)$-module structure of $\H^{*+2}(M\U(1),\Z_2)\otimes\H^{*+1}(M\O(1),\Z_2)\otimes\H^*(\B \SU(3),\Z_2)\otimes\H^*(\B \SU(3),\Z_2)$ below degree 6.}
\label{fig:A_2(1)MU1MO1SU3SU3}
\end{figure}

\begin{figure}[H]
\begin{center}
\begin{tikzpicture}
\node at (0,-1) {0};
\node at (1,-1) {1};
\node at (2,-1) {2};
\node at (3,-1) {3};
\node at (4,-1) {4};
\node at (5,-1) {5};
\node at (6,-1) {$t-s$};
\node at (-1,0) {0};
\node at (-1,1) {1};
\node at (-1,2) {2};
\node at (-1,3) {3};
\node at (-1,4) {4};
\node at (-1,5) {5};
\node at (-1,6) {$s$};

\draw[->] (-0.5,-0.5) -- (-0.5,6);
\draw[->] (-0.5,-0.5) -- (6,-0.5);

\draw[fill] (0,0) circle(.05);

\draw (2,0) -- (2,1);
\draw (4,0) -- (4,2);
\draw[fill] (4.1,0) circle(.05);

\draw[fill] (4.2,0) circle(.05);
\draw[fill] (4.3,0) circle(.05);
\end{tikzpicture}
\end{center}
\caption{$\Omega_*^{\Pin^c \times \SU(3)\times\SU(3)}$.}
\label{fig:E_2PincSU3SU3}
\end{figure}
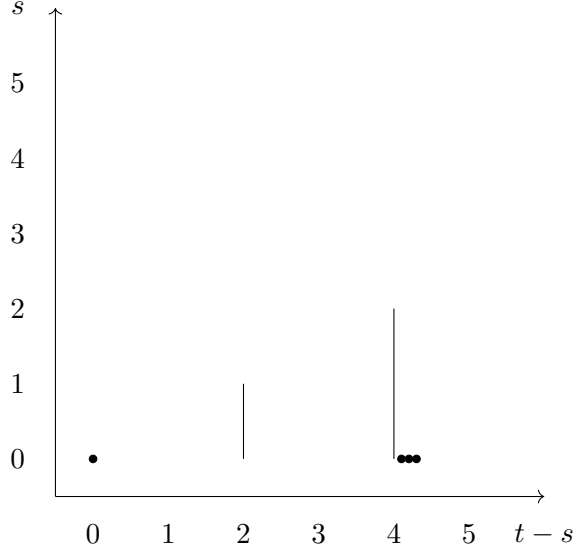

Combine the 2-torsion and 3-torsion results, we have
\begin{table}[H]
\centering
\begin{tabular}{ c c c}
\hline
\multicolumn{3}{c}{Bordism group}\\
\hline
$d$ & $\begin{matrix}
\Omega^{\Pin^c\times\SU(3)\times\SU(3) }_d\\=\Omega_d^{\Pin^+ \times_{\Z_2} \frac{\U(3) \times \U(3) }{(\Z_3 \times \Z_3 \times \U(1))}}\\=\Omega_d^{\Pin^- \times_{\Z_2} \frac{\U(3) \times \U(3) }{(\Z_3 \times \Z_3 \times \U(1))}}
\end{matrix}$
& generators \\
\hline
0& $\Z_2$\\
1& 0\\
2& $\Z_4$ & ABK mod 4\\
3 & 0\\
4 & $\Z_2^3\times\Z_8$ & $c_2\mod2,c_2'\mod2,c_1^2\mod2,(c_1\mod2)\text{ABK}$\\
5 & 0\\
\hline
\end{tabular}
\caption{Bordism group. Here $c_1$ is the Chern class of the $\U(1)$ bundle, $c_2$ ($c_2'$) is the Chern class of the $\SU(3)$ bundle, ABK is the Arf-Brown-Kervaire invariant.
}
\label{table:PincSU3SU3Bordism}
\end{table}

\section{Acknowledgements} 

The authors are listed in the alphabetical order by the standard convention.
JW thanks Kantaro Ohmori, Pavel Putrov, Nathan Seiberg, Edward Witten, and Yunqin Zheng for conversations.
JW are also grateful to many other colleagues for helpful conversations, 
at Institute for Advanced Study, MIT, Princeton University, Harvard University, National Taiwan University and University of Tokyo.
Part of this work was also reported by JW at the Aspen Center for Physics during
``Field Theory Dualities and Strongly Correlated Matter,'' March 18-24, 2018 under the title:
``Time Reversal, SU(N) Yang Mills, and Topological Phases'' 
at \href{http://www.its.caltech.edu/~motrunch/Aspen2018_Dualities/}{http://www.its.caltech.edu/$\sim$motrunch/Aspen2018\_Dualities/}.
ZW  acknowledges previous supports from NSFC grants 11431010 and 11571329. 
ZW is supported by the Shuimu Tsinghua Scholar Program.
JW was supported by
NSF Grant PHY-1606531 at IAS. 
This work is also supported by 
NSF Grant DMS-1607871 ``Analysis, Geometry and Mathematical Physics'' 
and Center for Mathematical Sciences and Applications at Harvard University.

\bibliographystyle{Yang-Mills.bst}
\bibliography{HAHS-III-QCD.bib}

\end{document}